\documentclass[superscriptaddress,onecolumn,english,floatfix,aps,pra,amsmath,amssymb,longbibliography]{revtex4-2}
\usepackage[T1]{fontenc}
\usepackage[utf8]{inputenc}
\usepackage{babel}
\usepackage{multirow}
\usepackage{comment}
\usepackage{color}
\usepackage{array}
\usepackage{enumitem}
\usepackage{times}
\usepackage[table,xcdraw]{xcolor}
\usepackage{xcolor}
\usepackage{epsfig}
\usepackage{subfigure}
\usepackage{amsmath}
\usepackage{amssymb}
\usepackage{natbib}
\usepackage{graphicx}
\usepackage{natbib}

\usepackage[colorlinks=true]{hyperref}
\hypersetup{citecolor=blue,linkcolor=blue,urlcolor=blue}

\begin{document}

\title{The time-dependent quantum harmonic oscillator: a pedagogical approach via the Lewis-Riesenfeld dynamical invariant method}

\author{Stanley S. Coelho}
\email{stanley.coelho@icen.ufpa.br}
\affiliation{Faculdade de F\'{i}sica, Universidade Federal do Par\'{a}, 66075-110, Bel\'{e}m, Par\'{a}, Brazil}	
\author{Lucas Queiroz}
\email{lucas.queiroz@ifpa.edu.br - Author to whom any correspondence should be addressed.}
\affiliation{Faculdade de F\'{i}sica, Universidade Federal do Par\'{a}, 66075-110, Bel\'{e}m, Par\'{a}, Brazil}
\affiliation{Instituto Federal de Educa\c{c}\~{a}o, Ci\^{e}ncia e Tecnologia do Par\'{a}, Campus Rural de Marab\'{a}, 68508-970, Marab\'{a}, Par\'{a}, Brazil}
\author{Danilo T. Alves}
\email{danilo@ufpa.br}
\affiliation{Faculdade de F\'{i}sica, Universidade Federal do Par\'{a}, 66075-110, Bel\'{e}m, Par\'{a}, Brazil}
	
\date{\today}

	
\begin{abstract}
In quantum mechanics courses, students often solve the Schrödinger equation for the harmonic oscillator with time-independent parameters.
However, time-dependent quantum harmonic oscillators are relevant in modeling several problems as, for instance, the description of quantum motion of particles in traps, shortcuts to adiabaticity, generation of squeezed states, as well as quantum scalar fields evolving in expanding universes.
In the present paper, we discuss, with a pedagogical approach, 
the quantum harmonic oscillator with time-dependent frequency via the Lewis-Riesenfeld dynamical invariant method, revisiting the main steps to obtain the wave function associated with this model, and briefly discussing the relation between this oscillator and the generation of squeezed states. 
As examples of didactic applications of time-dependent harmonic oscillators and the Lewis-Riesenfeld method in quantum mechanics courses, we solve the following problems: the calculation of the transition probability associated with a harmonic oscillator which undergoes jumps in its frequency, and the analysis of the dynamics of a quantum particle in a Paul trap. 
\end{abstract}
\maketitle

\section{Introduction}\label{sec:introduction}

The harmonic oscillator is one of the fundamental systems in physics, permeating from classical mechanics \cite{Landau-Lifishitz-Mechanics,Greiner-Classical-Mechanics,Marion-Classical-Dynamics,Goldstein-2011} to the quantum field theory \cite{Ryder-QFT,Greiner-Field-Quantization}. 
In the context of undergraduate courses, the harmonic oscillator with time-independent parameters is vastly
addressed in books in classical (e.g., in Refs. \cite{Landau-Lifishitz-Mechanics,Greiner-Classical-Mechanics,Marion-Classical-Dynamics}) and in quantum mechanics (e.g., in Refs. \cite{Sakurai-Quantum-Mechanics-2021,Cohen-Tannoudji-et-al-QM-vol-1,Griffiths-Quantum-Mechanics}). 
Regarding the harmonic oscillator with time-dependent parameters, here denominated as time-dependent harmonic oscillator (TDHO), although it is addressed in relatively few books of classical mechanics (e.g., in Refs. \cite{Landau-Lifishitz-Mechanics,Goldstein-2011}), it is still commonly discussed in classrooms \cite{Abe-EJP-2009}.
An example of a system of this nature is found in Ref. \cite{Nation-RMP-2012} (considering a motion with small oscillations), in which there is the illustrative situation of a child playing on a swing which, by bending its knees, periodically modulates its center of mass, resulting in amplification of the amplitude of the motion.
On the other hand, the TDHO
is more rarely covered in quantum mechanics textbooks (e.g.,  Refs. \cite{Griffiths-Quantum-Mechanics,Griffiths-Quantum-Mechanics-2018} include a problem involving the calculation of the transition probability of an oscillator that experiences a sudden jump in its frequency).

TDHOs are important in quantum physics \cite{Husimi-PTP-1953-II, Lewis-JMP-1969, Pedrosa-PRA-1997,Pedrosa-PRA-1997-singular-perturbation, Ciftja-JPA-1999, Nascimento-RBEF-2021}.
For example, TDHOs appear in the description of the interaction between a spinless charged quantum particle and a time-dependent external classical electromagnetic field \cite{Dodonov-PLA-1994, Aguiar-JMP-2016, Dodonov-JRLR-2018}, as well as in the quantization of the free electromagnetic field in nonstationary media \cite{Pedrosa-PRL-2009, Choi-PRA-2010, Pedrosa-PRA-2011, Unal-AP-2012, Lakehal-SR-2016}.
In atomic physics, the confinement of charged particles by oscillating electromagnetic fields can create effective potentials of the TDHO type \cite{Brown-PRL-1991,Alsing-PRL-2005, Menicucci-PRA-2007,Mihalcea-PS-2009, Mihalcea-AP-2022, Mihalcea-PR-2023,Mihalcea-Mathematics-2024}.
This device, known as a Paul trap, has many applications in modern physics \cite{Paul-RMP-1990}.
In this sense, TDHOs can also be used to study the quantum motion of particles in traps.
In the framework of quantum field theory in curved spacetimes, the modes of a scalar field, in an expanding universe described by the Friedmann-Robertson-Walker metric, behave like TDHOs whose frequency changes in time due to the evolution of spacetime (see Ref. \cite{Pedrosa-PRD-2004} and Refs. therein).
In quantum control techniques and shortcuts to adiabaticity, the oscillator frequency is deliberately manipulated to obtain the desired evolution of the system in finite time, without generating unwanted transitions \cite{Salamon-PCCP-2009, Chen-PRL-2010,Chen-PRA-2010,Stefanatos-PRA-2010, Choi-PRA-2012,Choi-PRA-2013,Kiely-JPB-2015,Odelin-RMP-2019,Beau-Entropy-2020, Huang-Chaos-2020,Dupays-PRR-2021,Grosso-Entropy-2023,Santos-EPJP-2024}.
In the context of quantum optics, squeezed states, used to overcome classical measurement limits (for example, in interferometry), are generated by manipulating systems modelled as TDHOs \cite{Janszky-OC-1986,Agarwal-PRL-1991,Janszky-PRA-1992,Degen-RMP-2017,Pezze-RMP-2018,Herrera-NJP-2023,Coelho-Entropy-2022,Coelho-PS-2024}. 
Therefore it is important to improve discussions on TDHOs in the context of undergraduate quantum mechanics courses, an effort that has already begun in some articles in physics teaching journals, such as those already mentioned \cite{Abe-EJP-2009,Andrews-AJP-1999,Nascimento-RBEF-2021}, and also in Refs. \cite{Castanos-AJP-2019,Leach-AJP-1978}.

While the energy is a constant of motion for the harmonic oscillator with time-independent parameters, it is not for the TDHO.
However, Lewis \cite{Lewis-PRL-1967,Lewis-JMP-1968} found a certain constant of motion for the TDHO, which is nowadays known
as the Lewis' dynamical invariant. 
Lutzky, in Ref. \cite{Lutzky-PLA-1978}, showed how to obtain such Lewis' invariant from Noether's theorem.
Considering that Lutzky's work was mostly known by specialists, but not by teachers, and also that the TDHO is an important issue in physics education, Abe \textit{et al.} \cite{Abe-EJP-2009} reviewed Lutzky's work in a didactic manner, interconnecting, in a simple way, the TDHO, the concept of invariant, and Noether’s theorem.

In the context of quantum mechanics, invariant operators, which can be viewed as generalizations of constants of motion, are useful for finding exact solutions of the Schrödinger equation for systems whose Hamiltonians explicitly depend on time \cite{Andrews-AJP-1999}.
As an example, one can cite the method proposed by Lewis and Riesenfeld \cite{Lewis-JMP-1969}, which uses invariant operators to solve the Schrödinger equation for generic time-dependent Hamiltonians and has already been applied in several problems in physics, including those which can be described by a Hamiltonian of a TDHO \cite{Lewis-JMP-1969, Malkin-PRD-1970, Hernandez-PLA-1980, Mendes-JPA-1980, Morales-JPA-1988-2, Gjaja-PRL-1992, Faccioli-PRL-1998, Pedrosa-PRL-2009, Chen-PRL-2010, Ban-PRL-2012, Qin-PRL-2013, Defenu-PRL-2018, Gu-AP-2024}.
We can also cite the method discussed by Andrews in Ref. \cite{Andrews-AJP-1999}, which obtained invariant operators similar to those used by Lewis and Reisenfeld \cite{Lewis-JMP-1969}, and, with a didactic approach, used them to solve the Schrödinger equation for time-dependent Hamiltonians that are quadratic in position and momentum, with the case of a Hamiltonian describing a TDHO being a particular application. 

There are also other methods, apart from the LR method, that make it possible to obtain exact solutions of the Schrödinger equation for TDHOs.
For example, Ref. \cite{Yeon-PRA-1993} analysed a TDHO with a general time-dependent quadratic Hamiltonian (which can represent both a TDHO with time-dependent mass and frequency, and one with only time-dependent frequency) via the path-integral method.
Something similar is done in Ref. \cite{Cheng-PLA-1985}.
In Ref. \cite{Brown-PRL-1991}, the wave function associated with a TDHO with time-dependent frequency was found using canonical transformations and, in Ref. \cite{Eguibar-JMP-2021}, was obtained using the Madelung-Bohm approach to quantum mechanics.
In Refs. \cite{Tibaduiza-BJP-2020,Tibaduiza-PS-2020,Tibaduiza-JPB-2021}, the TDHO with time-dependent frequency was studied using algebraic methods.
In this way, studying the TDHO also offers the opportunity to discuss important concepts, such as invariants operators, path-integral and algebraic methods, among others, which are increasingly relevant elements in contemporary physics, both theoretical and experimental.

In the present paper, aiming to contribute to the improvement of the discussions on TDHOs, especially in quantum mechanics courses, we didactically discuss the Lewis-Riesenfeld (LR) method \cite{Lewis-PRL-1967, Lewis-JMP-1968, Lewis-JMP-1969} and its application to the case of a harmonic oscillator with time-dependent frequency, describing the step-by-step use of the method.
As part of our didactic strategy, we map these steps into those already used by undergraduate students to solve the Schrödinger equation for the time-independent harmonic oscillator.
As particular applications, we discuss the calculation of the transition probability associated with a TDHO which undergoes a sudden jump in its frequency (the mentioned problem found in Refs. \cite{Griffiths-Quantum-Mechanics, Griffiths-Quantum-Mechanics-2018}), and we briefly discuss the dynamics of a quantum particle in a Paul trap, which is also modeled by a TDHO with a time-dependent frequency (from the classical point of view, the problem of particles in a Paul trap has been investigated in several physics teaching articles \cite{Winter-AJP-1991, Rueckner-AJP-1995, Ruby-AJP-1996, Nasse-EJP-2001, Johnson-AJP-2009, Wang-EJP-2013, Madsen-AJP-2014, Vinitsky-AJP-2015, Libbrecht-AJP-2018}). 
These calculations can be used as didactic examples of applications of the TDHOs and the LR method in quantum mechanics courses.

This paper is organized as follows. 
In Sec. \ref{sec:oscilador independente do tempo}, to prepare the reader for the subsequent sections, we briefly review the procedures for obtaining the wave function associated with the quantum harmonic oscillator with time-independent parameters, as is usually done in quantum mechanics textbooks.  
In Sec. \ref{sec:review}, we provide a detailed pedagogical review of the LR method for a generic Hamiltonian $\hat{H}(t)$. 
In Sec. \ref{sec:aplicação do método LR ao oscilador}, we provide a review of the application of the LR method to the case of a TDHO with a generic time-dependent frequency. More specifically:
In Sec. \ref{sec:wave-function}, we review the calculation of the wave function (via the LR method) associated with this model.
In Sec. \ref{sec:rel-tdho-se}, we discuss the generation of squeezed states by a TDHO with time-dependent frequency. 
In Sec. \ref{sec:griffiths}, we use these concepts to solve the problem of the calculation of the transition probability associated to a quantum harmonic oscillator that undergoes a sudden jump in its frequency.
In Sec. \ref{sec: Paul trap}, we also use all these tools to discuss the problem of the dynamics of a quantum particle in a Paul trap.  
In Sec. \ref{sec:final}, we present our final remarks.

\section{A brief review of the time-independent quantum harmonic oscillator}\label{sec:oscilador independente do tempo}

As will be seen later, although the solution of the time-dependent quantum harmonic oscillator needs a more robust method to lead with, it has several similarities with the time-independent correspondent problem.
In this way, we start by reviewing the problem of the time-independent quantum harmonic oscillator, which is a system that is widely studied in quantum mechanics courses \cite{Griffiths-Quantum-Mechanics,Sakurai-Quantum-Mechanics-2021}. 

First, we consider the Schrödinger equation for the one-dimensional time-independent harmonic oscillator,
\begin{eqnarray}
	\label{eq:eq-Schrodinger-est}
	i\hbar\frac{\partial}{\partial t}|\psi^{(0)}(t)\rangle=\hat{H}_{0}|\psi^{(0)}(t)\rangle,
\end{eqnarray}
with the Hamiltonian operator defined by
\begin{eqnarray}
	\label{eq:hamiltoniano-oscilador-TI}
	\hat{H}_{0}=\frac{\hat{p}^{2}}{2m_{0}}+\frac{1}{2}m_{0}\omega_{0}^{2}\hat{x}^{2},
\end{eqnarray}
where $m_0$ and $\omega_{0}$ are the time-independent mass and angular frequency, respectively. 
The position and momentum operators, given by $ \hat{x} $ and $ \hat{p}=-i\hbar\partial/\partial x $ (in position space), satisfy the canonical commutation relation $[\hat{x},\hat{p}]=i\hbar$ \cite{Griffiths-Quantum-Mechanics, Sakurai-Quantum-Mechanics-2021}.
Besides this, it is straight to note that, for the case of a time-independent harmonic oscillator, we have that $\hat{H}_{0}$ is an invariant operator, i.e., $d\hat{H}_{0}/dt=0$.

The solution of Eq. \eqref{eq:eq-Schrodinger-est} in the position space, given by $\Psi^{(0)}(x,t)=\langle x|\psi^{(0)}(t)\rangle$, can be written as \cite{Griffiths-Quantum-Mechanics, Sakurai-Quantum-Mechanics-2021}
\begin{eqnarray}
	\label{eq:Psi0}
	\Psi^{(0)}(x,t)=\sum_{n=0}^{\infty}C_{n}^{(0)}\Psi_{n}^{(0)}(x,t),
\end{eqnarray}
where $\Psi_{n}^{(0)}(x,t)$ are the wave functions associated to each allowed value of energy $ (E_n) $ of the system, and the coefficients $C_{n}^{(0)}$ depends only on the initial conditions. 
Using this solution into Eq. \eqref{eq:eq-Schrodinger-est} in the position space, and performing the method of separation of variables, one finds (see Appendix \ref{sec:append-time-indep-osci} for more details) \cite{Griffiths-Quantum-Mechanics, Sakurai-Quantum-Mechanics-2021}
\begin{eqnarray}
	\label{eq:Psi-n0-est-osc}
	\Psi_{n}^{(0)}(x,t)=\frac{1}{\sqrt{2^{n}n!}}\left(\frac{m_{0}\omega_{0}}{\pi\hbar}\right)^{\frac{1}{4}}\exp\left[-i\left(n+\frac{1}{2}\right)\omega_{0}t\right]\exp\left(-\frac{m_{0}\omega_{0}x^{2}}{2\hbar}\right){\cal H}_{n}\left(\sqrt{\frac{m_{0}\omega_{0}}{\hbar}}x\right).
\end{eqnarray}
We highlight that, for a TDHO [where $\hat{H}_{0}\to\hat{H}(t)$], the LR method prescribes that the wave function can be found by means of a sequence of steps similar to these ones used in the case of a time-independent harmonic oscillator (see Appendix \ref{sec:append-time-indep-osci}). 
In the time-independent case, $\hat{H}_{0}$ is an invariant operator, since  $d\hat{H}_{0}/dt=0$.
In the time-dependent case, the Hamiltonian is not an invariant operator anymore, but, according to the LR method, we can find another operator, denoted by $\hat{I}(t)$, so that this new operator is invariant, i.e., $d\hat{I}(t)/dt=0$. 
Thus, by finding $\hat{I}(t)$, the next steps are similar to those shown in Appendix \ref{sec:append-time-indep-osci}, as discussed next. 
%

\section{Lewis-Riesenfeld dynamical invariant method in quantum mechanics}
\label{sec:review}

Before discussing the solution of a TDHO, let us start reviewing the well known LR dynamical invariant method \cite{Lewis-PRL-1967, Lewis-JMP-1968, Lewis-JMP-1969}, which consists of a mechanism for finding exact solutions of the Schrödinger equation for systems whose Hamiltonians explicitly depend on time, i.e., for finding solutions of the following equation:
\begin{eqnarray}
	\label{eq:equação de Schrodinger}
	i\hbar\frac{\partial}{\partial t}|\psi(t)\rangle=\hat{H}(t)|\psi(t)\rangle.
\end{eqnarray}
The starting point for solving this equation consists in finding a Hermitian operator $ \hat{I}(t) $, called invariant operator.
Such operator has the property of being a constant of motion, i.e., $ d\hat{I}(t)/dt = 0 $, such that, from the Heisenberg equation \cite{Lewis-JMP-1969,Sakurai-Quantum-Mechanics-2021}, one has
\begin{eqnarray}
	\label{eq:equação de Heisenberg}
	\frac{\partial\hat{I}(t)}{\partial t}+\frac{1}{i\hbar}\bigl[\hat{I}(t),\hat{H}(t)\bigr]=0.
\end{eqnarray}
According to Ref. \cite{Lewis-JMP-1969}: ``The action of the invariant operator on a Schrödinger state vector produces another solution of the Schrödinger equation. This result is valid for any invariant, even if the latter involves the operation of time differentiation''.
To see this, one has to apply the operator $\partial\hat{I}(t)/\partial t$ on the state $ |\psi(t)\rangle $, so that, by using Eq. \eqref{eq:equação de Heisenberg}, one obtains
\begin{eqnarray}
	\frac{\partial\hat{I}(t)}{\partial t}|\psi(t)\rangle=-\frac{1}{i\hbar}\bigl[\hat{I}(t)\hat{H}(t)-\hat{H}(t)\hat{I}(t)\bigr]|\psi(t)\rangle.
	\label{eq:heisenberg-psi}
\end{eqnarray}
Using Eq. \eqref{eq:equação de Schrodinger} in this equation, one sees that
\begin{eqnarray}
	\label{eq:eq de Schrodinger I-psi}
	i\hbar\frac{\partial}{\partial t}\bigl[\hat{I}(t)|\psi(t)\rangle\bigr]=\hat{H}(t)\bigl[\hat{I}(t)|\psi(t)\rangle\bigr],
\end{eqnarray}
which shows that $ \hat{I}(t)|\psi(t)\rangle $ is also a solution of the Schrödinger equation.

Following Ref. \cite{Lewis-JMP-1969}: ``We  assume  that the invariant operator is  one of a  complete set of commuting observables, so that there  is  a  complete  set  of eigenstates  of $\hat{I}(t)$''.
In this way, let us denote the eigenvalues of $\hat{I}(t)$ as $\lambda $, which are time-independent (see Appendix \ref{sec:append-lambda} for more details) \cite{Lewis-JMP-1969}, and its orthonormal eigenstates associated to $\lambda $ as $ |\lambda,\zeta; t\rangle $, with $\zeta$ being all quantum numbers, other than $\lambda$, needed to specify the eigenstates. Therefore, one writes (for a discrete spectrum)
\begin{align}
	\label{eq:equação de autovalores para I}
	\hat{I}(t)|\lambda,\zeta;t\rangle= & \; \lambda|\lambda,\zeta;t\rangle,\\
	\label{eq:relação de completeza}
	\sum_{\lambda,\zeta}|\lambda,\zeta;t\rangle\langle\lambda,\zeta;t|= & \; 1,\\
	\label{eq:relação de ortonormalidade}
	\langle\lambda^{\prime},\zeta^{\prime};t|\lambda,\zeta;t\rangle= & \; \delta_{\lambda^{\prime},\lambda}\delta_{\zeta^{\prime},\zeta},
\end{align}
where $ \delta_{i,j} $ is the Kronecker delta (for a continuous spectrum, this is replaced by Dirac delta functions).

We remark that the eigenstates $ |\lambda,\zeta;t\rangle $ do not satisfy, in general, the Schrödinger equation, i.e.,
\begin{eqnarray}
	i\hbar\frac{\partial}{\partial t}|\lambda,\zeta;t\rangle\neq\hat{H}(t)|\lambda,\zeta;t\rangle.
\end{eqnarray}
Nevertheless, each state $ |\lambda,\zeta;t\rangle $ can be multiplied by a phase factor, resulting in a state $|\lambda,\zeta;t\rangle_{\alpha}$, such that
\begin{eqnarray}
	\label{eq:transformação de gauge}
	|\lambda,\zeta;t\rangle_{\alpha}=\exp[i\alpha_{\lambda,\zeta}(t)]|\lambda,\zeta;t\rangle,
\end{eqnarray}
with $ \alpha_{\lambda,\zeta}(t) $ being a real and time-dependent phase function. 
Furthermore, following Ref. \cite{Lewis-JMP-1969}: ``If the invariant does not involve  time  differentiation,  then we  are  able  to 	derive  a  simple  and  explicit  rule  for  choosing  the  phases of the eigenstates of $\hat{I}(t)$ such that these states  themselves  satisfy  the Schrödinger equation''.
In this way, assuming that the invariant operator does  not  involve  time differentiation \cite{Lewis-JMP-1969}, it is straightforward to see that the states $|\lambda,\zeta;t\rangle_{\alpha}$ are also eigenstates of $\hat{I}(t)$, and form an orthonormal basis. 
Besides this, if we require that the state vectors $|\lambda,\zeta;t\rangle_{\alpha} $ are solutions of the Schrödinger equation,
\begin{eqnarray}
	i\hbar\frac{\partial}{\partial t}|\lambda,\zeta;t\rangle_{\alpha}=\hat{H}(t)|\lambda,\zeta;t\rangle_{\alpha},
\end{eqnarray}
then, one obtains that the phase functions $ \alpha_{\lambda,\zeta}(t) $ have to obey the following equation:
\begin{eqnarray}
	\label{eq:forma geral função de fase}
	\frac{d\alpha_{\lambda,\zeta}(t)}{dt}=\langle\lambda,\zeta;t|\biggl[i\frac{\partial}{\partial t}-\frac{1}{\hbar}\hat{H}(t)\biggr]|\lambda,\zeta;t\rangle.
\end{eqnarray}
The solution $\alpha_{\lambda,\zeta}(t)$ of the equation above is known as the Lewis-Riesenfeld phase and, in turn, is composed of the dynamical and geometric phases. These phases have applications in different areas of physics (see, for example, Refs. \cite{Qian-PRL-1994,Shen-EPJD-2003,Dutta-PS-2022,Sen-PS-2024}).

The original problem of finding the eigenstates and eigenvalues of $\hat{H}(t)$ is replaced by the problem of finding the eigenstates and eigenvalues of $\hat{I}(t)$ and the phase functions $\alpha_{\lambda,\zeta}(t)$. 
By knowing the operator $\hat{I}(t)$ (which is, in general, obtained from an ansatz), one can calculate its eigenvalues and eigenstates by means of Eq. \eqref{eq:equação de autovalores para I}, and, with this, obtain the phase functions $ \alpha_{\lambda,\zeta}(t) $ from Eq. \eqref{eq:forma geral função de fase}. Therefore, the general solution $|\psi(t)\rangle$ of Eq. \eqref{eq:equação de Schrodinger} can be described as a linear combination of $ |\lambda,\zeta;t\rangle $ as 
\begin{eqnarray}
	\label{eq:solução geral da equação de Schrodinger}
	|\psi(t)\rangle=\sum_{\lambda,\zeta}C_{\lambda,\zeta}\exp[i\alpha_{\lambda,\zeta}(t)]|\lambda,\zeta;t\rangle,
\end{eqnarray}
in which the time-independent coefficients $ C_{\lambda,\zeta} $ depend on the initial conditions. 
For more details on the steps between Eqs. \eqref{eq:transformação de gauge}-\eqref{eq:solução geral da equação de Schrodinger}, see Appendix \ref{append: rel-I-H}.

In summary, the LR method consists of the following steps: 
\begin{enumerate}
	\item One proposes a Hermitian invariant operator $ \hat{I}(t) $, which satisfies Eq.  \eqref{eq:equação de Heisenberg}, and does not contain time derivatives;
	\item From Eq. \eqref{eq:equação de autovalores para I}, one calculates the eigenvalues $ \lambda $ and eigenstates $ |\lambda,\zeta; t\rangle $ associated to $ \hat{I}(t) $;
	\item Using Eq. \eqref{eq:forma geral função de fase}, one obtains the phase functions $ \alpha_{\lambda,\zeta}(t) $;
	\item Finally, by means of Eq. \eqref{eq:solução geral da equação de Schrodinger}, one calculates the state vector $ | \psi(t)\rangle $.
\end{enumerate}	

As a consistency check of Eq. \eqref{eq:solução geral da equação de Schrodinger}, let us discuss the situation in which we have a time-independent Hamiltonian $[\hat{H}(t)=\hat{H}_{\text{TI}}]$.
In this situation, one can propose that the invariant operator is given by $\hat{I}(t)=\hat{I}_{\text{TI}}$, with $\hat{I}_{\text{TI}}$ being a time-independent operator.
This results in $\partial\hat{I}(t)/\partial t = \partial\hat{I}_{\text{TI}}/\partial t=0$, which, from Eq. \eqref{eq:equação de Heisenberg}, leads to  $\bigl[\hat{I}(t),\hat{H}(t)\bigr] = \bigl[\hat{I}_{\text{TI}},\hat{H}_{\text{TI}}\bigr]=0$.  
As a consequence, the eigenstates of $ \hat{H}_{\text{TI}} $, denoted by $ |E_j\rangle $, are also eigenstates of $ \hat{I}_{\text{TI}} $ \cite{Sakurai-Quantum-Mechanics-2021}, which means that the state $|\lambda,\zeta;t\rangle$ coincides with $|E_j\rangle$, so that we can do the mapping $|\lambda,\zeta;t\rangle\to|E_j\rangle$ in Eq. \eqref{eq:solução geral da equação de Schrodinger} and write
\begin{eqnarray}
	\label{eq:psi-caso-estatico}
	|\psi(t)\rangle=\sum_{j}C_{j}\exp[i\alpha_{j}(t)]|E_j\rangle.
\end{eqnarray}
The same mapping has to be made in Eq. \eqref{eq:forma geral função de fase} in order to obtain the phase function for this case.
Since the state $ |E_j\rangle $ is time-independent $[\partial|E_j\rangle/\partial t=0]$ and using the eigenvalue equation $\hat{H}_{\text{TI}}|E_j\rangle = E_j|E_j\rangle$, where $E_j$ are the eigenvalues of $\hat{H}_{\text{TI}}$, one can find the solution of Eq. \eqref{eq:forma geral função de fase}, which is given by
\begin{eqnarray}
	\alpha_{j}(t)=-\frac{E_{j}t}{\hbar}.
\end{eqnarray}
By substituting this result into Eq. \eqref{eq:psi-caso-estatico}, one sees that we recover the general result for $|\psi(t)\rangle$ commonly presented in quantum mechanics textbooks \cite{Griffiths-Quantum-Mechanics, Sakurai-Quantum-Mechanics-2021} in the context of problems involving time-independent Hamiltonians.

By comparing the characteristics of a time-dependent problem within the context of the LR method with the corresponding time-independent case, we can find correlations, which are summarized in Table \ref{tab1}.
\begin{table}[h!]
	\centering
	\begin{tabular}{c|c|c|}
		\cline{2-3}
		\multicolumn{1}{l|}{}                               & \textbf{Time-independent case}                                          & \textbf{Time-dependent case}                                                                                                                                          \\ \hline
		\multicolumn{1}{|c|}{\textbf{Schrödinger equation}} & $i\hbar\partial_{t}|\psi(t)\rangle=\hat{H}_{\text{TI}}|\psi(t)\rangle$ & $i\hbar\partial_{t}|\psi(t)\rangle=\hat{H}(t)|\psi(t)\rangle$                                                                                                         \\ \hline
		\multicolumn{1}{|c|}{\textbf{Invariant operator}} & $\hat{H}_{\text{TI}}$ & $\hat{I}(t)$                                                                                                         \\ \hline
		\multicolumn{1}{|c|}{\textbf{Eigenvalue equation}}  & $\hat{H}_{\text{TI}}|E_{j}\rangle=E_{j}|E_{j}\rangle$                  & $\hat{I}(t)|\lambda,\zeta;t\rangle=\lambda|\lambda,\zeta;t\rangle$                                                                                                    \\ \hline
		\multicolumn{1}{|c|}{\textbf{General solution}}     & $|\psi(t)\rangle=\sum_{j}C_{j}\exp[i\alpha_{j}(t)]|E_j\rangle$          & $|\psi(t)\rangle=\sum_{\lambda,\zeta}C_{\lambda,\zeta}\exp[i\alpha_{\lambda,\zeta}(t)]|\lambda,\zeta;t\rangle$                                                        \\ \hline
		\multicolumn{1}{|c|}{\textbf{Phase function}}      & $\alpha_{j}(t)=-E_{j}t/\hbar$                                           & $\alpha_{\lambda,\zeta}(t)=\int_{0}^{t}dt^{\prime}\langle\lambda,\zeta;t^{\prime}|[i\partial_{t^{\prime}}-\hat{H}(t^{\prime})/\hbar]|\lambda,\zeta;t^{\prime}\rangle$ \\ \hline
	\end{tabular}
	\caption{Comparison between the time-independent and time-dependent cases.}
	\label{tab1}
\end{table}
%

\section{The solution of a quantum harmonic oscillator with time-dependent frequency via Lewis-Riesenfeld dynamical invariant method}
\label{sec:aplicação do método LR ao oscilador}

In the present section, we analyze the one-dimensional TDHO with time-dependent frequency $\omega(t)$, which is described by the Hamiltonian operator
\begin{eqnarray}
	\label{eq:hamiltoniano do oscilador}
	\hat{H}(t)=\frac{\hat{p}^{2}}{2m_{0}}+\frac{1}{2}m_{0}\omega(t)^{2}\hat{x}^{2}.
\end{eqnarray}
For didactic reasons, in this paper we focus only on the case of a time-dependent frequency, but, in the literature, one finds more general models that have been studied.
For example, in Refs. \cite{Pedrosa-PRA-1997,Pedrosa-PRA-1997-singular-perturbation}, one finds the discussion of the case where the frequency and the mass depend on time.
In Refs. \cite{Gao-PRA-1991,Kim-PRA-2001,Choi-JKPS-2004}, one finds the discussion of the driven generalized TDHO, where there are time-dependent parameters coupled to linear terms in $\hat{x}$ and $\hat{p}$, as well as external forces, in the Hamiltonian operator.
In Ref. \cite{Choi-JPSJ-2009}, one finds the discussion of the case where there are dependences of the type $1/\hat{x}$ and $1/\hat{x}^2$, multiplied by time-dependent factors, in the Hamiltonian operator. 
Finally, in Refs. \cite{Dutta-IJTP-2020,Dutta-PS-2021,Dutta-IJTP-2024}, there are cases where there are damped harmonic oscillators in a time-dependent noncommutative space, and, in Ref. \cite{Sobhani-JKPS-2016}, there is the case of systems with two coupled TDHOs.

\subsection{Calculation of the wave function}\label{subsec:wave function} \label{sec:wave-function}

\subsubsection{The invariant operator}\label{subsec:operador I}

Lewis and Reisenfeld \cite{Lewis-JMP-1969} assumed the existence, for a TDHO, of a Hermitian invariant operator $\hat{I}(t)$ of the quadratic form [similar to $\hat{H}(t)$] given by
\begin{eqnarray}
	\label{eq:operador I genérico}
	\hat{I}(t)=\frac{1}{2}[\eta_{1}(t)\hat{x}^{2}+\eta_{2}(t)\hat{p}^{2}+\eta_{3}(t)\{\hat{x},\hat{p}\}],
\end{eqnarray}
where $ \eta_{1}(t) $, $ \eta_{2}(t) $ and $ \eta_{3}(t) $ are real time-dependent functions [which ensures that $ \hat{I}(t) $ is Hermitian], and $ \left\{ \hat{x},\hat{p}\right\} $ is the anticommutator between the operators $ \hat{x} $ and $ \hat{p} $. 
(Note that, in the particular case where the mass and the frequency are independent of time,
the invariant operator becomes $\hat{I}(t)\to\hat{I}_{0}=\hat{H}_0/\omega_0
=\frac{1}{\omega_0}\left[\frac{1}{2}m_{0}\omega_{0}^{2}\hat{x}^{2}+\frac{1}{2m_{0}}\hat{p}^{2}
\right]$ \cite{Lewis-JMP-1969}.)
Since the invariant operator must be a constant of motion $ [d\hat{I}(t)/dt=0] $, then we require that this ansatz satisfies Eq. \eqref{eq:equação de Heisenberg}.
Thus, one finds that the invariant operator, associated to a TDHO with time-dependent frequency, is given by \cite{Lewis-PRL-1967,Lewis-JMP-1968,Lewis-JMP-1969,Pedrosa-PRA-1997,Pedrosa-PRA-1997-singular-perturbation} (for more details, see Appendix \ref{append:I})
\begin{eqnarray}
	\label{eq:I do oscilador}
	\hat{I}(t)=\frac{1}{2}\left\{ \left[\frac{\hat{x}}{\rho(t)}\right]^{2}+\left[\rho(t)\hat{p}-m_{0}\dot{\rho}(t)\hat{x}\right]^{2}\right\} ,
\end{eqnarray}
where the parameter $\rho(t)$ satisfies the so-called Ermakov-Pinney equation \cite{Prykarpatskyy-JMS-2018, Pinney-PAMS-1950,Lima-JMO-2009,Carinena-IJGMMP-2009}:
\begin{eqnarray}
	\label{eq:equação de Ermakov-Pinney}
	\ddot{\rho}(t)+\omega(t)^{2}\rho(t)=\frac{1}{m_{0}^{2}\rho(t)^{3}}.
\end{eqnarray}
In this way, by knowing how the frequency of the oscillator depend on time, one can obtain an expression for the invariant operator by only calculating the parameter $\rho (t)$ using Eq. \eqref{eq:equação de Ermakov-Pinney}.
We highlight that although Eq. \eqref{eq:equação de Ermakov-Pinney} is a nonlinear differential equation, it has a general exact solution (see Appendix \ref{append:rho}).

\subsubsection{The eigenvalues and eigenfunctions of the invariant operator}\label{subsec:autovalores e autofunções de I}

Following Ref. \cite{Lewis-JMP-1969}: ``The eigenstates and eigenvalues of the invariant operator $ \hat{I}(t) $ may be found by an operator technique that is completely analogous to the method introduced by Dirac for diagonalizing the Hamiltonian of a constant-frequency harmonic oscillator''.
In other words, from this point, the steps for finding the eigenstates and eigenvalues of $ \hat{I}(t) $ are similar to those discussed in Sec. \ref{sec:oscilador independente do tempo} and used to find the eigenstates and eigenvalues of $ \hat{H}_{0} $.
We start by defining the operators $\hat{a}(t)$ and $\hat{a}^{\dagger}(t)$ as \cite{Lewis-JMP-1969}
\begin{eqnarray}
	\label{eq:operador de aniquilação}
	\hat{a}(t)=\frac{1}{\sqrt{2\hbar}}\left\{ \frac{\hat{x}}{\rho(t)}+i\left[\rho(t)\hat{p}-m_{0}\dot{\rho}(t)\hat{x}\right]\right\} ,
\end{eqnarray}
and
\begin{eqnarray}
	\label{eq:operador de criação}
	\hat{a}^{\dagger}(t)=\frac{1}{\sqrt{2\hbar}}\left\{ \frac{\hat{x}}{\rho(t)}-i\left[\rho(t)\hat{p}-m_{0}\dot{\rho}(t)\hat{x}\right]\right\} .
\end{eqnarray}
From these expressions, one can obtain that $\hat{a}(t)$ and $\hat{a}^{\dagger}(t)$ satisfy the commutation relation $\left[\hat{a}(t),\hat{a}^{\dagger}(t)\right] = 1$. 
By knowing this, the next step is to express the operator $ \hat{I}(t) $ [Eq. \eqref{eq:I do oscilador}] in terms of $\hat{a}(t)$ and $\hat{a}^{\dagger}(t)$, which one obtains that
\begin{eqnarray}
	\label{eq:I do oscilador com a e a dagger}
	\hat{I}(t)=\hbar\left[\hat{a}^{\dagger}(t)\hat{a}(t)+\frac{1}{2}\right].
\end{eqnarray}

In the sequence, we define the number operator $\hat{N}(t)=\hat{a}^{\dagger}(t)\hat{a}(t)$, whose eigenvalue and eigenstate are given by $ n $ and $ |n;t\rangle $, respectively, so that one writes \cite{Lewis-JMP-1969} 
\begin{eqnarray}
	\label{eq:efeito da atuação do operador N em um autoestado do operador I}
	\hat{N}(t)|n;t\rangle=n|n;t\rangle,
\end{eqnarray}
with $ n \in \mathbb{N}$. 
Thus, by means of Eqs. \eqref{eq:equação de autovalores para I}, \eqref{eq:I do oscilador com a e a dagger} and \eqref{eq:efeito da atuação do operador N em um autoestado do operador I}, one sees that the eigenstate of the operator $ \hat{I}(t) $ can be simply written as $ |n;t\rangle $ [i.e., we can do the mapping $|\lambda,\zeta;t\rangle\to|n;t\rangle$], and its eigenvalues are given by $\lambda_{n}=\hbar(n+1/2) $.
Finally, defining the eigenstates of the invariant operator $\hat{I}(t)$ in position space, namely eigenfunctions, as $\Phi_{n}(x,t)=\langle x|n;t\rangle $, we have (see Appendix \ref{append:eq:phi_n} for more details) \cite{Pedrosa-PRA-1997,Pedrosa-PRA-1997-singular-perturbation}:
\begin{eqnarray}
	\label{eq:phi_n}
	\Phi_{n}(x,t)=\frac{1}{\sqrt{2^{n}n!}}\Phi_{0}(x,t){\cal H}_{n}\left[\frac{x}{\hbar^{\frac{1}{2}}\rho(t)}\right],
\end{eqnarray}
where ${\cal H}_{n}$ are the Hermite polynomials of order $n$, and
\begin{eqnarray}
	\label{eq:phi0}
	\Phi_{0}(x,t)=\biggl[\frac{1}{\pi\hbar\rho(t)^{2}}\biggr]^{\frac{1}{4}}\exp\left\{ \frac{im_{0}}{2\hbar}\left[\frac{\dot{\rho}(t)}{\rho(t)}+\frac{i}{m_{0}\rho(t)^{2}}\right]x^{2}\right\}.
\end{eqnarray}
Note that, like the operator $ \hat{I}(t) $, the eigenfunctions $ \Phi_{n}(x,t) $ are completely determined if we know the parameter $ \rho(t) $. 
Moreover, the eigenfunctions $ \Phi_{n}(x,t) $ clearly form a complete orthonormal set (as required), since the Hermite polynomials constitute a basis of Hilbert space \cite{Arfken-Mathematical-Physics-2003}.
The last step is to find the general solution of the Schrödinger equation [Eq. \eqref{eq:solução geral da equação de Schrodinger}].
For this, we have to calculate the phase functions $\alpha_{\lambda,\zeta}(t)$ for this case by means of Eq. \eqref{eq:forma geral função de fase}.

\subsubsection{The phase functions}

For the TDHO, we can do the mapping $\alpha_{\lambda,\zeta}(t) \to \alpha_{n}(t)$ in Eq. \eqref{eq:forma geral função de fase}, so that it can be written as
\begin{eqnarray}
	\label{eq:forma geral função de fase 2}
	\frac{d\alpha_{n}(t)}{dt}=\langle n;t|\biggl[i\frac{\partial}{\partial t}-\frac{1}{\hbar}\hat{H}(t)\biggr]|n;t\rangle.
\end{eqnarray}
Let us manipulate each term on the right side of the equation separately.
For the first term, $ i\langle n;t|\frac{\partial}{\partial t}|n;t\rangle $, we find (see Appendix \ref{append:alpha-n} for more details)
\begin{eqnarray}
	i\langle n;t|\frac{\partial}{\partial t}|n;t\rangle=-\frac{1}{2}\left(n+\frac{1}{2}\right)m_{0}\bigl[\rho(t)\ddot{\rho}(t)-\dot{\rho}(t)^{2}\bigr],
	\label{eq:sanduíche id/dt final}
\end{eqnarray}
and, for the second term on the right side of Eq. \eqref{eq:forma geral função de fase 2}, $ (1/\hbar)\langle n;t|\hat{H}(t)|n;t\rangle $, we obtain
\begin{eqnarray}
	\label{eq:sanduíche do H}	
	\frac{1}{\hbar}\langle n;t|\hat{H}(t)|n;t\rangle=\frac{1}{2m_{0}}\left(n+\frac{1}{2}\right)\bigl[\rho(t)^{-2}+m_{0}^{2}\dot{\rho}(t)^{2}+m_{0}^{2}\omega(t)^{2}\rho(t)^{2}\bigr].
\end{eqnarray}
Thus, substituting Eqs. \eqref{eq:sanduíche id/dt final} and \eqref{eq:sanduíche do H} into Eq. \eqref{eq:forma geral função de fase 2}, and using Eq. \eqref{eq:equação de Ermakov-Pinney}, one obtains that the phase function $\alpha_{n}(t)$ is given by \cite{Lewis-PRL-1967, Lewis-JMP-1969, Pedrosa-PRA-1997, Pedrosa-PRA-1997-singular-perturbation}:
\begin{eqnarray}
	\label{eq:alpha_n}
	\alpha_{n}(t)=-\left(n+\frac{1}{2}\right)\int_{0}^{t}\frac{dt^{\prime}}{m_{0}\rho\left(t^{\prime}\right)^{2}}.
\end{eqnarray}
%

\subsubsection{The general solution of the Schrödinger equation}

For the TDHO, Eq. \eqref{eq:solução geral da equação de Schrodinger} is rewritten as $|\psi(t)\rangle=\sum_{n=0}^{\infty}C_{n}\exp[i\alpha_{n}(t)]|n;t\rangle$.
By applying $ \langle x| $ in both sides of this equation, one obtains $\Psi(x,t)=\sum_{n=0}^{\infty}C_{n}\Psi_{n}(x,t)$, where $\Psi(x,t)=\langle x|\psi(t)\rangle$ and $\Psi_{n}(x,t)=\exp[i\alpha_{n}(t)]\Phi_{n}(x,t)$.
In this way, using Eqs. \eqref{eq:phi_n} and \eqref{eq:alpha_n}, one finally finds the wave function $ \Psi_{n}(x,t) $, which is given by \cite{Pedrosa-PRA-1997,Pedrosa-PRA-1997-singular-perturbation,Ciftja-JPA-1999,Pepore-ScienceAsia-2006}
\begin{align}
	\nonumber\Psi_{n}(x,t)= & \;\frac{1}{\sqrt{2^{n}n!\pi^{\frac{1}{2}}\hbar^{\frac{1}{2}}\rho(t)}}\exp\left[-i\left(n+\frac{1}{2}\right)\int_{0}^{t}\frac{dt^{\prime}}{m_{0}\rho\left(t^{\prime}\right)^{2}}\right]\\
	& \times\exp\left\{ \frac{im_{0}}{2\hbar}\left[\frac{\dot{\rho}(t)}{\rho(t)}+\frac{i}{m_{0}\rho(t)^{2}}\right]x^{2}\right\} {\cal H}_{n}\left[\frac{x}{\hbar^{\frac{1}{2}}\rho(t)}\right].
	\label{eq:função de onda do oscilador}
\end{align}
One can conclude that the initial problem is reduced to simply obtain the solution of the parameter $ \rho(t) $ by means of Eq. \eqref{eq:equação de Ermakov-Pinney}, which, although is a nonlinear differential equation, has an exact solution.
This is equivalent to solving a classical harmonic oscillator with time-dependent frequency (see Appendix \ref{append:rho} for more details), as Husimi concluded in 1953 \cite{Husimi-PTP-1953-II}.
%

\subsubsection{Static case}\label{sec:static case}

For the particular case where the mass and the frequency of the oscillator are independent of time [$m(t)=m_0$ and $\omega(t)=\omega_0$], the parameter $ \rho(t) $ takes the form \cite{Lewis-JMP-1969,Pedrosa-PRA-1997-singular-perturbation,Ciftja-JPA-1999}
\begin{eqnarray}
	\label{eq:rho0}
	\rho(t)=\rho_{0}=\frac{1}{\sqrt{m_{0}\omega_{0}}}.
\end{eqnarray}
In this case, it is immediate to see that the Hamiltonian $\hat{H}(t)$ falls back to the Hamiltonian $\hat{H}_0$ of a time-independent oscillator, and that the invariant operator becomes $\hat{I}(t)=\hat{I}_0=\hat{H}_0/\omega_0$ \cite{Lewis-JMP-1969}, and the operators $\hat{a}(t)$ and $\hat{a}^{\dagger}(t)$ become, respectively, $\hat{a}_0$ and $\hat{a}_0^{\dagger}$ [Eqs. \eqref{eq:a0} and \eqref{eq:a0 dagger}].
Thus, the wave function ${\Psi_{n}(x,t)}$ recover the wave function ${\Psi_{n}^{\left(0\right)}(x,t)}$ related to the time-independent oscillator, given by Eq. \eqref{eq:Psi-n0-est-osc}, as expected.
Similar to the previous section, by comparing the characteristics of a TDHO problem within the context of the LR method with the corresponding time-independent case, we can find correlations, which are summarized in Table \ref{tab2}.
\begin{table}[h!]
	\centering
	\begin{tabular}{c|c|c|}
		\cline{2-3}
		\multicolumn{1}{l|}{}                               & \textbf{Time-independent case}                                          & \textbf{Time-dependent case}                                                                                                                                          \\ \hline
		\multicolumn{1}{|c|}{\textbf{Schrödinger equation}} & $i\hbar\partial_{t}|\psi(t)\rangle=\hat{H}_{0}|\psi(t)\rangle$ & $i\hbar\partial_{t}|\psi(t)\rangle=\hat{H}(t)|\psi(t)\rangle$                                                                                                         \\ \hline
		\multicolumn{1}{|c|}{\textbf{Invariant operator}} & $\hat{H}_{0}$ given in Eq. \eqref{eq:hamiltoniano-oscilador-TI} & $\hat{I}(t)$ given in Eq. \eqref{eq:I do oscilador}                                                                                                         \\ \hline
		\multicolumn{1}{|c|}{\textbf{Eigenvalue equation}}  & $\hat{H}_{0}\Phi_{n}^{(0)}(x,t)=\hbar\omega_{0}\left(n+1/2\right)\Phi_{n}^{(0)}(x,t)$                  & $\hat{I}(t)\Phi_{n}(x,t)=\hbar\left(n+1/2\right)\Phi_{n}(x,t)$                                                                                                    \\ \hline
		\multicolumn{1}{|c|}{\textbf{Eigenfunction}}    & $\Phi_{n}^{(0)}(x,t)$ given by Eq. \eqref{eq:Phi-n0-est} & $\Phi_{n}(x,t)$ given by Eq. \eqref{eq:phi_n}          \\ \hline
		\multicolumn{1}{|c|}{\textbf{General solution}}     & $\Psi^{(0)}(x,t)=\sum_{n=0}^{\infty}C_{n}^{(0)}\exp[i\alpha_{n}^{(0)}(t)]\Phi_{n}^{(0)}(x,t)$          & $\Psi(x,t)=\sum_{n=0}^{\infty}C_{n}\exp[i\alpha_{n}(t)]\Phi_{n}(x,t)$                                                        \\ \hline
		\multicolumn{1}{|c|}{\textbf{Phase function}}   & $\alpha_{n}^{(0)}(t)=-(n+1/2)\omega_{0}t$                                    & $\alpha_{n}(t)=-(n+1/2)\int_{0}^{t}dt^{\prime}/[m_{0}\rho(t^{\prime})^{2}]$ \\ \hline
	\end{tabular}
	\caption{Comparison between the time-independent and time-dependent quantum harmonic oscillators.}
	\label{tab2}
\end{table}
%

\subsection{The time-dependent quantum harmonic oscillator and the generation of squeezed states}
\label{sec:rel-tdho-se}

In quantum physics, it is not possible to know the exact position and momentum of a particle at the same time, due to Heisenberg's uncertainty principle. 
This principle establishes a fundamental limit to the precision with which these two quantities can be measured simultaneously, which is given by $\sigma_{\hat{x}}\sigma_{\hat{p}}\geq\hbar/2$ (for the ground state) \cite{Griffiths-Quantum-Mechanics,Sakurai-Quantum-Mechanics-2021}, where $\sigma_{\hat{x}}$ and $\sigma_{\hat{p}}$ are the standard deviation associated with the position and momentum measurements, respectively.
In the case of the quantum harmonic oscillator, when its frequency is modulated in a nonadiabatic way (that is, depends explicitly on the time), the quantum state, which was initially in equilibrium, is transformed into a so-called “squeezed” state, as pointed out in Refs. \cite{Janszky-OC-1986, Pedrosa-PRA-1997, Pedrosa-PRA-1997-singular-perturbation, Pedrosa-PRA-2011, Tibaduiza-BJP-2020, Xin-PRL-2021}. 
In this sense, TDHOs provide a natural setting for studying these states.
The squeezed states occur when the variance of an observable (for example, position) becomes smaller than its value for a coherent state, leading to an increase in the variance of the complementary observable (for example, momentum) in such a way that the Heisenberg's uncertainty principle is still satisfied \cite{Guerry-Quantum-Optics-2005, Tibaduiza-BJP-2020}.
These states are fundamental for applications in quantum metrology, quantum sensing and quantum computing \cite{Pezze-RMP-2018,Degen-RMP-2017}. 

To illustrate how to quantify the squeezing in an oscillator, it can be defined the parameters that characterize when the oscillator states are squeezed, namely the squeezing parameter $r(t)$ and the squeezing phase $\varphi(t)$, given respectively by (for more details, see Refs.  \cite{Daneshmand-CTP-2017,Tibaduiza-JPB-2021,Coelho-PS-2024}) 
\begin{eqnarray}
	r(t)=\cosh^{-1}\{[\lambda(t)]^{\frac{1}{2}}\},
	\label{eq:r}
\end{eqnarray}
where
\begin{eqnarray}
	\lambda(t)=\frac{m_{0}^{2}\dot{\rho}(t)^{2}+\rho(t)^{-2}+m_{0}^{2}\omega(t)^{2}\rho(t)^{2}+2m_{0}\omega(t)}{4m_{0}\omega(t)},
	\label{eq:lambda(t)}
\end{eqnarray}
and
\begin{eqnarray}
	\varphi(t)=\cos^{-1}\left\{ \frac{1+m_{0}\omega(t)\rho(t)^{2}-2\cosh^{2}[r(t)]}{2\sinh[r(t)]\cosh[r(t)]}\right\}.
	\label{eq:phi}
\end{eqnarray}
Note that, when the oscillator is time-independent, the parameter $ \rho(t) $ is given by Eq. \eqref{eq:rho0} and when substituting it into Eqs. \eqref{eq:r}-\eqref{eq:phi}, one obtains that the squeezing parameter is null and the squeezing phase is undefined. 
Thus, for a time-independent harmonic oscillator, there are no squeezed states, as expected.

Taking into account the wave function $ \Psi_{n}(x,t) $ [Eq. \eqref{eq:função de onda do oscilador}], we can define the expected value of a given observable $\hat{{\cal O}}$ in the state $\Psi_{n}(x,t)$ for any $t\geq0$ as \cite{Sakurai-Quantum-Mechanics-2021, Griffiths-Quantum-Mechanics}
\begin{eqnarray}
	\label{eq:<O>}
	\langle\hat{{\cal O}}\rangle(n,t)=\int_{-\infty}^{+\infty}dx\Psi_{n}^{*}(x,t)\hat{{\cal O}}\Psi_{n}(x,t).
\end{eqnarray}
Using this equation and the definition of the variance associated with the operator $\hat{{\cal O}}$, namely  $\sigma_{\hat{{\cal O}}}^{2}(n,t)=[\langle\hat{{\cal O}}^{2}\rangle(n,t)]-[\langle\hat{{\cal O}}\rangle(n,t)]^{2}$, we have, for the position and momentum operators, that \cite{Coelho-PS-2024}
\begin{align}
	\label{eq:<x^{2}>}
	\sigma_{\hat{x}}^{2}(n,t)\;= & \;\,\bigl\{\cosh^{2}[r(t)]+2\sinh[r(t)]\cosh[r(t)]\cos[\varphi(t)]+\sinh^{2}[r(t)]\bigr\}\biggl(n+\frac{1}{2}\biggr)\frac{\hbar}{m_{0}\omega(t)},\\
	\label{eq:<p^{2}>}
	\sigma_{\hat{p}}^{2}(n,t)\;= & \;\,\bigl\{\cosh^{2}[r(t)]-2\sinh[r(t)]\cosh[r(t)]\cos[\varphi(t)]+\sinh^{2}[r(t)]\bigr\}\biggl(n+\frac{1}{2}\biggr)\hbar m_{0}\omega(t).
\end{align}
As discussed, in the time-independent case [$\omega(t)=\omega_{0}$] we have $ r(t)=0 $, so that Eqs. \eqref{eq:<x^{2}>} and \eqref{eq:<p^{2}>} recover the usual variances of a time-independent harmonic oscillator \cite{Sakurai-Quantum-Mechanics-2021}. 
If we choose, for example and simplicity, $\varphi(t)=0$ in Eqs. \eqref{eq:<x^{2}>} and \eqref{eq:<p^{2}>}, we obtain
\begin{eqnarray}
	\label{eq:var-x-p-phi=0}	
	\sigma_{\hat{x}}^{2}(n,t)|_{\varphi(t)=0}=\exp[2r(t)]\biggl(n+\frac{1}{2}\biggr)\frac{\hbar}{m_{0}\omega(t)},\;\;\;\sigma_{\hat{p}}^{2}(n,t)|_{\varphi(t)=0}=\exp[-2r(t)]\biggl(n+\frac{1}{2}\biggr)\hbar m_{0}\omega(t).
\end{eqnarray}
From Eqs. \eqref{eq:var-x-p-phi=0}, one can see that while the ratio $\sigma_{\hat{x}}^{2}(n,t)|_{\varphi(t)=0}/\left[(n+1/2)\frac{\hbar}{m_{0}\omega(t)}\right]$ (involving the variance in the position) enhances exponentially with time, the ratio $\sigma_{\hat{p}}^{2}(n,t)|_{\varphi(t)=0}/\left[(n+1/2)\hbar m_{0}\omega(t)\right]$ (involving the variance in the momentum) decreases (or is squeezed) exponentially. 

Furthermore, it is known that the system subjected to a TDHO potential can also make transitions between different states.
In this sense, the probability $P(t)_{0\to\nu}$ of the oscillator making a transition from its ground state to an excited state $\nu$ is defined by \cite{Kim-OC-1989,Tibaduiza-BJP-2020,Guerry-Quantum-Optics-2005}
\begin{eqnarray}
	\label{eq:dist-prob-fótons}	
	P(t)_{0\to\nu}=\frac{\nu!\tanh^{\nu}[r(t)]}{2^{\nu}[(\nu/2)!]^{2}\cosh[r(t)]},
\end{eqnarray} 
where $ \nu=0,2,4,... \;$. 
When we consider $\nu=0$ in this equation, one has the probability of persistence in the fundamental state, $ P_{p}(t)=1/\cosh[r(t)] $.
With this, one can also obtain the probability of excitation, which is given by $P_{e}(t)=1-P_{p}(t)$ \cite{Tibaduiza-BJP-2020}.
We will apply these concepts in the following sections.

Now, let us discuss an analogy between a quantum and a classical TDHO. More specifically,
the corresponding in the classical case of the generation of the squeezed states in the quantum.
To this end, let us consider the scenario of the child playing on a swing in a park,
as mentioned in Sec. \ref{sec:introduction}. 
If the child bends the knees and then extends the legs at the right time (see Fig. 1 of Ref. \cite{Nation-RMP-2012}), energy is transferred to the system child + swing (through the modulation of the center of mass). 
That situation create a pendulum with effective variable length over time.
Mathematically, for motions with small amplitudes, this system is represented by a classical harmonic oscillator with an effective time-dependent oscillation frequency, which satisfies the equation \cite{Landau-Lifishitz-Mechanics}
\begin{eqnarray}
	\ddot{\Theta}(t)+\Omega(t)^{2}\Theta(t)=0,
\end{eqnarray}
where $\Theta(t)$ is the angular displacement, and the function $\Omega(t)$ is the oscillation frequency, which is periodic in time [$\Omega(t+T)=\Omega(t)$, where $T$ is the oscillation period].
When the swing frequency is modulated as twice the natural oscillation frequency $\Omega_{0}\equiv\Omega(t)|_{t=0}$, for example \cite{Landau-Lifishitz-Mechanics},
\begin{eqnarray}
	\Omega(t)^{2}=\Omega_{0}^{2}\left\{ 1+h\cos\left[\left(2\Omega_{0}+\epsilon\right)t\right]\right\},\;\;\;h\ll1,\;\;\;\epsilon\ll\Omega_{0},
\end{eqnarray}
we have the so-called parametric resonance (or parametric amplification \cite{Nation-RMP-2012}), such that  $\Theta(t)$ assumes the form \cite{Landau-Lifishitz-Mechanics}
\begin{eqnarray}
	\label{eq:sol-Theta}	
	\Theta(t)=\Theta_{0}^{(1)}\exp(st)\cos\left[\left(\Omega_{0}+\frac{\epsilon}{2}\right)t\right]+\Theta_{0}^{(2)}\exp(-st)\sin\left[\left(\Omega_{0}+\frac{\epsilon}{2}\right)t\right],
\end{eqnarray}
in which $\Theta_{0}^{(1)}$ and $\Theta_{0}^{(2)}$ are constants (particularly, $\Theta_{0}^{(1)}$ is the initial amplitude of the motion), $s=\frac{1}{2}\sqrt{\left(\frac{h\Omega_{0}}{2}\right)^{2}-\epsilon^{2}}$, and $-\frac{h\Omega_{0}}{2}<\epsilon<\frac{h\Omega_{0}}{2}$ (for $\epsilon$ to be a real quantity).
In this way, one of the motion components [the first term on the right hand side of Eq. \eqref{eq:sol-Theta}] increases exponentially (the amplitude increases), while the other decreases exponentially (or is ``squeezed''). 
Note that this is analogous to what happens in the quantum case shown in Eqs. \eqref{eq:var-x-p-phi=0}, where the ratio related to the variance in the position increases exponentially with time, whereas the ratio related to the variance in momentum decreases (or is squeezed) exponentially.
For a more detailed discussion on this analogy, see Ref. \cite{Nation-RMP-2012}.

\subsection{Application to an undergraduate and postgraduate problem}\label{sec:griffiths}

In this section, by means of the results presented in previous sections, we solve the problem 2.14 of Ref. \cite{Griffiths-Quantum-Mechanics} (or the problem 11.19 of Ref. \cite{Griffiths-Quantum-Mechanics-2018}), which reads as follows: ``A particle is in the ground state of the harmonic oscillator with classical frequency $\omega_{0}$, when suddenly the spring constant quadruples, so $\omega_{1}=2\omega_{0}$, without initially changing the wave function. What is the probability that a measurement of the energy would still return the value $\hbar\omega_{0}/2$? What is the probability of getting $\hbar\omega_{0}$?''

\begin{figure}[h!]
	\epsfig{file=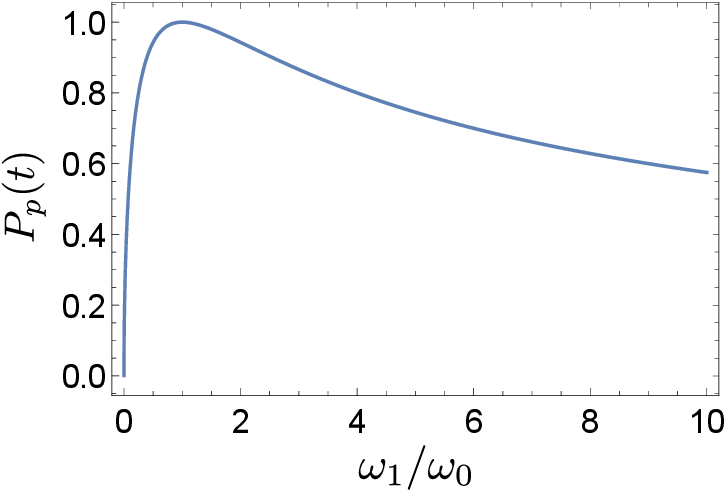, width=0.4 \linewidth}
	\caption{Behavior of $P_{p}(t)$ given by Eq. \eqref{eq:prob-p-griffiths}, as a function of $\omega_{1}/\omega_{0}$.}	
	\label{fig:behavior-prob-griffiths}
\end{figure}

Let us start discussing this problem considering a general value for $ \omega_{1} $.
In a general way, this problem can be modeled as a TDHO with a time-dependent frequency defined by
\begin{eqnarray}
	\omega(t)=\begin{cases}
		\omega_{0}, & t<0,\\
		\omega_{1}, & t\geq0.
	\end{cases}
\end{eqnarray}
In this case, the energy spectrum in the interval $t\geq0$ simply becomes $E_{n}^{\prime}=\hbar\omega_{1}\left(n+1/2\right)$.
Besides this, as discussed, quantum harmonic oscillators with time-dependent parameters describe squeezed states.
To calculate the squeezing parameter $ r(t) $, we first have to calculate the parameter $ \rho(t) $, which is calculated from the Ermakov-Pinney equation [Eq. \eqref{eq:equação de Ermakov-Pinney}].
One can find that the solution of the Ermakov-Pinney equation for the interval $t\geq0$, and subjected to the initial conditions $\rho(t)|_{t=0}=\rho_{0}$ and $\dot{\rho}(t)|_{t=0}=0$, is given by \cite{Coelho-Entropy-2022}
\begin{eqnarray}
	\label{eq:rho-griffiths}	
	\rho(t)=\sqrt{\frac{\omega_{0}\sin^{2}(\omega_{1}t)}{m_{0}\omega_{1}^{2}}+\frac{\cos^{2}(\omega_{1}t)}{m_{0}\omega_{0}}}.
\end{eqnarray}
Substituting this result into Eqs. \eqref{eq:r} and \eqref{eq:lambda(t)}, we obtain 
\begin{eqnarray}
	\label{eq:r-griffiths}	
	r(t)=\cosh^{-1}\left(\frac{\omega_{0}+\omega_{1}}{2\sqrt{\omega_{0}\omega_{1}}}\right).
\end{eqnarray}
Moreover, since the system is initially in the ground state, it is also interesting to calculate the probability of its persistence in the ground state.
For this purpose, substituting Eq. \eqref{eq:r-griffiths} in Eq. \eqref{eq:dist-prob-fótons} with $ \nu=0 $, one obtains
\begin{eqnarray}
	\label{eq:prob-p-griffiths}	
	P_{p}(t)=\frac{2\sqrt{\omega_{0}\omega_{1}}}{\omega_{0}+\omega_{1}}.
\end{eqnarray}
Note that in both limits $\omega_{1}/\omega_{0}\to0$ and $\omega_{1}/\omega_{0}\to\infty$, we have $P_{p}(t)\to 0$, which means that after the system changes its frequency in these limit situations, it will also be excited.
The behavior of $P_{p}(t)$ given by Eq. \eqref{eq:prob-p-griffiths} is illustrated in Fig. \ref{fig:behavior-prob-griffiths}.

For the problem 2.14 of Ref. \cite{Griffiths-Quantum-Mechanics} (or the problem 11.19 of Ref. \cite{Griffiths-Quantum-Mechanics-2018}), we have $\omega_{1}=2\omega_{0}$.
In this case, the energy spectrum in the interval $t\geq0$ becomes $E_{n}^{\prime}=2\hbar\omega_{0}\left(n+1/2\right)$.
Note that, in this interval, the energy associated to the ground state $ (n=0) $ is $E_{0}^{\prime}=\hbar\omega_{0}$. 
This means that, when $t\geq0$, the probability of a measurement of the energy resulting in any value less than $E_{0}^{\prime}$ is zero and, thus, such probability for $\hbar\omega_{0}/2$ is zero.
On the other hand, the probability of getting $\hbar\omega_{0}$ may not be zero and is simply given by Eq. \eqref{eq:prob-p-griffiths} with $\omega_{1}=2\omega_{0}$ (since $\hbar\omega_{0}$ is the energy corresponding to the ground state for $t\geq0$) which results in $P_{p}(t)=\frac{2\sqrt{2}}{3} \cong 0.943$.
This result is the same as that obtained from the elementary quantum mechanics methods discussed in undergraduate and postgraduate courses (see Refs. \cite{Griffiths-Quantum-Mechanics,Griffiths-Quantum-Mechanics-2018}). 

It is worth noting that alternative methods can be used to solve the same problem.
For example, in Refs. \cite{Janszky-OC-1986,Janszky-PRA-1989,Janszky-PRA-1992, Tibaduiza-BJP-2020}, instead of deriving the wave functions $\Psi_{n}(x,t)$ after the frequency change (as done here, as well as in Refs. \cite{Griffiths-Quantum-Mechanics,Griffiths-Quantum-Mechanics-2018}), in those articles, using the Heisenberg picture, through the Bogoliubov transformations relating annihilation and creation operators before and after the frequency changes, it becomes possible to write the squeezing parameter $r(t)$ in terms of these coefficients, and then, by means of Eq. \eqref{eq:dist-prob-fótons}, one can to obtain the same result as found in Eq. \eqref{eq:prob-p-griffiths}.

Next, we analyze a model where the usual methods taught at undergraduate and postgraduate level cannot be applied directly.

\subsection{Quantum dynamics of a particle in a Paul trap}\label{sec: Paul trap}

Paul traps, also known as radio-frequency traps, are devices used to capture and manipulate ions using oscillating electric fields \cite{Paul-RMP-1990, Brown-PRL-1991,Leibfried-RMP-2003}.
More specifically, these fields create an oscillating quadrupole potential that generates an average force capable of confining particles in all directions.
Developed by German physicist Wolfgang Paul (who received the Nobel Prize for physics in 1989 for this innovation), they have applications in different areas of physics as, for instance, mass spectroscopy, cooling processes, quantum computing, among others (for more details, see Refs. \cite{Paul-RMP-1990, Brown-PRL-1991,Leibfried-RMP-2003}). 

For a classic mechanical analogy (which can help visualize how the device functions), consider a ball in a potential that has the form of a  ``horse saddle'', in which the ball is fixed at the center of the saddle (see Ref. \cite{Lofgren-PT-2023}, where Fig. 1 provides an illustration of this model).
If the saddle is static, the ball does not stay at the center for a long time. Indeed, it rolls downwards in the unstable direction. In the other direction, the ball is in stable equilibrium because the upward curvature prevents it from falling. In other words, the static saddle is a stable potential in one direction and unstable in another.
On the other hand, if the saddle rotates rapidly around its vertical axis (like a carousel), the ball no longer falls: it remains ``trapped'' at the center of the saddle (exhibiting only small oscillations around this point), since the constant rotation changes the directions of stability and instability so quickly that the ball does not have time to escape.
For a clarifying discussion, see Ref. \cite{Lofgren-PT-2023} and references therein.

In this sense, due to the importance of this device for the modern physics, it could be very enriching to make a basic discussion on this topic in undergraduate and postgraduate quantum mechanics courses.
However, the tools usually taught in these courses are not enough to deal with this problem, since the dynamics of a particle in a Paul trap is effectively modeled, for each of the three Cartesian coordinates of the center of mass of the trapped particle, by a TDHO with time-dependent frequency of the form \cite{Brown-PRL-1991,Leibfried-RMP-2003,Lima-JMO-2009}
\begin{eqnarray}
	\label{eq:Paul trap}	
	\omega(t)=\frac{\omega_{0}}{\sqrt{\beta+\gamma}}\sqrt{\beta+\gamma\cos\left(\frac{2\pi t}{\tau}\right)},
\end{eqnarray}
where $\tau$ is the oscillation period [thus, $\omega(t+\tau)=\omega(t)$], and $\beta$ and $\gamma$ are numerical factors such that $\beta>\gamma$.
By means of the LR method, this problem can be solved exactly and, thus, it can be interesting to make a brief analyzes of this case (something similar has already been done, for example, in Ref. \cite{Lima-JMO-2009}). 

\begin{figure}[h!]
	\epsfig{file=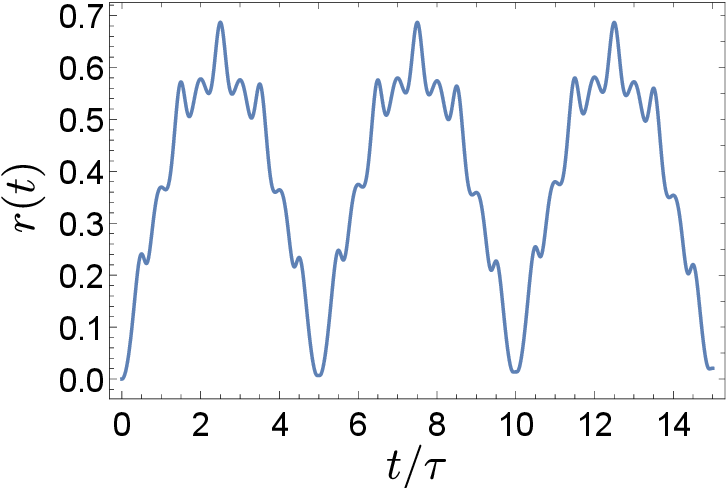, width=0.4 \linewidth}
	\caption{Behavior of the squeezing parameter $r(t)$ associated with the model defined by Eq. \eqref{eq:Paul trap}, as a function of $t/\tau$. In this figure we consider, in arbitrary units, $\beta=1$, $\gamma=1/2$ and $\omega_{0}\tau=3$.}	
	\label{fig:behavior-r-paul}
\end{figure}

\begin{figure}[h!]
	\centering
	\subfigure[\label{fig:varX-paul}]{\epsfig{file=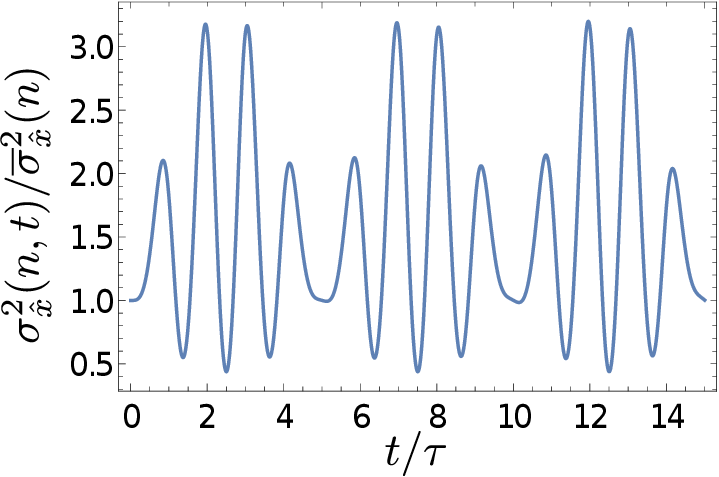,  width=0.4 \linewidth}}
	\hspace{1.0mm}
	\centering
	\subfigure[\label{fig:varP-paul}]{\epsfig{file=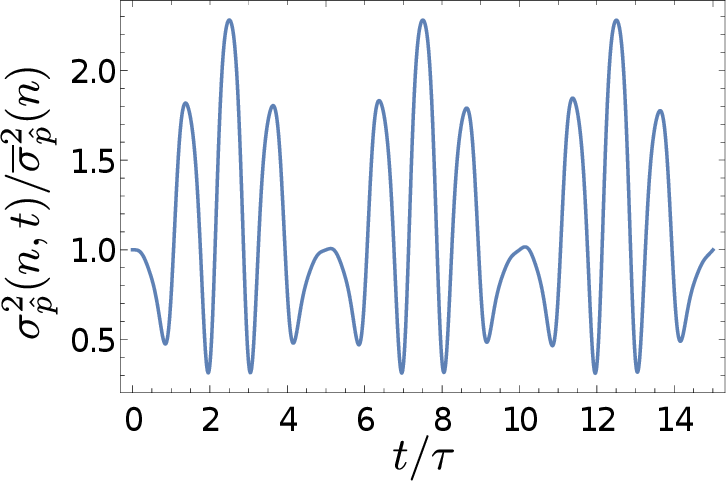, width=0.4 \linewidth}}
	\hspace{1.0mm}
	\caption{Behavior of the variances (a) $\sigma_{\hat{x}}^{2}(n,t)/\overline{\sigma}_{\hat{x}}^{2}(n)$ and (b) $\sigma_{\hat{p}}^{2}(n,t)/\overline{\sigma}_{\hat{p}}^{2}(n)$, with $ \overline{\sigma}_{\hat{x}}^{2}(n)= \tfrac{\hbar(n+{1}/{2})}{m_0\omega_0}$ and $ \overline{\sigma}_{\hat{p}}^{2}(n)=\hbar m_0\omega_0(n+{1}/{2}) $, as a function of $t/\tau$. In these figures we consider, in arbitrary units, $\beta=1$, $\gamma=1/2$ and $\omega_{0}\tau=3$.}	
	\label{fig:varX-P-paul}
\end{figure}

\begin{figure}[h!]
	\centering
	\subfigure[\label{fig:Pp-paul}]{\epsfig{file=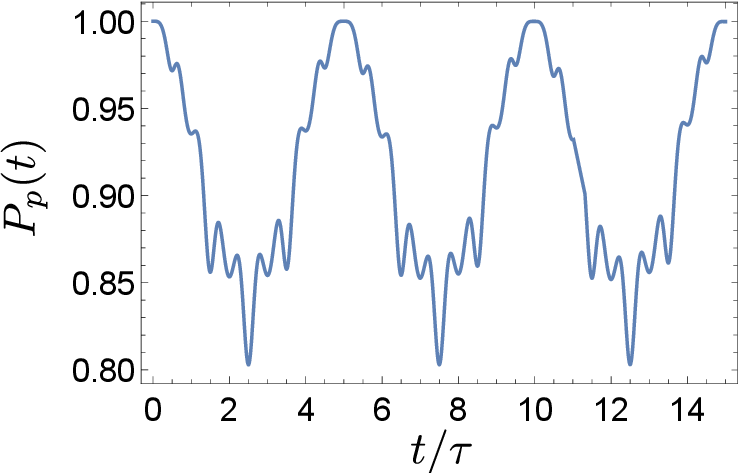,  width=0.4 \linewidth}}
	\hspace{1.0mm}
	\centering
	\subfigure[\label{fig:Pe-paul}]{\epsfig{file=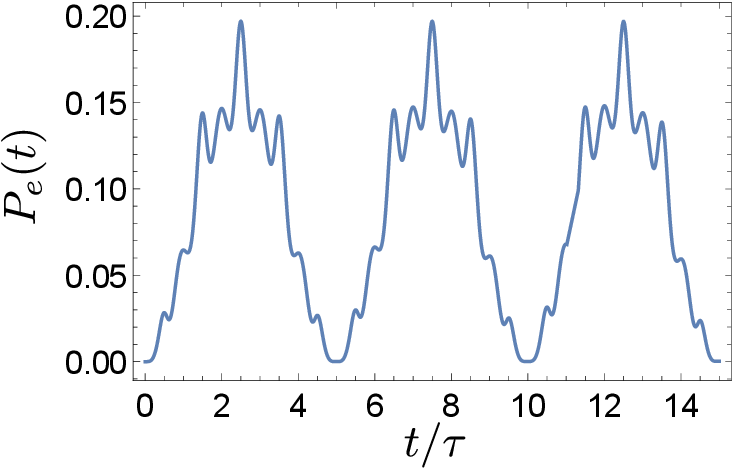, width=0.4 \linewidth}}
	\hspace{1.0mm}
	\caption{Behavior of the probabilities (a) $P_{p}(t)$ and (b) $P_{e}(t)$, as a function of $t/\tau$. In these figures we consider, in arbitrary units, $\beta=1$, $\gamma=1/2$ and $\omega_{0}\tau=3$.}	
	\label{fig:P-P-E-paul}
\end{figure}

Using Eqs. \eqref{eq:Paul trap} and \eqref{eq:rho-Paul} (see Appendix \ref{append:rho-Paul}) in Eqs. \eqref{eq:r} and \eqref{eq:lambda(t)}, we obtain the squeezing parameter $r(t)$ for this model, whose behavior is shown in Fig. \ref{fig:behavior-r-paul}.
Note the periodic oscillatory behavior of the squeezing parameter $r(t)$ \cite{Lima-JMO-2009}, which occurs naturally due to the form of the frequency $\omega(t)$ in Eq. \eqref{eq:Paul trap}.
This also influences the behavior of all the other associated quantities [see Eqs. \eqref{eq:<x^{2}>}-\eqref{eq:dist-prob-fótons}], as can be seen in Fig. \ref{fig:varX-P-paul} for the variances of position and momentum, and in Fig. \ref{fig:P-P-E-paul} for the probability of persistence in the ground state $P_{p}(t)$ and the probability of excitation $P_{e}(t)$.
Note that when the values of $\sigma_{\hat{x}}^{2}(n,t)$ are increasing in Fig. \ref{fig:varX-paul}, the corresponding values of $\sigma_{\hat{p}}^{2}(n,t)$ in Fig. \ref{fig:varP-paul} are decreasing, and vice-versa, being in agreement with Heisenberg's uncertainty principle.
In a similar way, when the values of $P_{p}(t)$ are increasing in Fig. \ref{fig:Pp-paul}, the corresponding values of $P_{e}(t)$ in Fig. \ref{fig:Pe-paul} are decreasing, and vice-versa, since $P_{p}(t)+P_{e}(t)=1$.
In addition, Figs. \ref{fig:behavior-r-paul} and \ref{fig:Pe-paul} show that $P_{e}(t)$ assumes the maximum values in the same times as $r(t)$ takes on its maximum values.
This occurs because $P_{e}(t)=1-1/\cosh[r(t)]$, i.e., the higher the value of $r(t)$, $P_{e}(t)$ approaches to $1$.

\section{Final Remarks}\label{sec:final}
 
In Sec. \ref{sec:oscilador independente do tempo}, we briefly reviewed the algebraic method to obtain the solution of the Schrödinger equation for the problem of a time-independent quantum harmonic oscillator, which is widely studied in quantum mechanics courses \cite{Griffiths-Quantum-Mechanics, Sakurai-Quantum-Mechanics-2021, Cohen-Tannoudji-et-al-QM-vol-1}. 
This section prepares the reader for the subsequent sections, since the application of the Lewis-Riesenfeld method to a time-dependent harmonic oscillator prescribes a sequence of steps similar to those already used in the time-independent case (and which were summarized in the end of Sec. \ref{sec:oscilador independente do tempo}).

In Sec. \ref{sec:review}, we discussed the general formulation of the Lewis-Riesenfeld method \cite{Lewis-PRL-1967,Lewis-JMP-1968,Lewis-JMP-1969}, step-by-step and with a didactic approach, and we presented a summary of the method, organized into four steps. 
We highlighted that the usual way to approach the problem of the time-independent quantum harmonic oscillator (looking for eigenfunctions of the time-independent Hamiltonian operator) is a particular case of the Lewis-Riesenfeld approach (looking for eigenfunctions of the Lewis' invariant operator).
We also presented Table \ref{tab1}, where we summarized a comparison between the characteristics of a general approach to obtain the solution of a time-independent case and the correspondent ones of a time-dependent situation.

In Sec. \ref{sec:wave-function}, we provided a didactic application of the Lewis-Riesenfeld method, obtaining the wave function for the case of a time-dependent harmonic oscillator with a generic time-dependent frequency.
We highlighted that the steps for finding the eigenstates and eigenvalues of the Lewis' invariant operator for this system (Lewis' invariant operator is written in terms of ladder operators) are similar to those of the usual algebraic method used to find the eigenstates and eigenvalues of the Hamiltonian of a time-independent quantum harmonic oscillator (the time-invariant Hamiltonian operator is written in terms of ladder operators).
Closing this section, we presented Table \ref{tab2}, where we summarized a comparison between the characteristics of the time-independent oscillator addressed via algebraic approach (discussed in Sec. \ref{sec:oscilador independente do tempo}) and the correspondent ones of a time-dependent oscillator addressed via Lewis-Riesenfeld method.

In Sec. \ref{sec:rel-tdho-se}, we briefly discussed the relation between the quantum time-dependent harmonic oscillator and the generation of squeezed states, using, in Sec. \ref{sec:griffiths}, these concepts to calculate the transition probability for a time-dependent harmonic oscillator experiencing a sudden jump in its frequency \cite{Griffiths-Quantum-Mechanics, Griffiths-Quantum-Mechanics-2018}.
In Sec. \ref{sec: Paul trap}, we also used all these tools to briefly study the dynamics of a quantum particle in a Paul trap \cite{Brown-PRL-1991,Lima-JMO-2009}.
These calculations performed in Secs. \ref{sec:griffiths} and \ref{sec: Paul trap} can be used as didactic examples of applications of the Lewis-Riesenfeld method, which exactly solves the Schrödinger equation with explicitly time-dependent Hamiltonians, in quantum mechanics courses.

The presented discussion aims to contribute to the effort already begun in articles in physics teaching journals \cite{Abe-EJP-2009, Andrews-AJP-1999, Nascimento-RBEF-2021, Castanos-AJP-2019, Leach-AJP-1978} to improve the discussions on time-dependent quantum harmonic oscillators in the context of undergraduate courses.

\section*{Declaration of competing interest}
The authors declare no conflict of interest.

\section*{Acknowledgements}
S.S.C. and  L.Q. were supported by Coordenação de Aperfeiçoamento de Pessoal de Nível Superior - Brazil (CAPES), Finance Code 001. 
This work was partially supported by CNPq - Brazil, Processo 408735/2023-6 CNPq/MCTI.

\appendix
\section{Time-independent harmonic oscillator: a brief review}\label{sec:append-time-indep-osci}

First, we write the function $\Psi_{n}^{(0)}(x,t)$ in Eq. \eqref{eq:Psi0} as 
\begin{eqnarray}
	\label{eq:Psi-n0-est}
	\Psi_{n}^{(0)}(x,t)=\Phi_{n}^{(0)}(x)\exp\left(-\frac{iE_{n}t}{\hbar}\right),
\end{eqnarray}
such that $ E_n $ and $ \Phi_{n}^{(0)}(x) $ have to be obtained by means of the eigenvalue equations
\begin{equation}
	\label{eq:H0-E0}	
	\hat{H}_{0}\Phi_{n}^{(0)}(x)=E_{n}\Phi_{n}^{(0)}(x).
\end{equation}
For our purposes, here we solve this equation by means of an algebraic method, using the so-called annihilation and creation operators (sometimes they are also denominated as lowering and raising operators, respectively).

The algebraic method for solving Eq. \eqref{eq:H0-E0} can be performed by introducing the annihilation ($\hat{a}_{0}$) and creation ($\hat{a}_{0}^{\dagger}$) operators, which are defined in terms of the operators $ \hat{x} $ and $ \hat{p} $ as \cite{Griffiths-Quantum-Mechanics, Sakurai-Quantum-Mechanics-2021}
\begin{align}
	\label{eq:a0}
	\hat{a}_{0}= & \; \sqrt{\frac{m_{0}\omega_{0}}{2\hbar}}\left(\hat{x}+\frac{i\hat{p}}{m_{0}\omega_{0}}\right),\\
	\label{eq:a0 dagger}
	\hat{a}_{0}^{\dagger}= & \; \sqrt{\frac{m_{0}\omega_{0}}{2\hbar}}\left(\hat{x}-\frac{i\hat{p}}{m_{0}\omega_{0}}\right),
\end{align}
where one notes that these operators are non-Hermitian.
From these expressions and from the commutation relation between $ \hat{x} $ and $ \hat{p} $, one can obtain that $[\hat{a}_{0},\hat{a}_{0}^{\dagger}]=1$, and we can rewrite the operator $ \hat{H}_0 $ as 
\begin{eqnarray}
	\hat{H}_{0}=\hbar\omega_{0}\left(\hat{a}_{0}^{\dagger}\hat{a}_{0}+\frac{1}{2}\right).
	\label{eq:hamiltoniano do oscilador independente do tempo}
\end{eqnarray}
The product $ \hat{a}_{0}^{\dagger}\hat{a}_{0}$ in this equation is denominated as the number operator $\hat{N}_0=\hat{a}_{0}^{\dagger}\hat{a}_{0}$. 
In this way, let us denote the eigenvalues of $\hat{N}_0$ as $ n $ and its normalized eigenstates as $ \left| n \right\rangle $ [whose representation in the position space, $ \left\langle x |n\right\rangle $, is the function $ \Phi_{n}^{(0)}(x) $], so that 
\begin{eqnarray}
	\label{eq:eq-autovalor-n0}
	\hat{N}_{0}|n\rangle=n|n\rangle.
\end{eqnarray}
From this, one can obtain that 
\begin{equation}
	\hat{H}_{0}|n\rangle=\hbar\omega_{0}\left(n+\frac{1}{2}\right)|n\rangle,
\end{equation}
from which we can conclude that the energy eigenvalues (that is, the eigenvalues of the operator $\hat{H}_0$) are given by \cite{Griffiths-Quantum-Mechanics, Sakurai-Quantum-Mechanics-2021}
\begin{eqnarray}
	\label{eq:E-n-est}
	E_{n}=\hbar\omega_{0}\left(n+\frac{1}{2}\right).
\end{eqnarray}

To calculate the eigenstates $ \left| n \right\rangle $, one has to obtain the action of the operators $ \hat{a}_{0} $ and $ \hat{a}_{0}^{\dagger} $ in this states, which, from the commutation relation between these operators and Eq. \eqref{eq:eq-autovalor-n0}, one finds that $ \hat{a}_{0} $ and $ \hat{a}_{0}^{\dagger} $ satisfy the following algebra \cite{Griffiths-Quantum-Mechanics,Sakurai-Quantum-Mechanics-2021}:
\begin{align}
	\hat{a}_{0}|n\rangle= & \;\sqrt{n}|n-1\rangle,   \label{eq:a0-n} \\
	\hat{a}_{0}^{\dagger}|n\rangle= & \;\sqrt{n+1}|n+1\rangle,   \label{eq:a0-dagger-n}
\end{align}
which justify the nomenclature annihilation and creation operators (or lowering and raising operators, respectively).
Note that repeatedly application of $ \hat{a}_{0} $ produces states with smaller and smaller values of $ n $.
In this way, the action of the operator $ \hat{a}_0 $ in the fundamental eigenstate $|0\rangle$ has to be null, i.e.,
\begin{eqnarray}
	\label{eq:a-ket-0}
	\hat{a}_0|0\rangle=0.
\end{eqnarray}
From the fundamental eigenstate, we can generate any excited state $|n\rangle$ by applying $ \hat{a}_{0}^{\dagger} $ in it repeatedly.
Thus, let us calculate it by applying $ \langle x| $ in Eq. \eqref{eq:a-ket-0}, and using Eq. \eqref{eq:a0} in position space, so that one obtains
\begin{eqnarray}
	\label{eq:phi-0-est}
	\left(x+\frac{\hbar}{m_{0}\omega_{0}}\frac{\partial}{\partial x}\right)\Phi_{0}^{(0)}(x)=0,
\end{eqnarray}
where $ \Phi_{0}^{(0)}(x)=\langle x|0\rangle $. 
By solving this differential equation, one obtains the normalized solution  \cite{Griffiths-Quantum-Mechanics,Sakurai-Quantum-Mechanics-2021}
\begin{eqnarray}
	\label{eq:phi0-est}
	\Phi_{0}^{(0)}(x)=\left(\frac{m_{0}\omega_{0}}{\pi\hbar}\right)^{\frac{1}{4}}\exp\left(-\frac{m_{0}\omega_{0}x^{2}}{2\hbar}\right).
\end{eqnarray}
By applying $\hat{a}_{0}^{\dagger} $ in $ \Phi_{0}^{(0)}(x) $ repeatedly, one obtains the following expression for $\Phi_{n}^{(0)}(x) $:
\begin{eqnarray}
	\Phi_{n}^{(0)}(x)=\frac{(a_{0}^{\dagger})^{n}}{\sqrt{n!}}\Phi_{0}^{(0)}(x),
	\label{eq:Phi-n0-geral}
\end{eqnarray}
which, using Eqs. \eqref{eq:a0 dagger} (in position space) and \eqref{eq:phi0-est}, one obtains that \cite{Griffiths-Quantum-Mechanics,Sakurai-Quantum-Mechanics-2021}
\begin{eqnarray}
	\label{eq:Phi-n0-est}
	\Phi_{n}^{(0)}(x)=\frac{1}{\sqrt{2^{n}n!}}\Phi_{0}^{(0)}(x){\cal H}_{n}\left(\sqrt{\frac{m_{0}\omega_{0}}{\hbar}}x\right),
\end{eqnarray}
where $ {\cal H}_{n} $ are the usual Hermite polynomials of order $ n $. 
In this sense, using Eqs. \eqref{eq:E-n-est} and \eqref{eq:Phi-n0-est} in Eq. \eqref{eq:Psi-n0-est}, we obtain the wave function of the time-independent quantum harmonic oscillator \cite{Sakurai-Quantum-Mechanics-2021, Griffiths-Quantum-Mechanics}.

In summary, the algebraic method used to find the solution of Eq. \eqref{eq:H0-E0} consists in the following steps: 
\begin{enumerate}
	\item Define the annihilation $(\hat{a}_{0})$ and creation $(\hat{a}_{0}^{\dagger})$ operators [Eqs. \eqref{eq:a0} and \eqref{eq:a0 dagger}, respectively];
	\item Express the Hamiltonian operator $\hat{H}_{0}$ in terms of $\hat{a}_{0}$ and $\hat{a}_{0}^{\dagger}$ [Eq. \eqref{eq:hamiltoniano do oscilador independente do tempo}];
	\item Define the number operator $ \hat{N} = \hat{a}_{0}^{\dagger}\hat{a}_{0} $ with eigenvalue and eigenstate denoted by $ n $ and $ \left| n \right\rangle $, respectively [Eq. \eqref{eq:eq-autovalor-n0}];
	\item Find the effect of $\hat{a}_{0}$ and $\hat{a}_{0}^{\dagger}$ on the state $ \left| n \right\rangle $ [Eqs. \eqref{eq:a0-n} and \eqref{eq:a0-dagger-n}];
	\item Define the ground state $ \left| 0 \right\rangle $ from the property that the application of the annihilation operator in this state has to be null [Eq. \eqref{eq:a-ket-0}], and calculate $ \Phi_{0}^{(0)}(x)=\langle x|0\rangle $;
	\item Calculate the excited states $\Phi_{n}^{(0)}(x)$ by applying repeatedly the creation operator in $\Phi_{0}^{(0)}(x)$ [Eq. \eqref{eq:Phi-n0-geral}];
	\item Find the general solution $\Psi^{(0)}(x,t)$ of the Schrödinger equation [Eq. \eqref{eq:Psi0}, with $ \Psi_{n}^{(0)}(x,t) $ given in Eq. \eqref{eq:Psi-n0-est-osc}].
\end{enumerate}

\section{Temporal independence of the eigenvalues $ \lambda $}
\label{sec:append-lambda}

Applying the operator $\partial\hat{I}(t)/\partial t$ in the state $|\lambda,\zeta;t\rangle$,
\begin{eqnarray}	
	\frac{\partial\hat{I}(t)}{\partial t}|\lambda,\zeta;t\rangle=-\frac{1}{i\hbar}\bigl[\hat{I}(t)\hat{H}(t)-\hat{H}(t)\hat{I}(t)\bigr]|\lambda,\zeta;t\rangle,
	\label{eq:eq de heisenberg atuando em |lambda,n;t>}
\end{eqnarray}
acting $ \langle\lambda^{\prime},\zeta^{\prime};t| $ in Eq. \eqref{eq:eq de heisenberg atuando em |lambda,n;t>} and using Eq. \eqref{eq:equação de autovalores para I}, we obtain
\begin{eqnarray}	
	\langle\lambda^{\prime},\zeta^{\prime};t|\frac{\partial\hat{I}(t)}{\partial t}|\lambda,\zeta;t\rangle=-\frac{\left(\lambda^{\prime}-\lambda\right)}{i\hbar}\langle\lambda^{\prime},\zeta^{\prime};t|\hat{H}(t)|\lambda,\zeta;t\rangle.
	\label{eq:sanduíche do dI/dt}
\end{eqnarray}
However, deriving Eq. (\ref{eq:equação de autovalores para I}) with respect to time,
\begin{eqnarray}
	\label{eq:derivada da equação de autovalores de I}
	\frac{\partial\hat{I}(t)}{\partial t}\bigl|\lambda,\zeta;t\rangle+\hat{I}(t)\frac{\partial}{\partial t}\bigl|\lambda,\zeta;t\rangle=\frac{\partial\lambda}{\partial t}\bigl|\lambda,\zeta;t\rangle+\lambda\frac{\partial}{\partial t}\bigl|\lambda,\zeta;t\rangle,
\end{eqnarray}
operating $\langle\lambda^{\prime},\zeta^{\prime};t|$ in Eq. \eqref{eq:derivada da equação de autovalores de I}, utilizing the Eq. \eqref{eq:relação de ortonormalidade} and noting that the operator $ \hat{I}(t) $ is Hermitian, we find
\begin{eqnarray}
	\label{eq:sanduíche do dI/dt com as deltas}	
	\langle\lambda^{\prime},\zeta^{\prime};t|\frac{\partial\hat{I}(t)}{\partial t}|\lambda,\zeta;t\rangle=\frac{\partial\lambda}{\partial t}\delta_{\lambda^{\prime},\lambda}\delta_{\zeta^{\prime},\zeta}-\left(\lambda^{\prime}-\lambda\right)\langle\lambda^{\prime},\zeta^{\prime};t|\frac{\partial}{\partial t}|\lambda,\zeta;t\rangle.
\end{eqnarray}
Since Eqs. \eqref{eq:sanduíche do dI/dt} and \eqref{eq:sanduíche do dI/dt com as deltas} are equal, consequently
\begin{eqnarray}
	\label{eq:B2=B4}
	\frac{\partial\lambda}{\partial t}\delta_{\lambda^{\prime},\lambda}\delta_{\zeta^{\prime},\zeta}-\left(\lambda^{\prime}-\lambda\right)\langle\lambda^{\prime},\zeta^{\prime};t|\frac{\partial}{\partial t}|\lambda,\zeta;t\rangle +\frac{\left(\lambda^{\prime}-\lambda\right)}{i\hbar}\langle\lambda^{\prime},\zeta^{\prime};t|\hat{H}(t)|\lambda,\zeta;t\rangle=0.
\end{eqnarray}
Hence, for $ \lambda^{\prime}= \lambda $ and $ \zeta^{\prime}= \zeta $, it will result from Eq. \eqref{eq:B2=B4} that
\begin{eqnarray}
	\label{eq:dlambda/dt=0}
	\frac{\partial\lambda}{\partial t}=0.
\end{eqnarray}
%

\section{The relation between the eigenstates of $ \hat{I}(t) $ and $ \hat{H}(t) $ operators}\label{append: rel-I-H}

According to Eq. \eqref{eq:eq de Schrodinger I-psi}, the action of the operator $\hat{I}(t)$ on any state ket, which is a solution of the Schrödinger equation, produces a new solution of it. From Eqs. \eqref{eq:derivada da equação de autovalores de I} and \eqref{eq:dlambda/dt=0}, we have
\begin{eqnarray}
	\label{dI/dt|lambda>}
	\frac{\partial\hat{I}(t)}{\partial t}|\lambda,\zeta;t\rangle=\bigl[\lambda-\hat{I}(t)\bigr]\frac{\partial}{\partial t}|\lambda,\zeta;t\rangle,
\end{eqnarray}
then, multiplying Eq. \eqref{dI/dt|lambda>} by bra $\langle\lambda^{\prime},\zeta^{\prime};t|$, we get
\begin{eqnarray}
	\langle\lambda^{\prime},\zeta^{\prime};t|\frac{\partial\hat{I}(t)}{\partial t}|\lambda,\zeta;t\rangle=\langle\lambda^{\prime},\zeta^{\prime};t|\bigl[\lambda-\hat{I}(t)\bigr]\frac{\partial}{\partial t}|\lambda,\zeta;t\rangle,
	\label{eq:derivada de I braketeada}
\end{eqnarray}
and, equating Eqs. \eqref{eq:sanduíche do dI/dt} and \eqref{eq:derivada de I braketeada}, we obtain
\begin{eqnarray}
	\label{eq:relação entre I e H}
	i\hbar\left(\lambda-\lambda^{\prime}\right)\langle\lambda^{\prime},\zeta^{\prime};t|\frac{\partial}{\partial t}|\lambda,\zeta;t\rangle=\left(\lambda-\lambda^{\prime}\right)\langle\lambda^{\prime},\zeta^{\prime};t|\hat{H}(t)|\lambda,\zeta;t\rangle.
\end{eqnarray}
In the general case where $ \lambda^{\prime}\neq \lambda $ and $ \zeta^{\prime}\neq \zeta $, we find
\begin{eqnarray}
	\label{eq:lambda=/lambda'}
	i\hbar\langle\lambda^{\prime},\zeta^{\prime};t|\frac{\partial}{\partial t}|\lambda,\zeta;t\rangle=\langle\lambda^{\prime},\zeta^{\prime};t|\hat{H}(t)|\lambda,\zeta;t\rangle.
\end{eqnarray}
To analyze the case $ \lambda^{\prime}=\lambda $ and $ \zeta^{\prime}=\zeta $, we will proceed as follows.
Knowing that multiplying the state vector by a phase factor of the form $\exp[i\alpha_{\lambda,\zeta}(t)]$ [$\alpha_{\lambda,\zeta}(t)$ is a real and time-dependent function] will not change the probability amplitude because the scalar product will remain unchanged, so we can define a new set of state vectors, which relate to the initial state ket $|\lambda,\zeta;t\rangle$ by means of the gauge transformation \cite{Lewis-JMP-1969}
\begin{eqnarray}
	\label{eq:transf de calibre append}
	|\lambda,\zeta;t\rangle_{\alpha}=\exp[i\alpha_{\lambda,\zeta}(t)]|\lambda,\zeta;t\rangle.
\end{eqnarray}

According to Ref. \cite{Lewis-JMP-1969}, because the operator $ \hat{I}(t) $ contains no temporal derivatives, one can immediately conclude that
\begin{eqnarray}
	\hat{I}(t)|\lambda,\zeta;t\rangle_{\alpha}=\lambda|\lambda,\zeta;t\rangle_{\alpha},
\end{eqnarray}
that is, the states $ |\lambda,\zeta;t\rangle_{\alpha} $ are also eigenstates of the operator $ \hat{I}(t) $, so that $ |\lambda,\zeta;t\rangle_{\alpha} $ and $ |\lambda,\zeta;t\rangle $ share the same eigenvalue spectrum and, consequently, the same completeness and orthonormality properties.
Thus, if we require that the eigenstates $|\lambda,\zeta;t\rangle_{\alpha}$ satisfy the Schrödinger equation:  
\begin{eqnarray}
	\label{eq:eq de Schrodinger lambda-alpha}
	i\hbar\frac{\partial}{\partial t}|\lambda,\zeta;t\rangle_{\alpha}=\hat{H}(t)|\lambda,\zeta;t\rangle_{\alpha},
\end{eqnarray}
then it will be possible to find the phases $ \alpha_{\lambda,\zeta}(t) $. For this, deriving Eq. \eqref{eq:transf de calibre append} with respect to time and substituting the result in Eq. \eqref{eq:eq de Schrodinger lambda-alpha}, we arrive at
\begin{eqnarray}
	i\frac{\partial}{\partial t}|\lambda,\zeta;t\rangle-\frac{d\alpha_{\lambda,\zeta}(t)}{dt}|\lambda,\zeta;t\rangle=\frac{1}{\hbar}\hat{H}(t)|\lambda,\zeta;t\rangle.
	\label{eq:f1}
\end{eqnarray}
Next, multiplying Eq. \eqref{eq:f1} by bra $\langle\lambda^{\prime},\zeta^{\prime};t|$ and using Eq. \eqref{eq:relação de ortonormalidade}, we have
\begin{eqnarray}
	\frac{d\alpha_{\lambda,\zeta}(t)}{dt}\delta_{\lambda^{\prime},\lambda}\delta_{\zeta^{\prime},\zeta}=\langle\lambda^{\prime},\zeta^{\prime};t|\biggl[i\frac{\partial}{\partial t}-\frac{1}{\hbar}\hat{H}(t)\biggr]|\lambda,\zeta;t\rangle.
	\label{eq:da/dt-deltas}
\end{eqnarray}
When $ \lambda^{\prime}\neq\lambda $ and $ \zeta^{\prime}\neq \zeta $, Eq. \eqref{eq:da/dt-deltas} will be [see Eq. \eqref{eq:lambda=/lambda'}]
\begin{eqnarray}
	\langle\lambda^{\prime},\zeta^{\prime};t|\left[i\frac{\partial}{\partial t}-\frac{1}{\hbar}\hat{H}(t)\right]|\lambda,\zeta;t\rangle=0,
\end{eqnarray}
while for $ \lambda^{\prime}= \lambda $ and $ \zeta^{\prime}= \zeta $, Eq. \eqref{eq:da/dt-deltas} will fall to
\begin{eqnarray}
	\label{eq:eq-dif-alpha}
	\frac{d\alpha_{\lambda,\zeta}(t)}{dt}=\langle\lambda,\zeta;t|\left[i\frac{\partial}{\partial t}-\frac{1}{\hbar}\hat{H}(t)\right]|\lambda,\zeta;t\rangle,
\end{eqnarray}
whose formal solution is given by
\begin{eqnarray}
	\alpha_{\lambda,\zeta}(t)=\int_{0}^{t}dt^{\prime}\langle\lambda,\zeta;t^{\prime}|\left[i\frac{\partial}{\partial t^{\prime}}-\frac{1}{\hbar}\hat{H}\left(t^{\prime}\right)\right]|\lambda,\zeta;t^{\prime}\rangle.
	\label{eq:alpha-append}
\end{eqnarray}
Having the phase functions in hand, we can determine the eigenstates $|\lambda,\zeta;t\rangle_{\alpha}$ since, in principle, we know how to calculate the eigenstates $|\lambda,\zeta;t\rangle$ from Eq. \eqref{eq:equação de autovalores para I}.
Therefore, with the eigenstates $|\lambda,\zeta;t\rangle_{\alpha}$ being solutions of the Schrödinger equation, any other solution can be described as a linear combination of these states. Accordingly, given that $|\psi(t)\rangle$ is a solution of the Schrödinger equation, then we can describe $|\psi(t)\rangle$ as a superposition of $|\lambda,\zeta,t\rangle_{\alpha}$ in the form
\begin{eqnarray}
	|\psi(t)\rangle=\sum_{\lambda,\zeta}C_{\lambda,\zeta}\exp\left[i\alpha_{\lambda,\zeta}(t)\right]|\lambda,\zeta;t\rangle,
\end{eqnarray}
where the coefficients $ C_{\lambda,\zeta} $ depend on the initial conditions and are given by
\begin{eqnarray}
	C_{\lambda,\zeta}=\langle\lambda,\zeta;0|\psi\left(0\right)\rangle.
\end{eqnarray}
%

\section{Main steps to obtain the invariant operator $\hat{I}(t)$}\label{append:I}

Starting from Eq. \eqref{eq:operador I genérico}, by substituting it into Eq. \eqref{eq:equação de Heisenberg}, one finds a system of coupled differential equations involving the functions $ \eta_{1}(t) $, $ \eta_{2}(t) $ and $ \eta_{3}(t) $:
\begin{align}
	\label{eq:equação auxiliar 1}
	\dot{\eta}_{1}(t)= & \;2m_{0}\omega(t)^{2}\eta_{3}(t),\\
	\label{eq:equação auxiliar 2}
	\dot{\eta}_{2}(t)= & -\frac{2\eta_{3}(t)}{m_{0}},\\
	\label{eq:equação auxiliar 3}
	\dot{\eta}_{3}(t)= & \;m_{0}\omega(t)^{2}\eta_{2}(t)-\frac{\eta_{1}(t)}{m_{0}}.
\end{align}
According to Lewis and Reisenfeld \cite{Lewis-JMP-1969}, to decouple this system 
``it is convenient'' (using their words) to introduce a real function $\rho(t)$, such that
\begin{eqnarray}
	\label{eq:mudança de variáveis}
	\eta_{2}(t)=\rho(t)^{2}.	
\end{eqnarray}
This leads $ \rho(t) $ to obey
the Ermakov-Pinney equation [Eq. \eqref{eq:equação de Ermakov-Pinney}], 
which, although it is a nonlinear equation, has a known solution \cite{Pinney-PAMS-1950} (more details later).
Then, from Eqs. \eqref{eq:equação auxiliar 2} and \eqref{eq:equação auxiliar 3}, we have
\begin{align}
	\label{eq:equação auxiliar 2 com o parâmetro rho}
	\eta_{3}(t)= & -m_{0}\rho(t)\dot{\rho}(t),\\
	\eta_{1}(t)= & \; m_{0}^{2}\dot{\rho}(t)^{2}+m_{0}^{2}\rho(t)[\ddot{\rho}(t)+\omega(t)^{2}\rho(t)].
	\label{eq:eta}
\end{align}
Using these expressions for $\eta_{1}(t)$ and $\eta_{3}(t)$ in Eq. \eqref{eq:equação auxiliar 1},
Lewis and Reisenfeld obtain the Ermakov-Pinney equation.
From Eqs. \eqref{eq:mudança de variáveis}-\eqref{eq:eta}, one sees that the functions $ \eta_{1}(t) $, $ \eta_{2}(t) $ and $ \eta_{3}(t) $ are completely determined by the parameter $ \rho(t) $. 
Thus, substituting these equations into Eq. \eqref{eq:operador I genérico}, one obtains Eq. \eqref{eq:I do oscilador}.

\section{A brief discuss about the parameter $\rho(t)$}\label{append:rho}

The equation \eqref{eq:equação de Ermakov-Pinney} originally appeared in 1880 with the Russian mathematician V. P. Ermakov (see Ref. \cite{Leach-AADM-2008} and Refs. therein).
In his work, Ermakov introduced a concept of invariant, today known as Ermakov's invariant.
In 1950, in a short article, the American mathematician Edmund Pinney presented the general solution of the nonlinear Eq. \eqref{eq:equação de Ermakov-Pinney} as 
\begin{eqnarray}
	\rho(t)=\sqrt{Au(t)^{2}+Bv(t)^{2}+2Cu(t)v(t)},
	\label{eq:sol-E-P-m0}
\end{eqnarray}
where $u(t)$ and $v(t)$ are two linearly independent solutions of the homogeneous equation (which has the same form of the equation of a classical TDHO with time-dependent frequency)
\begin{eqnarray}
	\label{eq:Ermakov-Pinney-mj-H}
	\ddot{k}(t)+\omega(t)^{2}k(t)=0,
\end{eqnarray}
and $A$, $B$ and $C$ are integration constants that depend on the boundary conditions of the problem at a certain time instant $t=t_{0}$ and satisfy the relation
\begin{eqnarray}
	\label{eq:relação entre as constantes e o Wronskiano m0}
	AB-C^{2}=\frac{1}{m_{0}^{2}\,W\left[u\left(t=t_{0}\right),v\left(t=t_{0}\right)\right]^{2}},
\end{eqnarray}
with $W\left[u\left(t=t_{0}\right),v\left(t=t_{0}\right)\right]$ being the Wronskian of $u(t=t_{0})$ and $v(t=t_{0})$.
Despite its theoretical value, Ermakov's work remained virtually unknown in the west until the late 1960s, gaining considerable prominence after the works of Lewis and Riesenfeld \cite{Lewis-PRL-1967,Lewis-JMP-1968,Lewis-JMP-1969}, as well as others (see Ref. \cite{Leach-AADM-2008} and Refs. therein).

A possible physical interpretation of the parameter $\rho(t)$ can be obtained when we write the variances $\sigma_{\hat{x}}^{2}(n,t)$ and $\sigma_{\hat{p}}^{2}(n,t)$ in terms of $\rho(t)$ [by means of Eqs. \eqref{eq:r}-\eqref{eq:phi} and \eqref{eq:<x^{2}>}-\eqref{eq:<p^{2}>}], namely \cite{Pedrosa-PRA-1997}
\begin{align}
	\sigma_{\hat{x}}^{2}(n,t)\;= & \;\left(n+\frac{1}{2}\right)\hbar\rho(t)^{2},\\
	\sigma_{\hat{p}}^{2}(n,t)\;= & \;\left(n+\frac{1}{2}\right)\hbar\left[\frac{1}{\rho(t)^{2}}+m_{0}^{2}\dot{\rho}(t)^{2}\right],
\end{align}
which allows us to conclude, for example, that the behavior of $\rho(t)^2$, unless a constant, coincides with the behavior of $\sigma_{\hat{x}}^{2}(n,t)$.

It is important to mention that the Ermakov-Pinney equation is also present in the description of the classical harmonic oscillator with time-dependent frequency. 
Starting from the classical Hamiltonian,
\begin{eqnarray}
H(t)=\frac{p(t)^{2}}{2m_{0}}+\frac{1}{2}m_{0}\omega(t)^{2}x(t)^{2},
\end{eqnarray}
where the coordinate $x(t)$ satisfies 
\begin{eqnarray}
\label{eq:x-mov}	
\ddot{x}(t)+\omega(t)^{2}x(t)=0,
\end{eqnarray}
it follows that the solution for $x(t)$ has the form \cite{Fiore-JPA-2025}
\begin{eqnarray}
\label{eq:x-b}	
x(t)=s(t)\sin[\theta(t)],
\end{eqnarray}
where $\theta(t)$ is a phase function. 
Replacing Eq. \eqref{eq:x-b} in Eq. \eqref{eq:x-mov}, we obtain \cite{Fiore-JPA-2025}
\begin{eqnarray}
s(t)\ddot{\theta}(t)+2\dot{s}(t)\dot{\theta}(t)=0,\;\;\;\ddot{s}(t)=-\omega(t)^{2}s(t)+s(t)\dot{\theta}(t)^{2}.
\end{eqnarray}
Noting that $s(t)^{2}\dot{\theta}(t)\equiv L$, with $L$ being a constant (see Ref. \cite{Fiore-JPA-2025} for more details), we have
\begin{eqnarray}
\ddot{s}(t)+\omega(t)^{2}s(t)=\frac{L^{2}}{s(t)^{3}},
\end{eqnarray}
which means that the function $s(t)$ satisfies an Ermakov-Pinney type equation.
Thus, in both quantum and classical TDHO with time-dependent frequency, one has to solve an Ermakov-Pinney equation. 
In this sense, to solve the problem of a quantum TDHO is equivalent to solve a classical one.
Moreover, we can also explore this equivalence considering, for example, 
the position operator written in the Heisenberg picture [$\hat{x}\to\hat{x}(t)$] \cite{Nation-RMP-2012}:
\begin{eqnarray}
\hat{x}(t)=k(t)\hat{a}_{0}+k^{*}(t)\hat{a}_{0}^{\dagger},
\label{eq:x-t}
\end{eqnarray}
where $k(t)$ obeys Eq. \eqref{eq:Ermakov-Pinney-mj-H} (which, as already mentioned, has the same form of the equation of a classical TDHO with time-dependent frequency), and satisfy the relation
\begin{eqnarray}
k(t)\dot{k}^{*}(t)-k^{*}(t)\dot{k}(t)=\frac{i\hbar}{m_{0}}.
\label{eq:kkkk}
\end{eqnarray}
The solution of Eq. \eqref{eq:Ermakov-Pinney-mj-H} has the form \cite{Dabrowski-PRD-2016}
\begin{eqnarray}
k(t)=\rho(t)\exp[-i\gamma(t)],
\label{eq:k-rho-gamma}
\end{eqnarray}
and using this equation into Eq. \eqref{eq:kkkk}, one obtains
$\gamma(t)$ written in terms of $\rho(t)$. Using this into Eq. 
\eqref{eq:k-rho-gamma} and this result into Eq. \eqref{eq:Ermakov-Pinney-mj-H},
one has that the parameter $\rho(t)$ satisfies the Ermakov-Pinney equation \eqref{eq:equação de Ermakov-Pinney}. Thus, the problem of finding the quantum operator in Eq. \eqref{eq:x-t} requires
the solution of a classical TDHO [Eq. \eqref{eq:Ermakov-Pinney-mj-H}].
%

\section{Eigenstates of the invariant operator $\hat{I}(t)$ in position space}\label{append:eq:phi_n}

From the commutation relation between operators $\hat{a}(t)$ and $\hat{a}^{\dagger}(t)$, we find the effect of these operators in the state $ |n;t\rangle $:
\begin{align}
	\label{eq:atuação do operador de aniquilação num autoestado do operador I}
	\hat{a}(t)|n;t\rangle= & \; \sqrt{n}|n-1;t\rangle,\\
	\label{eq:atuação do operador de criação num autoestado do operador I}
	\hat{a}^{\dagger}(t)|n;t\rangle= & \; \sqrt{n+1}|n+1;t\rangle.
\end{align}
In the sequence, by knowing Eq. \eqref{eq:atuação do operador de aniquilação num autoestado do operador I}, we can define the ground state of $ \hat{I}(t) $, denoted by $ |0;t\rangle $, which has to satisfy
\begin{eqnarray}
	\label{eq:atuação de a no estado 0}
	\hat{a}(t)|0;t\rangle=0.
\end{eqnarray}
By applying $ \langle x| $ in this equation, and using Eq. \eqref{eq:operador de aniquilação}, one can obtain that \cite{Lewis-JMP-1968}
\begin{eqnarray}
	\biggl\{\frac{x}{\rho(t)}-i\Bigl[i\hbar\rho(t)\frac{\partial}{\partial x}+m_{0}\dot{\rho}(t)x\Bigr]\biggr\}\Phi_{0}(x,t)=0,
	\label{eq:eq para phi0}
\end{eqnarray}
where $ \Phi_{0}(x,t)=\langle x|0;t\rangle $, whose solution is given by 
\begin{eqnarray}
	\label{eq:phi0-ap}
	\Phi_{0}(x,t)=\biggl[\frac{1}{\pi\hbar\rho(t)^{2}}\biggr]^{\frac{1}{4}}\exp\left\{ \frac{im_{0}}{2\hbar}\left[\frac{\dot{\rho}(t)}{\rho(t)}+\frac{i}{m_{0}\rho(t)^{2}}\right]x^{2}\right\}.
\end{eqnarray}
From this solution, we can calculate the excited states $\Phi_{n}(x,t)=\langle x|n;t\rangle $ by applying repeatedly the operator $ \hat{a}^{\dagger}(t) $ in $\Phi_{0}(x,t)$, so that one finds
\begin{eqnarray}
	\label{eq:atuação de a dagger em 0}
	\Phi_{n}(x,t)=\frac{\bigl[a(t)^{\dagger}\bigr]^{n}}{\sqrt{n!}}\Phi_{0}(x,t).
\end{eqnarray}

To obtain an explicit expression for $\Phi_{n}(x,t)$, we will do the following procedure. Introducing the unitary operator $\hat{{\cal U}}(t)$, defined by \cite{Pedrosa-PRA-1997,Pedrosa-PRA-1997-singular-perturbation}
\begin{eqnarray}
	\label{eq:U-dagger}
	\hat{{\cal U}}(t)=\exp\left[-\frac{im(t)\dot{\rho}(t)\hat{x}^{2}}{2\hbar\rho(t)}\right],
\end{eqnarray}
we can rewrite Eq. \eqref{eq:phi0-ap} as 
\begin{eqnarray}
	\label{eq:Phi_0-U}
	\Phi_{0}(x,t)=\hat{{\cal U}}^{\dagger}(t)\biggl[\frac{1}{\pi\hbar\rho(t)^{2}}\biggr]^{\frac{1}{4}}\exp\left[-\frac{x^{2}}{2\hbar\rho(t)^{2}}\right].
\end{eqnarray}
Defining
\begin{eqnarray}
	\label{eq:Phi-0-linha-2}
	\Phi_{0}^{\prime}(x,t)=\biggl[\frac{1}{\pi\hbar\rho(t)^{2}}\biggr]^{\frac{1}{4}}\exp\left[-\frac{x^{2}}{2\hbar\rho(t)^{2}}\right],
\end{eqnarray}
and substituting the above equation into Eq. \eqref{eq:Phi_0-U}, we have
\begin{eqnarray}
	\label{eq:Phi-0-1}
	\Phi_{0}(x,t)=\hat{{\cal U}}^{\dagger}(t)\Phi_{0}^{\prime}(x,t).
\end{eqnarray}
Furthermore, using the Baker-Campbell-Hausdorff relation \cite{Sakurai-Quantum-Mechanics-2021,Tibaduiza-BJP-2020}:
\begin{eqnarray}
	\exp(\hat{{\cal G}})\hat{{\cal O}}\exp(-\hat{{\cal G}})=\hat{{\cal O}}+[\hat{{\cal G}},\hat{{\cal O}}]+\frac{1}{2!}[\hat{{\cal G}},[\hat{{\cal G}},\hat{{\cal O}}]]+\ldots+\frac{1}{k!}[\hat{{\cal G}},[\hat{{\cal G}},[\hat{{\cal G}},\ldots[\hat{{\cal G}},\hat{{\cal O}}]]]\ldots]+\ldots,
	\label{eq:BCH-relations}
\end{eqnarray} 
we have
\begin{eqnarray}
	\hat{a}^{\dagger}(t)=\hat{{\cal U}}^{\dagger}(t)\hat{a}^{\prime\dagger}(t)\,\hat{{\cal U}}(t),
\end{eqnarray}
where (remember that we are in position space)
\begin{eqnarray}
	\label{eq:a-dagger-linha}
	\hat{a}^{\prime\dagger}(t)=\frac{1}{\sqrt{2\hbar}}\left[\frac{x}{\rho(t)}-\hbar\rho(t)\frac{\partial}{\partial x}\right].
\end{eqnarray}

As such, by analogy with Eq. \eqref{eq:atuação de a dagger em 0}, we have 
\begin{eqnarray}
	\label{eq:Phi-n-linha}
	\Phi_{n}^{\prime}(x,t)=\frac{\bigl[\hat{a}^{\prime\dagger}(t)\bigr]^{n}}{\sqrt{n!}}\,\Phi_{0}^{\prime}(x,t).
\end{eqnarray}
Consequently, using Eqs. \eqref{eq:Phi-0-linha-2} and \eqref{eq:a-dagger-linha} in Eq. \eqref{eq:Phi-n-linha}, we find
\begin{eqnarray}
	\Phi_{n}^{\prime}(x,t)=\frac{1}{\sqrt{2^{n}n!}}\biggl[\frac{1}{\pi\hbar\rho(t)^{2}}\biggr]^{\frac{1}{4}}\left[\frac{x}{\hbar^{\frac{1}{2}}\rho(t)}-\hbar^{\frac{1}{2}}\rho(t)\frac{\partial}{\partial x}\right]^{n}\exp\left[-\frac{x^{2}}{2\hbar\rho(t)^{2}}\right].
	\label{eq:Phi-n-linha-2}
\end{eqnarray}
However, knowing the following relation involving the Hermite polynomials \cite{Arfken-Mathematical-Physics-2003}:
\begin{eqnarray}
	\label{eq:rel-H}
	{\cal H}_{n}(z)=\exp\left(\frac{z^{2}}{2}\right)\left(z-\frac{\partial}{\partial z}\right)^{n}\exp\left(-\frac{z^{2}}{2}\right),
\end{eqnarray}
so, by changing variables $z=x/[\hbar^{\frac{1}{2}}\rho(t)]$, we get
\begin{eqnarray}
	{\cal H}_{n}\left[\frac{x}{\hbar^{\frac{1}{2}}\rho(t)}\right]=\exp\left[\frac{x^{2}}{2\hbar\rho(t)^{2}}\right]\left[\frac{x}{\hbar^{\frac{1}{2}}\rho(t)}-\hbar^{\frac{1}{2}}\rho(t)\frac{\partial}{\partial x}\right]^{n}\exp\left[-\frac{x^{2}}{2\hbar\rho(t)^{2}}\right].
	\label{eq:H-n-rho}
\end{eqnarray}
Hence, rewriting Eq. \eqref{eq:Phi-n-linha-2} as
\begin{eqnarray}
	\Phi_{n}^{\prime}(x,t)=\frac{1}{\sqrt{2^{n}n!}}\biggl[\frac{1}{\pi\hbar\rho(t)^{2}}\biggr]^{\frac{1}{4}}\left\{ \exp\left[\frac{x^{2}}{2\hbar\rho(t)^{2}}\right]\left[\frac{x}{\hbar^{\frac{1}{2}}\rho(t)}-\hbar^{\frac{1}{2}}\rho(t)\frac{\partial}{\partial x}\right]^{n}\exp\left[-\frac{x^{2}}{2\hbar\rho(t)^{2}}\right]\right\} \exp\left[-\frac{x^{2}}{2\hbar\rho(t)^{2}}\right],
	\label{eq:Phi-n-linha-3}
\end{eqnarray}
substituting Eq. \eqref{eq:H-n-rho} into Eq. \eqref{eq:Phi-n-linha-3} and returning to the original function by \cite{Pedrosa-PRA-1997,Pedrosa-PRA-1997-singular-perturbation}
\begin{eqnarray}
	\Phi_{n}(x,t)=\hat{{\cal U}}^{\dagger}(t)\Phi_{n}^{\prime}(x,t),
\end{eqnarray}
finally
\begin{eqnarray}
	\Phi_{n}(x,t)=\frac{1}{\sqrt{2^{n}n!}}\Phi_{0}(x,t){\cal H}_{n}\left[\frac{x}{\hbar^{\frac{1}{2}}\rho(t)}\right].
\end{eqnarray}
%


\section{Main steps to obtain the phase functions $\alpha_{n}(t)$}\label{append:alpha-n}

For the first term, $ i\langle n;t|\frac{\partial}{\partial t}|n;t\rangle $, by writing $ |n;t\rangle=[\hat{a}^{\dagger}(t)/\sqrt{n} \;]|n-1;t\rangle $ and performing the time derivative, one obtains 
\begin{eqnarray}
	\langle n;t|\frac{\partial}{\partial t}|n;t\rangle=\frac{1}{\sqrt{n}}\langle n;t|\frac{\partial \hat{a}^{\dagger}(t)}{\partial t}|n-1;t\rangle+\langle n-1;t|\frac{\partial}{\partial t}|n-1;t\rangle.
	\label{eq:sanduíche de d/dt em termos de a dagger}
\end{eqnarray}
Using recursively this equation one obtains, by induction, that \cite{Lewis-JMP-1969}
\begin{eqnarray}
	\label{eq:d/dt sanduichado com 0}
	\langle n;t|\frac{\partial}{\partial t}|n;t\rangle=\frac{in}{2}m_{0}[\rho(t)\ddot{\rho}(t)-\dot{\rho}(t)^{2}]+\langle0;t|\frac{\partial}{\partial t}|0;t\rangle.
\end{eqnarray}
Noting that we can write the scalar product between two state vectors $|A\rangle$ and $|B\rangle$, in position space, as \cite{Sakurai-Quantum-Mechanics-2021}
\begin{eqnarray}
	\label{eq:prod-esc-x}
	\langle B|A\rangle=\int_{-\infty}^{+\infty}dx\langle B|x\rangle\langle x|A\rangle,
\end{eqnarray}
making
\begin{eqnarray}
	|A\rangle=\frac{\partial}{\partial t}|0;t\rangle,\;\;\langle B|=\langle0;t|,
\end{eqnarray}
we arrive in
\begin{eqnarray}
	\label{eq:d/dt-x}
	\langle 0;t|\frac{\partial}{\partial t}|0;t\rangle=\int_{-\infty}^{+\infty}dx\langle 0;t|x\rangle\langle x|\frac{\partial}{\partial t}|0;t\rangle.
\end{eqnarray}
However, realizing that
\begin{eqnarray}
	\langle x|\frac{\partial}{\partial t}|0;t\rangle=\frac{\partial\Phi_{0}(x,t)}{\partial t},\;\;\langle 0;t|x\rangle=\Phi_{0}^{*}(x,t),
\end{eqnarray}
the Eq. \eqref{eq:d/dt-x} will become
\begin{eqnarray}
	\label{eq:d/dt-x-2}	
	\langle 0;t|\frac{\partial}{\partial t}|0;t\rangle=\int_{-\infty}^{+\infty}dx\,\Phi_{0}^{*}(x,t)\frac{\partial\Phi_{0}(x,t)}{\partial t}.
\end{eqnarray}

Given this, rewriting Eq. \eqref{eq:phi0} as
\begin{eqnarray}
	\label{eq:phi0-u-v}
	\Phi_{0}(x,t)=b(t)\exp\left[-c(t)x^{2}\right],
\end{eqnarray}
where
\begin{eqnarray}	
	b(t)=\biggl[\frac{1}{\pi\hbar\rho(t)^{2}}\biggr]^{\frac{1}{4}},\;\;c(t)=-\frac{im_{0}}{2\hbar}\biggl[\frac{\dot{\rho}(t)}{\rho(t)}+\frac{i}{m_{0}\rho(t)^{2}}\biggr],
	\label{eq:u-v}
\end{eqnarray}
deriving Eq. \eqref{eq:phi0-u-v} with respect to time, we find
\begin{eqnarray}
	\label{eq:dPhi0/dt}	
	\frac{\partial\Phi_{0}(x,t)}{\partial t}=\left[F(t)x^{2}-\frac{\dot{\rho}(t)}{2\rho(t)}\right]b(t)\exp\left[-c(t)x^{2}\right],
\end{eqnarray}
in which
\begin{eqnarray}
	F(t)=\frac{im_{0}\ddot{\rho}(t)}{2\hbar\rho(t)}-\frac{im_{0}\dot{\rho}(t)^{2}}{2\hbar\rho(t)^{2}}+\frac{\dot{\rho}(t)}{\hbar\rho(t)^{3}},
	\label{eq:F(t)}
\end{eqnarray}
so, substituting the complex conjugate of Eq. \eqref{eq:phi0-u-v} and Eq. \eqref{eq:dPhi0/dt} into Eq. \eqref{eq:d/dt-x-2}, we arrive at
\begin{eqnarray}
	\langle0;t|\frac{\partial}{\partial t}|0;t\rangle=b(t)^{2}\left\{ F(t)\int_{-\infty}^{+\infty}dx\exp\left[-\frac{x^{2}}{\hbar\rho(t)^{2}}\right]x^{2}-\frac{\dot{\rho}(t)}{2\rho(t)}\int_{-\infty}^{+\infty}dx\exp\left[-\frac{x^{2}}{\hbar\rho(t)^{2}}\right]\right\} .
	\label{sanduíche de d/dt pro estado de vácuo 2}
\end{eqnarray}
Solving the Gaussian integrals and returning to the original variables, finally \cite{Lewis-JMP-1969}
\begin{equation}
	\label{eq:append-sanduíche d/dt em 0}
	\langle0;t|\frac{\partial}{\partial t}|0;t\rangle=\frac{im_{0}}{4}[\rho(t)\ddot{\rho}(t)-\dot{\rho}(t)^{2}],
\end{equation}
which, substituting in Eq. \eqref{eq:d/dt sanduichado com 0}, results in
\begin{eqnarray}
	i\langle n;t|\frac{\partial}{\partial t}|n;t\rangle=-\frac{1}{2}\left(n+\frac{1}{2}\right)m_{0}\bigl[\rho(t)\ddot{\rho}(t)-\dot{\rho}(t)^{2}\bigr].
	\label{eq:sanduíche id/dt final-apend}
\end{eqnarray}

Now, for the term $ \langle n;t|\hat{H}(t)|n;t\rangle $, from Eq. \eqref{eq:hamiltoniano do oscilador}, we have
\begin{eqnarray}
	\label{eq:elemento de matriz de H na base de I}	
	\langle n;t|\hat{H}(t)|n;t\rangle=\frac{1}{2m_{0}}\langle n;t|\hat{p}^{2}|n;t\rangle+\frac{1}{2}m_{0}\omega(t)^{2}\langle n;t|\hat{x}^{2}|n;t\rangle.
\end{eqnarray}
From Eqs. \eqref{eq:operador de aniquilação} and \eqref{eq:operador de criação}, we can express the operators $ \hat{x} $ and $ \hat{p} $ in terms of $ \hat{a}(t) $ and $ \hat{a}^{\dagger}(t)$, and obtain that 
\begin{align}
	\label{eq:elemento de matrix x^2}
	\langle n;t|\hat{x}^{2}|n;t\rangle= & \left(n+\frac{1}{2}\right)\hbar\rho(t)^{2},\\
	\label{eq:elemento de matrix p^2}
	\langle n;t|\hat{p}^{2}|n;t\rangle= & \left(n+\frac{1}{2}\right)\hbar\bigl[m_{0}^{2}\dot{\rho}(t)^{2}+\rho(t)^{-2}\bigr],
\end{align}
which implies in
\begin{eqnarray}
	\label{eq:sanduíche do H-apend}	
	\langle n;t|\hat{H}(t)|n;t\rangle=\frac{\hbar}{2m_{0}}\left(n+\frac{1}{2}\right)\bigl[\rho(t)^{-2}+m_{0}^{2}\dot{\rho}(t)^{2}+m_{0}^{2}\omega(t)^{2}\rho(t)^{2}\bigr].
\end{eqnarray}
%


\section{Solution of the Ermakov-Pinney equation for the frequency model given by Eq. \eqref{eq:Paul trap}}\label{append:rho-Paul}

Let us consider the Ermakov-Pinney equation for the frequency model given by Eq. \eqref{eq:Paul trap}:
\begin{eqnarray}
\ddot{\rho}(t)+\frac{\omega_{0}^{2}}{\beta+\gamma}\left[\beta+\gamma\cos\left(\frac{2\pi t}{\tau}\right)\right]\rho(t)=\frac{1}{m_{0}^{2}\rho(t)^{3}}.
\end{eqnarray}
Following the steps described in Appendix \ref{append:rho} and considering the initial conditions $\rho(t)|_{t=0}=\rho_{0}$ and $\dot{\rho}(t)|_{t=0}=0$, we obtain
\begin{eqnarray}
	\label{eq:rho-Paul}	
	\rho(t)=\sqrt{\frac{M_{C}\left[\frac{\beta\omega_{0}^{2}\tau^{2}}{\pi^{2}(\beta+\gamma)},-\frac{\gamma\omega_{0}^{2}\tau^{2}}{2\pi^{2}(\beta+\gamma)},\frac{\pi t}{\tau}\right]^{2}}{m_{0}\omega_{0}M_{C}\left[\frac{\beta\omega_{0}^{2}\tau^{2}}{\pi^{2}(\beta+\gamma)},-\frac{\gamma\omega_{0}^{2}\tau^{2}}{2\pi^{2}(\beta+\gamma)},0\right]^{2}}+\frac{\omega_{0}\tau^{2}M_{S}\left[\frac{\beta\omega_{0}^{2}\tau^{2}}{\pi^{2}(\beta+\gamma)},-\frac{\gamma\omega_{0}^{2}\tau^{2}}{2\pi^{2}(\beta+\gamma)},\frac{\pi t}{\tau}\right]^{2}}{\pi^{2}m_{0}\dot{M}_{S}\left[\frac{\beta\omega_{0}^{2}\tau^{2}}{\pi^{2}(\beta+\gamma)},-\frac{\gamma\omega_{0}^{2}\tau^{2}}{2\pi^{2}(\beta+\gamma)},\frac{\pi t}{\tau}\right]^{2}\Bigr|_{t=0}}},
\end{eqnarray}
in which $M_{C}$ and $M_{S}$ are the even and odd Mathieu functions, respectively.


\begin{thebibliography}{115}%
	\makeatletter
	\providecommand \@ifxundefined [1]{%
		\@ifx{#1\undefined}
	}%
	\providecommand \@ifnum [1]{%
		\ifnum #1\expandafter \@firstoftwo
		\else \expandafter \@secondoftwo
		\fi
	}%
	\providecommand \@ifx [1]{%
		\ifx #1\expandafter \@firstoftwo
		\else \expandafter \@secondoftwo
		\fi
	}%
	\providecommand \natexlab [1]{#1}%
	\providecommand \enquote  [1]{``#1''}%
	\providecommand \bibnamefont  [1]{#1}%
	\providecommand \bibfnamefont [1]{#1}%
	\providecommand \citenamefont [1]{#1}%
	\providecommand \href@noop [0]{\@secondoftwo}%
	\providecommand \href [0]{\begingroup \@sanitize@url \@href}%
	\providecommand \@href[1]{\@@startlink{#1}\@@href}%
	\providecommand \@@href[1]{\endgroup#1\@@endlink}%
	\providecommand \@sanitize@url [0]{\catcode `\\12\catcode `\$12\catcode
		`\&12\catcode `\#12\catcode `\^12\catcode `\_12\catcode `\%12\relax}%
	\providecommand \@@startlink[1]{}%
	\providecommand \@@endlink[0]{}%
	\providecommand \url  [0]{\begingroup\@sanitize@url \@url }%
	\providecommand \@url [1]{\endgroup\@href {#1}{\urlprefix }}%
	\providecommand \urlprefix  [0]{URL }%
	\providecommand \Eprint [0]{\href }%
	\providecommand \doibase [0]{https://doi.org/}%
	\providecommand \selectlanguage [0]{\@gobble}%
	\providecommand \bibinfo  [0]{\@secondoftwo}%
	\providecommand \bibfield  [0]{\@secondoftwo}%
	\providecommand \translation [1]{[#1]}%
	\providecommand \BibitemOpen [0]{}%
	\providecommand \bibitemStop [0]{}%
	\providecommand \bibitemNoStop [0]{.\EOS\space}%
	\providecommand \EOS [0]{\spacefactor3000\relax}%
	\providecommand \BibitemShut  [1]{\csname bibitem#1\endcsname}%
	\let\auto@bib@innerbib\@empty
	\bibitem [{\citenamefont {Landau}\ and\ \citenamefont
		{Lifshitz}(1976)}]{Landau-Lifishitz-Mechanics}%
	\BibitemOpen
	\bibfield  {author} {\bibinfo {author} {\bibfnamefont {L.~D.}\ \bibnamefont
			{Landau}}\ and\ \bibinfo {author} {\bibfnamefont {E.~M.}\ \bibnamefont
			{Lifshitz}},\ }\href@noop {} {\emph {\bibinfo {title} {Mechanics: Volume
				1}}},\ \bibinfo {edition} {3rd}\ ed.,\ Vol.~\bibinfo {volume} {1}\ (\bibinfo
	{publisher} {Oxford: Pergamon},\ \bibinfo {year} {1976})\BibitemShut
	{NoStop}%
	\bibitem [{\citenamefont {Greiner}(2003)}]{Greiner-Classical-Mechanics}%
	\BibitemOpen
	\bibfield  {author} {\bibinfo {author} {\bibfnamefont {W.}~\bibnamefont
			{Greiner}},\ }\href@noop {} {\emph {\bibinfo {title} {Classical mechanics:
				systems of particles and Hamiltonian dynamics}}}\ (\bibinfo  {publisher}
	{Springer},\ \bibinfo {year} {2003})\BibitemShut {NoStop}%
	\bibitem [{\citenamefont {Marion}(2013)}]{Marion-Classical-Dynamics}%
	\BibitemOpen
	\bibfield  {author} {\bibinfo {author} {\bibfnamefont {J.~B.}\ \bibnamefont
			{Marion}},\ }\href@noop {} {\emph {\bibinfo {title} {Classical dynamics of
				particles and systems}}}\ (\bibinfo  {publisher} {Academic Press},\ \bibinfo
	{year} {2013})\BibitemShut {NoStop}%
	\bibitem [{\citenamefont {Goldstein}\ \emph {et~al.}(2011)\citenamefont
		{Goldstein}, \citenamefont {Poole},\ and\ \citenamefont
		{Safko}}]{Goldstein-2011}%
	\BibitemOpen
	\bibfield  {author} {\bibinfo {author} {\bibfnamefont {H.}~\bibnamefont
			{Goldstein}}, \bibinfo {author} {\bibfnamefont {C.~P.}\ \bibnamefont
			{Poole}},\ and\ \bibinfo {author} {\bibfnamefont {J.~L.}\ \bibnamefont
			{Safko}},\ }\href@noop {} {\emph {\bibinfo {title} {{Classical
					Mechanics}}}},\ \bibinfo {edition} {3rd}\ ed.\ (\bibinfo  {publisher}
	{Dorling Kindersley (India) Pvt. Ltd},\ \bibinfo {year} {2011})\BibitemShut
	{NoStop}%
	\bibitem [{\citenamefont {Ryder}(1996)}]{Ryder-QFT}%
	\BibitemOpen
	\bibfield  {author} {\bibinfo {author} {\bibfnamefont {L.~H.}\ \bibnamefont
			{Ryder}},\ }\href@noop {} {\emph {\bibinfo {title} {Quantum field theory}}}\
	(\bibinfo  {publisher} {Cambridge university press},\ \bibinfo {year}
	{1996})\BibitemShut {NoStop}%
	\bibitem [{\citenamefont {Greiner}\ and\ \citenamefont
		{Reinhardt}(1996)}]{Greiner-Field-Quantization}%
	\BibitemOpen
	\bibfield  {author} {\bibinfo {author} {\bibfnamefont {W.}~\bibnamefont
			{Greiner}}\ and\ \bibinfo {author} {\bibfnamefont {J.}~\bibnamefont
			{Reinhardt}},\ }\href {https://doi.org/10.1007/978-3-642-61485-9} {\emph
		{\bibinfo {title} {{Field Quantization}}}}\ (\bibinfo  {publisher} {Springer
		Berlin Heidelberg},\ \bibinfo {address} {Berlin, Heidelberg},\ \bibinfo
	{year} {1996})\BibitemShut {NoStop}%
	\bibitem [{\citenamefont {Sakurai}\ and\ \citenamefont
		{Napolitano}(2020)}]{Sakurai-Quantum-Mechanics-2021}%
	\BibitemOpen
	\bibfield  {author} {\bibinfo {author} {\bibfnamefont {J.~J.}\ \bibnamefont
			{Sakurai}}\ and\ \bibinfo {author} {\bibfnamefont {J.}~\bibnamefont
			{Napolitano}},\ }\href@noop {} {\emph {\bibinfo {title} {{Modern Quantum
					Mechanics}}}},\ \bibinfo {edition} {3rd}\ ed.\ (\bibinfo  {publisher}
	{Cambridge University Press},\ \bibinfo {address} {Cambridge, UK},\ \bibinfo
	{year} {2020})\ pp.\ \bibinfo {pages} {83--88}\BibitemShut {NoStop}%
	\bibitem [{\citenamefont {Cohen-Tannoudji}\ \emph {et~al.}(2019)\citenamefont
		{Cohen-Tannoudji}, \citenamefont {Diu},\ and\ \citenamefont
		{Lalo{\"e}}}]{Cohen-Tannoudji-et-al-QM-vol-1}%
	\BibitemOpen
	\bibfield  {author} {\bibinfo {author} {\bibfnamefont {C.}~\bibnamefont
			{Cohen-Tannoudji}}, \bibinfo {author} {\bibfnamefont {B.}~\bibnamefont
			{Diu}},\ and\ \bibinfo {author} {\bibfnamefont {F.}~\bibnamefont
			{Lalo{\"e}}},\ }\href@noop {} {\emph {\bibinfo {title} {Quantum mechanics,
				volume 1: Basic Concepts, Tools, and Applications}}}\ (\bibinfo  {publisher}
	{John Wiley \& Sons},\ \bibinfo {year} {2019})\BibitemShut {NoStop}%
	\bibitem [{\citenamefont {Griffiths}(2005)}]{Griffiths-Quantum-Mechanics}%
	\BibitemOpen
	\bibfield  {author} {\bibinfo {author} {\bibfnamefont {D.~J.}\ \bibnamefont
			{Griffiths}},\ }\href@noop {} {\emph {\bibinfo {title} {{Introduction to
					Quantum Mechanics}}}},\ \bibinfo {edition} {2nd}\ ed.\ (\bibinfo  {publisher}
	{Pearson Education},\ \bibinfo {year} {2005})\BibitemShut {NoStop}%
	\bibitem [{\citenamefont {Abe}\ \emph {et~al.}(2009)\citenamefont {Abe},
		\citenamefont {Itto},\ and\ \citenamefont {Matsunaga}}]{Abe-EJP-2009}%
	\BibitemOpen
	\bibfield  {author} {\bibinfo {author} {\bibfnamefont {S.}~\bibnamefont
			{Abe}}, \bibinfo {author} {\bibfnamefont {Y.}~\bibnamefont {Itto}},\ and\
		\bibinfo {author} {\bibfnamefont {M.}~\bibnamefont {Matsunaga}},\ }\bibfield
	{title} {\bibinfo {title} {{On Noether's theorem for the invariant of the
				time-dependent harmonic oscillator}},\ }\href
	{https://doi.org/10.1088/0143-0807/30/6/011} {\bibfield  {journal} {\bibinfo
			{journal} {Eur. J. Phys.}\ }\textbf {\bibinfo {volume} {30}},\ \bibinfo
		{pages} {1337} (\bibinfo {year} {2009})}\BibitemShut {NoStop}%
	\bibitem [{\citenamefont {Nation}\ \emph {et~al.}(2012)\citenamefont {Nation},
		\citenamefont {Johansson}, \citenamefont {Blencowe},\ and\ \citenamefont
		{Nori}}]{Nation-RMP-2012}%
	\BibitemOpen
	\bibfield  {author} {\bibinfo {author} {\bibfnamefont {P.~D.}\ \bibnamefont
			{Nation}}, \bibinfo {author} {\bibfnamefont {J.~R.}\ \bibnamefont
			{Johansson}}, \bibinfo {author} {\bibfnamefont {M.~P.}\ \bibnamefont
			{Blencowe}},\ and\ \bibinfo {author} {\bibfnamefont {F.}~\bibnamefont
			{Nori}},\ }\bibfield  {title} {\bibinfo {title} {{Colloquium : Stimulating
				uncertainty: Amplifying the quantum vacuum with superconducting circuits}},\
	}\href {https://doi.org/10.1103/RevModPhys.84.1} {\bibfield  {journal}
		{\bibinfo  {journal} {Rev. Mod. Phys.}\ }\textbf {\bibinfo {volume} {84}},\
		\bibinfo {pages} {1} (\bibinfo {year} {2012})}\BibitemShut {NoStop}%
	\bibitem [{\citenamefont {Griffiths}\ and\ \citenamefont
		{Schroeter}(2018)}]{Griffiths-Quantum-Mechanics-2018}%
	\BibitemOpen
	\bibfield  {author} {\bibinfo {author} {\bibfnamefont {D.~J.}\ \bibnamefont
			{Griffiths}}\ and\ \bibinfo {author} {\bibfnamefont {D.~F.}\ \bibnamefont
			{Schroeter}},\ }\href {https://doi.org/10.1017/9781316995433} {\emph
		{\bibinfo {title} {{Introduction to Quantum Mechanics}}}},\ \bibinfo
	{edition} {3rd}\ ed.\ (\bibinfo  {publisher} {Cambridge University Press},\
	\bibinfo {address} {Cambridge, UK},\ \bibinfo {year} {2018})\ p.\ \bibinfo
	{pages} {550}\BibitemShut {NoStop}%
	\bibitem [{\citenamefont {Husimi}(1953)}]{Husimi-PTP-1953-II}%
	\BibitemOpen
	\bibfield  {author} {\bibinfo {author} {\bibfnamefont {K.}~\bibnamefont
			{Husimi}},\ }\bibfield  {title} {\bibinfo {title} {{Miscellanea in Elementary
				Quantum Mechanics, II}},\ }\href {https://doi.org/10.1143/ptp/9.4.381}
	{\bibfield  {journal} {\bibinfo  {journal} {Prog. Theor. Phys.}\ }\textbf
		{\bibinfo {volume} {9}},\ \bibinfo {pages} {381} (\bibinfo {year}
		{1953})}\BibitemShut {NoStop}%
	\bibitem [{\citenamefont {Lewis}\ and\ \citenamefont
		{Riesenfeld}(1969)}]{Lewis-JMP-1969}%
	\BibitemOpen
	\bibfield  {author} {\bibinfo {author} {\bibfnamefont {H.~R.}\ \bibnamefont
			{Lewis}}\ and\ \bibinfo {author} {\bibfnamefont {W.~B.}\ \bibnamefont
			{Riesenfeld}},\ }\bibfield  {title} {\bibinfo {title} {{An Exact Quantum
				Theory of the Time‐Dependent Harmonic Oscillator and of a Charged Particle
				in a Time‐Dependent Electromagnetic Field}},\ }\href
	{https://doi.org/10.1063/1.1664991} {\bibfield  {journal} {\bibinfo
			{journal} {J. Math. Phys.}\ }\textbf {\bibinfo {volume} {10}},\ \bibinfo
		{pages} {1458} (\bibinfo {year} {1969})}\BibitemShut {NoStop}%
	\bibitem [{\citenamefont {Pedrosa}(1997)}]{Pedrosa-PRA-1997}%
	\BibitemOpen
	\bibfield  {author} {\bibinfo {author} {\bibfnamefont {I.~A.}\ \bibnamefont
			{Pedrosa}},\ }\bibfield  {title} {\bibinfo {title} {{Exact wave functions of
				a harmonic oscillator with time-dependent mass and frequency}},\ }\href
	{https://doi.org/10.1103/PhysRevA.55.3219} {\bibfield  {journal} {\bibinfo
			{journal} {Phys. Rev. A}\ }\textbf {\bibinfo {volume} {55}},\ \bibinfo
		{pages} {3219} (\bibinfo {year} {1997})}\BibitemShut {NoStop}%
	\bibitem [{\citenamefont {Pedrosa}\ \emph {et~al.}(1997)\citenamefont
		{Pedrosa}, \citenamefont {Serra},\ and\ \citenamefont
		{Guedes}}]{Pedrosa-PRA-1997-singular-perturbation}%
	\BibitemOpen
	\bibfield  {author} {\bibinfo {author} {\bibfnamefont {I.~A.}\ \bibnamefont
			{Pedrosa}}, \bibinfo {author} {\bibfnamefont {G.~P.}\ \bibnamefont {Serra}},\
		and\ \bibinfo {author} {\bibfnamefont {I.}~\bibnamefont {Guedes}},\
	}\bibfield  {title} {\bibinfo {title} {{Wave functions of a time-dependent
				harmonic oscillator with and without a singular perturbation}},\ }\href
	{https://doi.org/10.1103/PhysRevA.56.4300} {\bibfield  {journal} {\bibinfo
			{journal} {Phys. Rev. A}\ }\textbf {\bibinfo {volume} {56}},\ \bibinfo
		{pages} {4300} (\bibinfo {year} {1997})}\BibitemShut {NoStop}%
	\bibitem [{\citenamefont {Ciftja}(1999)}]{Ciftja-JPA-1999}%
	\BibitemOpen
	\bibfield  {author} {\bibinfo {author} {\bibfnamefont {O.}~\bibnamefont
			{Ciftja}},\ }\bibfield  {title} {\bibinfo {title} {{A simple derivation of
				the exact wavefunction of a harmonic oscillator with time-dependent mass and
				frequency}},\ }\href {https://doi.org/10.1088/0305-4470/32/36/303} {\bibfield
		{journal} {\bibinfo  {journal} {J. Phys. A. Math. Gen.}\ }\textbf {\bibinfo
			{volume} {32}},\ \bibinfo {pages} {6385} (\bibinfo {year}
		{1999})}\BibitemShut {NoStop}%
	\bibitem [{\citenamefont {Nascimento}\ \emph {et~al.}(2021)\citenamefont
		{Nascimento}, \citenamefont {Aguiar}, \citenamefont {Cortez},\ and\
		\citenamefont {Guedes}}]{Nascimento-RBEF-2021}%
	\BibitemOpen
	\bibfield  {author} {\bibinfo {author} {\bibfnamefont {J.~P.~G.}\
			\bibnamefont {Nascimento}}, \bibinfo {author} {\bibfnamefont
			{V.}~\bibnamefont {Aguiar}}, \bibinfo {author} {\bibfnamefont {D.~S.}\
			\bibnamefont {Cortez}},\ and\ \bibinfo {author} {\bibfnamefont
			{I.}~\bibnamefont {Guedes}},\ }\bibfield  {title} {\bibinfo {title}
		{{Dissipative Dynamics and Uncertainty Measures of a Charged Oscillator in
				the Presence of the Aharonov-Bohm Effect}},\ }\href
	{https://doi.org/10.1590/1806-9126-rbef-2020-0464} {\bibfield  {journal}
		{\bibinfo  {journal} {Rev. Bras. Ensino F{\'{i}}sica}\ }\textbf {\bibinfo
			{volume} {43}},\ \bibinfo {pages} {1} (\bibinfo {year} {2021})}\BibitemShut
	{NoStop}%
	\bibitem [{\citenamefont {Dodonov}\ \emph {et~al.}(1994)\citenamefont
		{Dodonov}, \citenamefont {Man'ko},\ and\ \citenamefont
		{Polynkin}}]{Dodonov-PLA-1994}%
	\BibitemOpen
	\bibfield  {author} {\bibinfo {author} {\bibfnamefont {V.}~\bibnamefont
			{Dodonov}}, \bibinfo {author} {\bibfnamefont {V.}~\bibnamefont {Man'ko}},\
		and\ \bibinfo {author} {\bibfnamefont {P.}~\bibnamefont {Polynkin}},\
	}\bibfield  {title} {\bibinfo {title} {{Geometrical squeezed states of a
				charged particle in a time-dependent magnetic field}},\ }\href
	{https://doi.org/10.1016/0375-9601(94)90444-8} {\bibfield  {journal}
		{\bibinfo  {journal} {Phys. Lett. A}\ }\textbf {\bibinfo {volume} {188}},\
		\bibinfo {pages} {232} (\bibinfo {year} {1994})}\BibitemShut {NoStop}%
	\bibitem [{\citenamefont {Aguiar}\ and\ \citenamefont
		{Guedes}(2016)}]{Aguiar-JMP-2016}%
	\BibitemOpen
	\bibfield  {author} {\bibinfo {author} {\bibfnamefont {V.}~\bibnamefont
			{Aguiar}}\ and\ \bibinfo {author} {\bibfnamefont {I.}~\bibnamefont
			{Guedes}},\ }\bibfield  {title} {\bibinfo {title} {{Entropy and information
				of a spinless charged particle in time-varying magnetic fields}},\ }\href
	{https://doi.org/10.1063/1.4962923} {\bibfield  {journal} {\bibinfo
			{journal} {J. Math. Phys.}\ }\textbf {\bibinfo {volume} {57}},\ \bibinfo
		{pages} {092103} (\bibinfo {year} {2016})}\BibitemShut {NoStop}%
	\bibitem [{\citenamefont {Dodonov}\ and\ \citenamefont
		{Horovits}(2018)}]{Dodonov-JRLR-2018}%
	\BibitemOpen
	\bibfield  {author} {\bibinfo {author} {\bibfnamefont {V.~V.}\ \bibnamefont
			{Dodonov}}\ and\ \bibinfo {author} {\bibfnamefont {M.~B.}\ \bibnamefont
			{Horovits}},\ }\bibfield  {title} {\bibinfo {title} {{Squeezing of Relative
				and Center-of-Orbit Coordinates of a Charged Particle by Step-Wise Variations
				of a Uniform Magnetic Field with an Arbitrary Linear Vector Potential}},\
	}\href {https://doi.org/10.1007/s10946-018-9733-1} {\bibfield  {journal}
		{\bibinfo  {journal} {J. Russ. Laser Res.}\ }\textbf {\bibinfo {volume}
			{39}},\ \bibinfo {pages} {389} (\bibinfo {year} {2018})}\BibitemShut
	{NoStop}%
	\bibitem [{\citenamefont {Pedrosa}\ and\ \citenamefont
		{Rosas}(2009)}]{Pedrosa-PRL-2009}%
	\BibitemOpen
	\bibfield  {author} {\bibinfo {author} {\bibfnamefont {I.~A.}\ \bibnamefont
			{Pedrosa}}\ and\ \bibinfo {author} {\bibfnamefont {A.}~\bibnamefont
			{Rosas}},\ }\bibfield  {title} {\bibinfo {title} {{Electromagnetic Field
				Quantization in Time-Dependent Linear Media}},\ }\href
	{https://doi.org/10.1103/PhysRevLett.103.010402} {\bibfield  {journal}
		{\bibinfo  {journal} {Phys. Rev. Lett.}\ }\textbf {\bibinfo {volume} {103}},\
		\bibinfo {pages} {010402} (\bibinfo {year} {2009})}\BibitemShut {NoStop}%
	\bibitem [{\citenamefont {Choi}(2010)}]{Choi-PRA-2010}%
	\BibitemOpen
	\bibfield  {author} {\bibinfo {author} {\bibfnamefont {J.~R.}\ \bibnamefont
			{Choi}},\ }\bibfield  {title} {\bibinfo {title} {{Interpreting quantum states
				of electromagnetic field in time-dependent linear media}},\ }\href
	{https://doi.org/10.1103/PhysRevA.82.055803} {\bibfield  {journal} {\bibinfo
			{journal} {Phys. Rev. A}\ }\textbf {\bibinfo {volume} {82}},\ \bibinfo
		{pages} {055803} (\bibinfo {year} {2010})}\BibitemShut {NoStop}%
	\bibitem [{\citenamefont {Pedrosa}(2011)}]{Pedrosa-PRA-2011}%
	\BibitemOpen
	\bibfield  {author} {\bibinfo {author} {\bibfnamefont {I.~A.}\ \bibnamefont
			{Pedrosa}},\ }\bibfield  {title} {\bibinfo {title} {{Quantum electromagnetic
				waves in nonstationary linear media}},\ }\href
	{https://doi.org/10.1103/PhysRevA.83.032108} {\bibfield  {journal} {\bibinfo
			{journal} {Phys. Rev. A}\ }\textbf {\bibinfo {volume} {83}},\ \bibinfo
		{pages} {032108} (\bibinfo {year} {2011})}\BibitemShut {NoStop}%
	\bibitem [{\citenamefont {{\"{U}}nal}(2012)}]{Unal-AP-2012}%
	\BibitemOpen
	\bibfield  {author} {\bibinfo {author} {\bibfnamefont {N.}~\bibnamefont
			{{\"{U}}nal}},\ }\bibfield  {title} {\bibinfo {title} {{Quasi-coherent states
				for a photon in time varying dielectric media}},\ }\href
	{https://doi.org/10.1016/j.aop.2012.05.005} {\bibfield  {journal} {\bibinfo
			{journal} {Ann. Phys. (N. Y).}\ }\textbf {\bibinfo {volume} {327}},\ \bibinfo
		{pages} {2177} (\bibinfo {year} {2012})}\BibitemShut {NoStop}%
	\bibitem [{\citenamefont {Lakehal}\ \emph {et~al.}(2016)\citenamefont
		{Lakehal}, \citenamefont {Maamache},\ and\ \citenamefont
		{Choi}}]{Lakehal-SR-2016}%
	\BibitemOpen
	\bibfield  {author} {\bibinfo {author} {\bibfnamefont {H.}~\bibnamefont
			{Lakehal}}, \bibinfo {author} {\bibfnamefont {M.}~\bibnamefont {Maamache}},\
		and\ \bibinfo {author} {\bibfnamefont {J.~R.}\ \bibnamefont {Choi}},\
	}\bibfield  {title} {\bibinfo {title} {{Novel quantum description for
				nonadiabatic evolution of light wave propagation in time-dependent linear
				media}},\ }\href {https://doi.org/10.1038/srep19860} {\bibfield  {journal}
		{\bibinfo  {journal} {Sci. Rep.}\ }\textbf {\bibinfo {volume} {6}},\ \bibinfo
		{pages} {19860} (\bibinfo {year} {2016})}\BibitemShut {NoStop}%
	\bibitem [{\citenamefont {Brown}(1991)}]{Brown-PRL-1991}%
	\BibitemOpen
	\bibfield  {author} {\bibinfo {author} {\bibfnamefont {L.~S.}\ \bibnamefont
			{Brown}},\ }\bibfield  {title} {\bibinfo {title} {{Quantum motion in a Paul
				trap}},\ }\href {https://doi.org/10.1103/PhysRevLett.66.527} {\bibfield
		{journal} {\bibinfo  {journal} {Phys. Rev. Lett.}\ }\textbf {\bibinfo
			{volume} {66}},\ \bibinfo {pages} {527} (\bibinfo {year} {1991})}\BibitemShut
	{NoStop}%
	\bibitem [{\citenamefont {Alsing}\ \emph {et~al.}(2005)\citenamefont {Alsing},
		\citenamefont {Dowling},\ and\ \citenamefont {Milburn}}]{Alsing-PRL-2005}%
	\BibitemOpen
	\bibfield  {author} {\bibinfo {author} {\bibfnamefont {P.~M.}\ \bibnamefont
			{Alsing}}, \bibinfo {author} {\bibfnamefont {J.~P.}\ \bibnamefont
			{Dowling}},\ and\ \bibinfo {author} {\bibfnamefont {G.~J.}\ \bibnamefont
			{Milburn}},\ }\bibfield  {title} {\bibinfo {title} {{Ion Trap Simulations of
				Quantum Fields in an Expanding Universe}},\ }\href
	{https://doi.org/10.1103/PhysRevLett.94.220401} {\bibfield  {journal}
		{\bibinfo  {journal} {Phys. Rev. Lett.}\ }\textbf {\bibinfo {volume} {94}},\
		\bibinfo {pages} {220401} (\bibinfo {year} {2005})}\BibitemShut {NoStop}%
	\bibitem [{\citenamefont {Menicucci}\ and\ \citenamefont
		{Milburn}(2007)}]{Menicucci-PRA-2007}%
	\BibitemOpen
	\bibfield  {author} {\bibinfo {author} {\bibfnamefont {N.~C.}\ \bibnamefont
			{Menicucci}}\ and\ \bibinfo {author} {\bibfnamefont {G.~J.}\ \bibnamefont
			{Milburn}},\ }\bibfield  {title} {\bibinfo {title} {{Single trapped ion as a
				time-dependent harmonic oscillator}},\ }\href
	{https://doi.org/10.1103/PhysRevA.76.052105} {\bibfield  {journal} {\bibinfo
			{journal} {Phys. Rev. A}\ }\textbf {\bibinfo {volume} {76}},\ \bibinfo
		{pages} {052105} (\bibinfo {year} {2007})}\BibitemShut {NoStop}%
	\bibitem [{\citenamefont {Mihalcea}(2009)}]{Mihalcea-PS-2009}%
	\BibitemOpen
	\bibfield  {author} {\bibinfo {author} {\bibfnamefont {B.~M.}\ \bibnamefont
			{Mihalcea}},\ }\bibfield  {title} {\bibinfo {title} {{A quantum parametric
				oscillator in a radiofrequency trap}},\ }\href
	{https://doi.org/10.1088/0031-8949/2009/T135/014006} {\bibfield  {journal}
		{\bibinfo  {journal} {Phys. Scr.}\ }\textbf {\bibinfo {volume} {T135}},\
		\bibinfo {pages} {014006} (\bibinfo {year} {2009})}\BibitemShut {NoStop}%
	\bibitem [{\citenamefont {Mihalcea}(2022)}]{Mihalcea-AP-2022}%
	\BibitemOpen
	\bibfield  {author} {\bibinfo {author} {\bibfnamefont {B.~M.}\ \bibnamefont
			{Mihalcea}},\ }\bibfield  {title} {\bibinfo {title} {{Quasienergy operators
				and generalized squeezed states for systems of trapped ions}},\ }\href
	{https://doi.org/10.1016/j.aop.2022.168926} {\bibfield  {journal} {\bibinfo
			{journal} {Ann. Phys. (N. Y).}\ }\textbf {\bibinfo {volume} {442}},\ \bibinfo
		{pages} {168926} (\bibinfo {year} {2022})}\BibitemShut {NoStop}%
	\bibitem [{\citenamefont {Mihalcea}\ \emph {et~al.}(2023)\citenamefont
		{Mihalcea}, \citenamefont {Filinov}, \citenamefont {Syrovatka},\ and\
		\citenamefont {Vasilyak}}]{Mihalcea-PR-2023}%
	\BibitemOpen
	\bibfield  {author} {\bibinfo {author} {\bibfnamefont {B.~M.}\ \bibnamefont
			{Mihalcea}}, \bibinfo {author} {\bibfnamefont {V.~S.}\ \bibnamefont
			{Filinov}}, \bibinfo {author} {\bibfnamefont {R.~A.}\ \bibnamefont
			{Syrovatka}},\ and\ \bibinfo {author} {\bibfnamefont {L.~M.}\ \bibnamefont
			{Vasilyak}},\ }\bibfield  {title} {\bibinfo {title} {{The physics and
				applications of strongly coupled Coulomb systems (plasmas) levitated in
				electrodynamic traps}},\ }\href
	{https://doi.org/10.1016/j.physrep.2023.03.004} {\bibfield  {journal}
		{\bibinfo  {journal} {Phys. Rep.}\ }\textbf {\bibinfo {volume} {1016}},\
		\bibinfo {pages} {1} (\bibinfo {year} {2023})}\BibitemShut {NoStop}%
	\bibitem [{\citenamefont {Mihalcea}(2024)}]{Mihalcea-Mathematics-2024}%
	\BibitemOpen
	\bibfield  {author} {\bibinfo {author} {\bibfnamefont {B.~M.}\ \bibnamefont
			{Mihalcea}},\ }\bibfield  {title} {\bibinfo {title} {{Solutions of the
				Mathieu–Hill Equation for a Trapped-Ion Harmonic Oscillator—A Qualitative
				Discussion}},\ }\href {https://doi.org/10.3390/math12192963} {\bibfield
		{journal} {\bibinfo  {journal} {Mathematics}\ }\textbf {\bibinfo {volume}
			{12}},\ \bibinfo {pages} {2963} (\bibinfo {year} {2024})}\BibitemShut
	{NoStop}%
	\bibitem [{\citenamefont {Paul}(1990)}]{Paul-RMP-1990}%
	\BibitemOpen
	\bibfield  {author} {\bibinfo {author} {\bibfnamefont {W.}~\bibnamefont
			{Paul}},\ }\bibfield  {title} {\bibinfo {title} {{Electromagnetic traps for
				charged and neutral particles}},\ }\href
	{https://doi.org/10.1103/RevModPhys.62.531} {\bibfield  {journal} {\bibinfo
			{journal} {Rev. Mod. Phys.}\ }\textbf {\bibinfo {volume} {62}},\ \bibinfo
		{pages} {531} (\bibinfo {year} {1990})}\BibitemShut {NoStop}%
	\bibitem [{\citenamefont {Carvalho}\ \emph {et~al.}(2004)\citenamefont
		{Carvalho}, \citenamefont {Furtado},\ and\ \citenamefont
		{Pedrosa}}]{Pedrosa-PRD-2004}%
	\BibitemOpen
	\bibfield  {author} {\bibinfo {author} {\bibfnamefont {A.~M. d.~M.}\
			\bibnamefont {Carvalho}}, \bibinfo {author} {\bibfnamefont {C.}~\bibnamefont
			{Furtado}},\ and\ \bibinfo {author} {\bibfnamefont {I.~A.}\ \bibnamefont
			{Pedrosa}},\ }\bibfield  {title} {\bibinfo {title} {{Scalar fields and exact
				invariants in a Friedmann-Robertson-Walker spacetime}},\ }\href
	{https://doi.org/10.1103/PhysRevD.70.123523} {\bibfield  {journal} {\bibinfo
			{journal} {Phys. Rev. D}\ }\textbf {\bibinfo {volume} {70}},\ \bibinfo
		{pages} {123523} (\bibinfo {year} {2004})}\BibitemShut {NoStop}%
	\bibitem [{\citenamefont {Salamon}\ \emph {et~al.}(2009)\citenamefont
		{Salamon}, \citenamefont {Hoffmann}, \citenamefont {Rezek},\ and\
		\citenamefont {Kosloff}}]{Salamon-PCCP-2009}%
	\BibitemOpen
	\bibfield  {author} {\bibinfo {author} {\bibfnamefont {P.}~\bibnamefont
			{Salamon}}, \bibinfo {author} {\bibfnamefont {K.~H.}\ \bibnamefont
			{Hoffmann}}, \bibinfo {author} {\bibfnamefont {Y.}~\bibnamefont {Rezek}},\
		and\ \bibinfo {author} {\bibfnamefont {R.}~\bibnamefont {Kosloff}},\
	}\bibfield  {title} {\bibinfo {title} {{Maximum work in minimum time from a
				conservative quantum system}},\ }\href {https://doi.org/10.1039/B816102J}
	{\bibfield  {journal} {\bibinfo  {journal} {Phys. Chem. Chem. Phys.}\
		}\textbf {\bibinfo {volume} {11}},\ \bibinfo {pages} {1027} (\bibinfo {year}
		{2009})}\BibitemShut {NoStop}%
	\bibitem [{\citenamefont {Chen}\ \emph {et~al.}(2010)\citenamefont {Chen},
		\citenamefont {Ruschhaupt}, \citenamefont {Schmidt}, \citenamefont {del
			Campo}, \citenamefont {Gu{\'{e}}ry-Odelin},\ and\ \citenamefont
		{Muga}}]{Chen-PRL-2010}%
	\BibitemOpen
	\bibfield  {author} {\bibinfo {author} {\bibfnamefont {X.}~\bibnamefont
			{Chen}}, \bibinfo {author} {\bibfnamefont {A.}~\bibnamefont {Ruschhaupt}},
		\bibinfo {author} {\bibfnamefont {S.}~\bibnamefont {Schmidt}}, \bibinfo
		{author} {\bibfnamefont {A.}~\bibnamefont {del Campo}}, \bibinfo {author}
		{\bibfnamefont {D.}~\bibnamefont {Gu{\'{e}}ry-Odelin}},\ and\ \bibinfo
		{author} {\bibfnamefont {J.~G.}\ \bibnamefont {Muga}},\ }\bibfield  {title}
	{\bibinfo {title} {{Fast Optimal Frictionless Atom Cooling in Harmonic Traps:
				Shortcut to Adiabaticity}},\ }\href
	{https://doi.org/10.1103/PhysRevLett.104.063002} {\bibfield  {journal}
		{\bibinfo  {journal} {Phys. Rev. Lett.}\ }\textbf {\bibinfo {volume} {104}},\
		\bibinfo {pages} {063002} (\bibinfo {year} {2010})}\BibitemShut {NoStop}%
	\bibitem [{\citenamefont {Chen}\ and\ \citenamefont
		{Muga}(2010)}]{Chen-PRA-2010}%
	\BibitemOpen
	\bibfield  {author} {\bibinfo {author} {\bibfnamefont {X.}~\bibnamefont
			{Chen}}\ and\ \bibinfo {author} {\bibfnamefont {J.~G.}\ \bibnamefont
			{Muga}},\ }\bibfield  {title} {\bibinfo {title} {{Transient energy excitation
				in shortcuts to adiabaticity for the time-dependent harmonic oscillator}},\
	}\href {https://doi.org/10.1103/PhysRevA.82.053403} {\bibfield  {journal}
		{\bibinfo  {journal} {Phys. Rev. A}\ }\textbf {\bibinfo {volume} {82}},\
		\bibinfo {pages} {053403} (\bibinfo {year} {2010})}\BibitemShut {NoStop}%
	\bibitem [{\citenamefont {Stefanatos}\ \emph {et~al.}(2010)\citenamefont
		{Stefanatos}, \citenamefont {Ruths},\ and\ \citenamefont
		{Li}}]{Stefanatos-PRA-2010}%
	\BibitemOpen
	\bibfield  {author} {\bibinfo {author} {\bibfnamefont {D.}~\bibnamefont
			{Stefanatos}}, \bibinfo {author} {\bibfnamefont {J.}~\bibnamefont {Ruths}},\
		and\ \bibinfo {author} {\bibfnamefont {J.-S.}\ \bibnamefont {Li}},\
	}\bibfield  {title} {\bibinfo {title} {{Frictionless atom cooling in harmonic
				traps: A time-optimal approach}},\ }\href
	{https://doi.org/10.1103/PhysRevA.82.063422} {\bibfield  {journal} {\bibinfo
			{journal} {Phys. Rev. A}\ }\textbf {\bibinfo {volume} {82}},\ \bibinfo
		{pages} {063422} (\bibinfo {year} {2010})}\BibitemShut {NoStop}%
	\bibitem [{\citenamefont {Choi}\ \emph {et~al.}(2012)\citenamefont {Choi},
		\citenamefont {Onofrio},\ and\ \citenamefont {Sundaram}}]{Choi-PRA-2012}%
	\BibitemOpen
	\bibfield  {author} {\bibinfo {author} {\bibfnamefont {S.}~\bibnamefont
			{Choi}}, \bibinfo {author} {\bibfnamefont {R.}~\bibnamefont {Onofrio}},\ and\
		\bibinfo {author} {\bibfnamefont {B.}~\bibnamefont {Sundaram}},\ }\bibfield
	{title} {\bibinfo {title} {{Squeezing and robustness of frictionless cooling
				strategies}},\ }\href {https://doi.org/10.1103/PhysRevA.86.043436} {\bibfield
		{journal} {\bibinfo  {journal} {Phys. Rev. A}\ }\textbf {\bibinfo {volume}
			{86}},\ \bibinfo {pages} {043436} (\bibinfo {year} {2012})}\BibitemShut
	{NoStop}%
	\bibitem [{\citenamefont {Choi}\ \emph {et~al.}(2013)\citenamefont {Choi},
		\citenamefont {Onofrio},\ and\ \citenamefont {Sundaram}}]{Choi-PRA-2013}%
	\BibitemOpen
	\bibfield  {author} {\bibinfo {author} {\bibfnamefont {S.}~\bibnamefont
			{Choi}}, \bibinfo {author} {\bibfnamefont {R.}~\bibnamefont {Onofrio}},\ and\
		\bibinfo {author} {\bibfnamefont {B.}~\bibnamefont {Sundaram}},\ }\bibfield
	{title} {\bibinfo {title} {{Ehrenfest dynamics and frictionless cooling
				methods}},\ }\href {https://doi.org/10.1103/PhysRevA.88.053401} {\bibfield
		{journal} {\bibinfo  {journal} {Phys. Rev. A}\ }\textbf {\bibinfo {volume}
			{88}},\ \bibinfo {pages} {053401} (\bibinfo {year} {2013})}\BibitemShut
	{NoStop}%
	\bibitem [{\citenamefont {Kiely}\ \emph {et~al.}(2015)\citenamefont {Kiely},
		\citenamefont {McGuinness}, \citenamefont {Muga},\ and\ \citenamefont
		{Ruschhaupt}}]{Kiely-JPB-2015}%
	\BibitemOpen
	\bibfield  {author} {\bibinfo {author} {\bibfnamefont {A.}~\bibnamefont
			{Kiely}}, \bibinfo {author} {\bibfnamefont {J.~P.~L.}\ \bibnamefont
			{McGuinness}}, \bibinfo {author} {\bibfnamefont {J.~G.}\ \bibnamefont
			{Muga}},\ and\ \bibinfo {author} {\bibfnamefont {A.}~\bibnamefont
			{Ruschhaupt}},\ }\bibfield  {title} {\bibinfo {title} {{Fast and stable
				manipulation of a charged particle in a Penning trap}},\ }\href
	{https://doi.org/10.1088/0953-4075/48/7/075503} {\bibfield  {journal}
		{\bibinfo  {journal} {J. Phys. B At. Mol. Opt. Phys.}\ }\textbf {\bibinfo
			{volume} {48}},\ \bibinfo {pages} {075503} (\bibinfo {year}
		{2015})}\BibitemShut {NoStop}%
	\bibitem [{\citenamefont {Gu{\'{e}}ry-Odelin}\ \emph
		{et~al.}(2019)\citenamefont {Gu{\'{e}}ry-Odelin}, \citenamefont {Ruschhaupt},
		\citenamefont {Kiely}, \citenamefont {Torrontegui}, \citenamefont
		{Mart{\'{i}}nez-Garaot},\ and\ \citenamefont {Muga}}]{Odelin-RMP-2019}%
	\BibitemOpen
	\bibfield  {author} {\bibinfo {author} {\bibfnamefont {D.}~\bibnamefont
			{Gu{\'{e}}ry-Odelin}}, \bibinfo {author} {\bibfnamefont {A.}~\bibnamefont
			{Ruschhaupt}}, \bibinfo {author} {\bibfnamefont {A.}~\bibnamefont {Kiely}},
		\bibinfo {author} {\bibfnamefont {E.}~\bibnamefont {Torrontegui}}, \bibinfo
		{author} {\bibfnamefont {S.}~\bibnamefont {Mart{\'{i}}nez-Garaot}},\ and\
		\bibinfo {author} {\bibfnamefont {J.~G.}\ \bibnamefont {Muga}},\ }\bibfield
	{title} {\bibinfo {title} {{Shortcuts to adiabaticity: Concepts, methods, and
				applications}},\ }\href {https://doi.org/10.1103/RevModPhys.91.045001}
	{\bibfield  {journal} {\bibinfo  {journal} {Rev. Mod. Phys.}\ }\textbf
		{\bibinfo {volume} {91}},\ \bibinfo {pages} {045001} (\bibinfo {year}
		{2019})}\BibitemShut {NoStop}%
	\bibitem [{\citenamefont {Beau}\ and\ \citenamefont {del
			Campo}(2020)}]{Beau-Entropy-2020}%
	\BibitemOpen
	\bibfield  {author} {\bibinfo {author} {\bibfnamefont {M.}~\bibnamefont
			{Beau}}\ and\ \bibinfo {author} {\bibfnamefont {A.}~\bibnamefont {del
				Campo}},\ }\bibfield  {title} {\bibinfo {title} {{Nonadiabatic Energy
				Fluctuations of Scale-Invariant Quantum Systems in a Time-Dependent Trap}},\
	}\href {https://doi.org/10.3390/e22050515} {\bibfield  {journal} {\bibinfo
			{journal} {Entropy}\ }\textbf {\bibinfo {volume} {22}},\ \bibinfo {pages}
		{515} (\bibinfo {year} {2020})}\BibitemShut {NoStop}%
	\bibitem [{\citenamefont {Huang}\ \emph {et~al.}(2020)\citenamefont {Huang},
		\citenamefont {Malomed},\ and\ \citenamefont {Chen}}]{Huang-Chaos-2020}%
	\BibitemOpen
	\bibfield  {author} {\bibinfo {author} {\bibfnamefont {T.-Y.}\ \bibnamefont
			{Huang}}, \bibinfo {author} {\bibfnamefont {B.~A.}\ \bibnamefont {Malomed}},\
		and\ \bibinfo {author} {\bibfnamefont {X.}~\bibnamefont {Chen}},\ }\bibfield
	{title} {\bibinfo {title} {{Shortcuts to adiabaticity for an interacting
				Bose–Einstein condensate via exact solutions of the generalized Ermakov
				equation}},\ }\href {https://doi.org/10.1063/5.0004309} {\bibfield  {journal}
		{\bibinfo  {journal} {Chaos}\ }\textbf {\bibinfo {volume} {30}},\ \bibinfo
		{pages} {053131} (\bibinfo {year} {2020})}\BibitemShut {NoStop}%
	\bibitem [{\citenamefont {Dupays}\ \emph {et~al.}(2021)\citenamefont {Dupays},
		\citenamefont {Spierings}, \citenamefont {Steinberg},\ and\ \citenamefont
		{del Campo}}]{Dupays-PRR-2021}%
	\BibitemOpen
	\bibfield  {author} {\bibinfo {author} {\bibfnamefont {L.}~\bibnamefont
			{Dupays}}, \bibinfo {author} {\bibfnamefont {D.~C.}\ \bibnamefont
			{Spierings}}, \bibinfo {author} {\bibfnamefont {A.~M.}\ \bibnamefont
			{Steinberg}},\ and\ \bibinfo {author} {\bibfnamefont {A.}~\bibnamefont {del
				Campo}},\ }\bibfield  {title} {\bibinfo {title} {{Delta-kick cooling,
				time-optimal control of scale-invariant dynamics, and shortcuts to
				adiabaticity assisted by kicks}},\ }\href
	{https://doi.org/10.1103/PhysRevResearch.3.033261} {\bibfield  {journal}
		{\bibinfo  {journal} {Phys. Rev. Res.}\ }\textbf {\bibinfo {volume} {3}},\
		\bibinfo {pages} {033261} (\bibinfo {year} {2021})}\BibitemShut {NoStop}%
	\bibitem [{\citenamefont {{Del Grosso}}\ \emph {et~al.}(2023)\citenamefont
		{{Del Grosso}}, \citenamefont {Lombardo}, \citenamefont {Mazzitelli},\ and\
		\citenamefont {Villar}}]{Grosso-Entropy-2023}%
	\BibitemOpen
	\bibfield  {author} {\bibinfo {author} {\bibfnamefont {N.~F.}\ \bibnamefont
			{{Del Grosso}}}, \bibinfo {author} {\bibfnamefont {F.~C.}\ \bibnamefont
			{Lombardo}}, \bibinfo {author} {\bibfnamefont {F.~D.}\ \bibnamefont
			{Mazzitelli}},\ and\ \bibinfo {author} {\bibfnamefont {P.~I.}\ \bibnamefont
			{Villar}},\ }\bibfield  {title} {\bibinfo {title} {{Adiabatic Shortcuts
				Completion in Quantum Field Theory: Annihilation of Created Particles}},\
	}\href {https://doi.org/10.3390/e25091249} {\bibfield  {journal} {\bibinfo
			{journal} {Entropy}\ }\textbf {\bibinfo {volume} {25}},\ \bibinfo {pages}
		{1249} (\bibinfo {year} {2023})}\BibitemShut {NoStop}%
	\bibitem [{\citenamefont {Santos}(2024)}]{Santos-EPJP-2024}%
	\BibitemOpen
	\bibfield  {author} {\bibinfo {author} {\bibfnamefont {J.~F.~G.}\
			\bibnamefont {Santos}},\ }\bibfield  {title} {\bibinfo {title}
		{{Shortcut-to-adiabaticity for coupled harmonic oscillators}},\ }\href
	{https://doi.org/10.1140/epjp/s13360-024-05718-7} {\bibfield  {journal}
		{\bibinfo  {journal} {Eur. Phys. J. Plus}\ }\textbf {\bibinfo {volume}
			{139}},\ \bibinfo {pages} {909} (\bibinfo {year} {2024})}\BibitemShut
	{NoStop}%
	\bibitem [{\citenamefont {Janszky}\ and\ \citenamefont
		{Yushin}(1986)}]{Janszky-OC-1986}%
	\BibitemOpen
	\bibfield  {author} {\bibinfo {author} {\bibfnamefont {J.}~\bibnamefont
			{Janszky}}\ and\ \bibinfo {author} {\bibfnamefont {Y.}~\bibnamefont
			{Yushin}},\ }\bibfield  {title} {\bibinfo {title} {{Squeezing via frequency
				jump}},\ }\href {https://doi.org/10.1016/0030-4018(86)90468-2} {\bibfield
		{journal} {\bibinfo  {journal} {Opt. Commun.}\ }\textbf {\bibinfo {volume}
			{59}},\ \bibinfo {pages} {151} (\bibinfo {year} {1986})}\BibitemShut
	{NoStop}%
	\bibitem [{\citenamefont {Agarwal}\ and\ \citenamefont
		{Kumar}(1991)}]{Agarwal-PRL-1991}%
	\BibitemOpen
	\bibfield  {author} {\bibinfo {author} {\bibfnamefont {G.~S.}\ \bibnamefont
			{Agarwal}}\ and\ \bibinfo {author} {\bibfnamefont {S.~A.}\ \bibnamefont
			{Kumar}},\ }\bibfield  {title} {\bibinfo {title} {{Exact quantum-statistical
				dynamics of an oscillator with time-dependent frequency and generation of
				nonclassical states}},\ }\href {https://doi.org/10.1103/PhysRevLett.67.3665}
	{\bibfield  {journal} {\bibinfo  {journal} {Phys. Rev. Lett.}\ }\textbf
		{\bibinfo {volume} {67}},\ \bibinfo {pages} {3665} (\bibinfo {year}
		{1991})}\BibitemShut {NoStop}%
	\bibitem [{\citenamefont {Janszky}\ and\ \citenamefont
		{Adam}(1992)}]{Janszky-PRA-1992}%
	\BibitemOpen
	\bibfield  {author} {\bibinfo {author} {\bibfnamefont {J.}~\bibnamefont
			{Janszky}}\ and\ \bibinfo {author} {\bibfnamefont {P.}~\bibnamefont {Adam}},\
	}\bibfield  {title} {\bibinfo {title} {{Strong squeezing by repeated
				frequency jumps}},\ }\href {https://doi.org/10.1103/PhysRevA.46.6091}
	{\bibfield  {journal} {\bibinfo  {journal} {Phys. Rev. A}\ }\textbf {\bibinfo
			{volume} {46}},\ \bibinfo {pages} {6091} (\bibinfo {year}
		{1992})}\BibitemShut {NoStop}%
	\bibitem [{\citenamefont {Degen}\ \emph {et~al.}(2017)\citenamefont {Degen},
		\citenamefont {Reinhard},\ and\ \citenamefont {Cappellaro}}]{Degen-RMP-2017}%
	\BibitemOpen
	\bibfield  {author} {\bibinfo {author} {\bibfnamefont {C.~L.}\ \bibnamefont
			{Degen}}, \bibinfo {author} {\bibfnamefont {F.}~\bibnamefont {Reinhard}},\
		and\ \bibinfo {author} {\bibfnamefont {P.}~\bibnamefont {Cappellaro}},\
	}\bibfield  {title} {\bibinfo {title} {{Quantum sensing}},\ }\href
	{https://doi.org/10.1103/RevModPhys.89.035002} {\bibfield  {journal}
		{\bibinfo  {journal} {Rev. Mod. Phys.}\ }\textbf {\bibinfo {volume} {89}},\
		\bibinfo {pages} {035002} (\bibinfo {year} {2017})}\BibitemShut {NoStop}%
	\bibitem [{\citenamefont {Pezz{\`{e}}}\ \emph {et~al.}(2018)\citenamefont
		{Pezz{\`{e}}}, \citenamefont {Smerzi}, \citenamefont {Oberthaler},
		\citenamefont {Schmied},\ and\ \citenamefont {Treutlein}}]{Pezze-RMP-2018}%
	\BibitemOpen
	\bibfield  {author} {\bibinfo {author} {\bibfnamefont {L.}~\bibnamefont
			{Pezz{\`{e}}}}, \bibinfo {author} {\bibfnamefont {A.}~\bibnamefont {Smerzi}},
		\bibinfo {author} {\bibfnamefont {M.~K.}\ \bibnamefont {Oberthaler}},
		\bibinfo {author} {\bibfnamefont {R.}~\bibnamefont {Schmied}},\ and\ \bibinfo
		{author} {\bibfnamefont {P.}~\bibnamefont {Treutlein}},\ }\bibfield  {title}
	{\bibinfo {title} {{Quantum metrology with nonclassical states of atomic
				ensembles}},\ }\href {https://doi.org/10.1103/RevModPhys.90.035005}
	{\bibfield  {journal} {\bibinfo  {journal} {Rev. Mod. Phys.}\ }\textbf
		{\bibinfo {volume} {90}},\ \bibinfo {pages} {035005} (\bibinfo {year}
		{2018})}\BibitemShut {NoStop}%
	\bibitem [{\citenamefont {{Garc{\'{i}}a Herrera}}\ \emph
		{et~al.}(2023)\citenamefont {{Garc{\'{i}}a Herrera}}, \citenamefont
		{Torres-Leal},\ and\ \citenamefont
		{Rodr{\'{i}}guez-Lara}}]{Herrera-NJP-2023}%
	\BibitemOpen
	\bibfield  {author} {\bibinfo {author} {\bibfnamefont {E.}~\bibnamefont
			{{Garc{\'{i}}a Herrera}}}, \bibinfo {author} {\bibfnamefont {F.}~\bibnamefont
			{Torres-Leal}},\ and\ \bibinfo {author} {\bibfnamefont {B.~M.}\ \bibnamefont
			{Rodr{\'{i}}guez-Lara}},\ }\bibfield  {title} {\bibinfo {title}
		{{Continuous-time quantum harmonic oscillator state engineering}},\ }\href
	{https://doi.org/10.1088/1367-2630/ad149c} {\bibfield  {journal} {\bibinfo
			{journal} {New J. Phys.}\ }\textbf {\bibinfo {volume} {25}},\ \bibinfo
		{pages} {123045} (\bibinfo {year} {2023})}\BibitemShut {NoStop}%
	\bibitem [{\citenamefont {Coelho}\ \emph {et~al.}(2022)\citenamefont {Coelho},
		\citenamefont {Queiroz},\ and\ \citenamefont {Alves}}]{Coelho-Entropy-2022}%
	\BibitemOpen
	\bibfield  {author} {\bibinfo {author} {\bibfnamefont {S.~S.}\ \bibnamefont
			{Coelho}}, \bibinfo {author} {\bibfnamefont {L.}~\bibnamefont {Queiroz}},\
		and\ \bibinfo {author} {\bibfnamefont {D.~T.}\ \bibnamefont {Alves}},\
	}\bibfield  {title} {\bibinfo {title} {{Exact Solution of a Time-Dependent
				Quantum Harmonic Oscillator with Two Frequency Jumps via the
				Lewis–Riesenfeld Dynamical Invariant Method}},\ }\href
	{https://doi.org/10.3390/e24121851} {\bibfield  {journal} {\bibinfo
			{journal} {Entropy}\ }\textbf {\bibinfo {volume} {24}},\ \bibinfo {pages}
		{1851} (\bibinfo {year} {2022})}\BibitemShut {NoStop}%
	\bibitem [{\citenamefont {{S. Coelho}}\ \emph {et~al.}(2024)\citenamefont {{S.
				Coelho}}, \citenamefont {Queiroz},\ and\ \citenamefont
		{Alves}}]{Coelho-PS-2024}%
	\BibitemOpen
	\bibfield  {author} {\bibinfo {author} {\bibfnamefont {S.}~\bibnamefont {{S.
					Coelho}}}, \bibinfo {author} {\bibfnamefont {L.}~\bibnamefont {Queiroz}},\
		and\ \bibinfo {author} {\bibfnamefont {D.~T.}\ \bibnamefont {Alves}},\
	}\bibfield  {title} {\bibinfo {title} {{Squeezing equivalence of quantum
				harmonic oscillators under different frequency modulations}},\ }\href
	{https://doi.org/10.1088/1402-4896/ad56d6} {\bibfield  {journal} {\bibinfo
			{journal} {Phys. Scr.}\ }\textbf {\bibinfo {volume} {99}},\ \bibinfo {pages}
		{085104} (\bibinfo {year} {2024})}\BibitemShut {NoStop}%
	\bibitem [{\citenamefont {Andrews}(1999)}]{Andrews-AJP-1999}%
	\BibitemOpen
	\bibfield  {author} {\bibinfo {author} {\bibfnamefont {M.}~\bibnamefont
			{Andrews}},\ }\bibfield  {title} {\bibinfo {title} {{Invariant operators for
				quadratic Hamiltonians}},\ }\href {https://doi.org/10.1119/1.19259}
	{\bibfield  {journal} {\bibinfo  {journal} {Am. J. Phys.}\ }\textbf {\bibinfo
			{volume} {67}},\ \bibinfo {pages} {336} (\bibinfo {year} {1999})}\BibitemShut
	{NoStop}%
	\bibitem [{\citenamefont {Casta{\~{n}}os}\ and\ \citenamefont
		{Zu{\~{n}}iga-Segundo}(2019)}]{Castanos-AJP-2019}%
	\BibitemOpen
	\bibfield  {author} {\bibinfo {author} {\bibfnamefont {L.~O.}\ \bibnamefont
			{Casta{\~{n}}os}}\ and\ \bibinfo {author} {\bibfnamefont {A.}~\bibnamefont
			{Zu{\~{n}}iga-Segundo}},\ }\bibfield  {title} {\bibinfo {title} {{The forced
				harmonic oscillator: Coherent states and the RWA}},\ }\href
	{https://doi.org/10.1119/1.5115395} {\bibfield  {journal} {\bibinfo
			{journal} {Am. J. Phys.}\ }\textbf {\bibinfo {volume} {87}},\ \bibinfo
		{pages} {815} (\bibinfo {year} {2019})}\BibitemShut {NoStop}%
	\bibitem [{\citenamefont {Leach}(1978)}]{Leach-AJP-1978}%
	\BibitemOpen
	\bibfield  {author} {\bibinfo {author} {\bibfnamefont {P.~G.~L.}\
			\bibnamefont {Leach}},\ }\bibfield  {title} {\bibinfo {title} {{Note on the
				time-dependent damped and forced harmonic oscillator}},\ }\href
	{https://doi.org/10.1119/1.11388} {\bibfield  {journal} {\bibinfo  {journal}
			{Am. J. Phys.}\ }\textbf {\bibinfo {volume} {46}},\ \bibinfo {pages} {1247}
		(\bibinfo {year} {1978})}\BibitemShut {NoStop}%
	\bibitem [{\citenamefont {Lewis}(1967)}]{Lewis-PRL-1967}%
	\BibitemOpen
	\bibfield  {author} {\bibinfo {author} {\bibfnamefont {H.~R.}\ \bibnamefont
			{Lewis}},\ }\bibfield  {title} {\bibinfo {title} {{Classical and Quantum
				Systems with Time-Dependent Harmonic-Oscillator-Type Hamiltonians}},\ }\href
	{https://doi.org/10.1103/PhysRevLett.18.636.2} {\bibfield  {journal}
		{\bibinfo  {journal} {Phys. Rev. Lett.}\ }\textbf {\bibinfo {volume} {18}},\
		\bibinfo {pages} {636} (\bibinfo {year} {1967})}\BibitemShut {NoStop}%
	\bibitem [{\citenamefont {Lewis}(1968)}]{Lewis-JMP-1968}%
	\BibitemOpen
	\bibfield  {author} {\bibinfo {author} {\bibfnamefont {H.~R.}\ \bibnamefont
			{Lewis}},\ }\bibfield  {title} {\bibinfo {title} {{Class of Exact Invariants
				for Classical and Quantum Time‐Dependent Harmonic Oscillators}},\ }\href
	{https://doi.org/10.1063/1.1664532} {\bibfield  {journal} {\bibinfo
			{journal} {J. Math. Phys.}\ }\textbf {\bibinfo {volume} {9}},\ \bibinfo
		{pages} {1976} (\bibinfo {year} {1968})}\BibitemShut {NoStop}%
	\bibitem [{\citenamefont {Lutzky}(1978)}]{Lutzky-PLA-1978}%
	\BibitemOpen
	\bibfield  {author} {\bibinfo {author} {\bibfnamefont {M.}~\bibnamefont
			{Lutzky}},\ }\bibfield  {title} {\bibinfo {title} {{Noether's theorem and the
				time-dependent harmonic oscillator}},\ }\href
	{https://doi.org/10.1016/0375-9601(78)90738-7} {\bibfield  {journal}
		{\bibinfo  {journal} {Phys. Lett. A}\ }\textbf {\bibinfo {volume} {68}},\
		\bibinfo {pages} {3} (\bibinfo {year} {1978})}\BibitemShut {NoStop}%
	\bibitem [{\citenamefont {Malkin}\ \emph {et~al.}(1970)\citenamefont {Malkin},
		\citenamefont {Man'ko},\ and\ \citenamefont {Trifonov}}]{Malkin-PRD-1970}%
	\BibitemOpen
	\bibfield  {author} {\bibinfo {author} {\bibfnamefont {I.~A.}\ \bibnamefont
			{Malkin}}, \bibinfo {author} {\bibfnamefont {V.~I.}\ \bibnamefont {Man'ko}},\
		and\ \bibinfo {author} {\bibfnamefont {D.~A.}\ \bibnamefont {Trifonov}},\
	}\bibfield  {title} {\bibinfo {title} {{Coherent States and Transition
				Probabilities in a Time-Dependent Electromagnetic Field}},\ }\href
	{https://doi.org/10.1103/PhysRevD.2.1371} {\bibfield  {journal} {\bibinfo
			{journal} {Phys. Rev. D}\ }\textbf {\bibinfo {volume} {2}},\ \bibinfo {pages}
		{1371} (\bibinfo {year} {1970})}\BibitemShut {NoStop}%
	\bibitem [{\citenamefont {Hernandez}\ and\ \citenamefont
		{Remaud}(1980)}]{Hernandez-PLA-1980}%
	\BibitemOpen
	\bibfield  {author} {\bibinfo {author} {\bibfnamefont {E.}~\bibnamefont
			{Hernandez}}\ and\ \bibinfo {author} {\bibfnamefont {B.}~\bibnamefont
			{Remaud}},\ }\bibfield  {title} {\bibinfo {title} {{Quantal fluctuations and
				invariant operators for a general time-dependent harmonic oscillator}},\
	}\href {https://doi.org/10.1016/0375-9601(80)90560-5} {\bibfield  {journal}
		{\bibinfo  {journal} {Phys. Lett. A}\ }\textbf {\bibinfo {volume} {75}},\
		\bibinfo {pages} {269} (\bibinfo {year} {1980})}\BibitemShut {NoStop}%
	\bibitem [{\citenamefont {Agayeva}(1980)}]{Mendes-JPA-1980}%
	\BibitemOpen
	\bibfield  {author} {\bibinfo {author} {\bibfnamefont {R.~G.}\ \bibnamefont
			{Agayeva}},\ }\bibfield  {title} {\bibinfo {title} {{Non-adiabatic parametric
				excitation of oscillator-type systems}},\ }\href
	{https://doi.org/10.1088/0305-4470/13/5/026} {\bibfield  {journal} {\bibinfo
			{journal} {J. Phys. A. Math. Gen.}\ }\textbf {\bibinfo {volume} {13}},\
		\bibinfo {pages} {1685} (\bibinfo {year} {1980})}\BibitemShut {NoStop}%
	\bibitem [{\citenamefont {Morales}(1988)}]{Morales-JPA-1988-2}%
	\BibitemOpen
	\bibfield  {author} {\bibinfo {author} {\bibfnamefont {D.~A.}\ \bibnamefont
			{Morales}},\ }\bibfield  {title} {\bibinfo {title} {{Correspondence between
				Berry's phase and Lewis's phase for quadratic Hamiltonians}},\ }\href
	{https://doi.org/10.1088/0305-4470/21/18/004} {\bibfield  {journal} {\bibinfo
			{journal} {J. Phys. A. Math. Gen.}\ }\textbf {\bibinfo {volume} {21}},\
		\bibinfo {pages} {L889} (\bibinfo {year} {1988})}\BibitemShut {NoStop}%
	\bibitem [{\citenamefont {Gjaja}\ and\ \citenamefont
		{Bhattacharjee}(1992)}]{Gjaja-PRL-1992}%
	\BibitemOpen
	\bibfield  {author} {\bibinfo {author} {\bibfnamefont {I.}~\bibnamefont
			{Gjaja}}\ and\ \bibinfo {author} {\bibfnamefont {A.}~\bibnamefont
			{Bhattacharjee}},\ }\bibfield  {title} {\bibinfo {title} {{Asymptotics of
				reflectionless potentials}},\ }\href
	{https://doi.org/10.1103/PhysRevLett.68.2413} {\bibfield  {journal} {\bibinfo
			{journal} {Phys. Rev. Lett.}\ }\textbf {\bibinfo {volume} {68}},\ \bibinfo
		{pages} {2413} (\bibinfo {year} {1992})}\BibitemShut {NoStop}%
	\bibitem [{\citenamefont {Faccioli}\ \emph {et~al.}(1998)\citenamefont
		{Faccioli}, \citenamefont {Finelli}, \citenamefont {Vacca},\ and\
		\citenamefont {Venturi}}]{Faccioli-PRL-1998}%
	\BibitemOpen
	\bibfield  {author} {\bibinfo {author} {\bibfnamefont {L.}~\bibnamefont
			{Faccioli}}, \bibinfo {author} {\bibfnamefont {F.}~\bibnamefont {Finelli}},
		\bibinfo {author} {\bibfnamefont {G.~P.}\ \bibnamefont {Vacca}},\ and\
		\bibinfo {author} {\bibfnamefont {G.}~\bibnamefont {Venturi}},\ }\bibfield
	{title} {\bibinfo {title} {{Comment on “Semiquantum Chaos”}},\ }\href
	{https://doi.org/10.1103/PhysRevLett.81.240} {\bibfield  {journal} {\bibinfo
			{journal} {Phys. Rev. Lett.}\ }\textbf {\bibinfo {volume} {81}},\ \bibinfo
		{pages} {240} (\bibinfo {year} {1998})}\BibitemShut {NoStop}%
	\bibitem [{\citenamefont {Ban}\ \emph {et~al.}(2012)\citenamefont {Ban},
		\citenamefont {Chen}, \citenamefont {Sherman},\ and\ \citenamefont
		{Muga}}]{Ban-PRL-2012}%
	\BibitemOpen
	\bibfield  {author} {\bibinfo {author} {\bibfnamefont {Y.}~\bibnamefont
			{Ban}}, \bibinfo {author} {\bibfnamefont {X.}~\bibnamefont {Chen}}, \bibinfo
		{author} {\bibfnamefont {E.~Y.}\ \bibnamefont {Sherman}},\ and\ \bibinfo
		{author} {\bibfnamefont {J.~G.}\ \bibnamefont {Muga}},\ }\bibfield  {title}
	{\bibinfo {title} {{Fast and robust spin manipulation in a quantum dot by
				electric fields}},\ }\href {https://doi.org/10.1103/PhysRevLett.109.206602}
	{\bibfield  {journal} {\bibinfo  {journal} {Phys. Rev. Lett.}\ }\textbf
		{\bibinfo {volume} {109}},\ \bibinfo {pages} {1} (\bibinfo {year}
		{2012})}\BibitemShut {NoStop}%
	\bibitem [{\citenamefont {Qin}\ \emph {et~al.}(2013)\citenamefont {Qin},
		\citenamefont {Davidson}, \citenamefont {Chung},\ and\ \citenamefont
		{Burby}}]{Qin-PRL-2013}%
	\BibitemOpen
	\bibfield  {author} {\bibinfo {author} {\bibfnamefont {H.}~\bibnamefont
			{Qin}}, \bibinfo {author} {\bibfnamefont {R.~C.}\ \bibnamefont {Davidson}},
		\bibinfo {author} {\bibfnamefont {M.}~\bibnamefont {Chung}},\ and\ \bibinfo
		{author} {\bibfnamefont {J.~W.}\ \bibnamefont {Burby}},\ }\bibfield  {title}
	{\bibinfo {title} {{Generalized courant-snyder theory for charged-particle
				dynamics in general focusing lattices}},\ }\href
	{https://doi.org/10.1103/PhysRevLett.111.104801} {\bibfield  {journal}
		{\bibinfo  {journal} {Phys. Rev. Lett.}\ }\textbf {\bibinfo {volume} {111}},\
		\bibinfo {pages} {1} (\bibinfo {year} {2013})}\BibitemShut {NoStop}%
	\bibitem [{\citenamefont {Defenu}\ \emph {et~al.}(2018)\citenamefont {Defenu},
		\citenamefont {Enss}, \citenamefont {Kastner},\ and\ \citenamefont
		{Morigi}}]{Defenu-PRL-2018}%
	\BibitemOpen
	\bibfield  {author} {\bibinfo {author} {\bibfnamefont {N.}~\bibnamefont
			{Defenu}}, \bibinfo {author} {\bibfnamefont {T.}~\bibnamefont {Enss}},
		\bibinfo {author} {\bibfnamefont {M.}~\bibnamefont {Kastner}},\ and\ \bibinfo
		{author} {\bibfnamefont {G.}~\bibnamefont {Morigi}},\ }\bibfield  {title}
	{\bibinfo {title} {{Dynamical Critical Scaling of Long-Range Interacting
				Quantum Magnets}},\ }\href {https://doi.org/10.1103/PhysRevLett.121.240403}
	{\bibfield  {journal} {\bibinfo  {journal} {Phys. Rev. Lett.}\ }\textbf
		{\bibinfo {volume} {121}},\ \bibinfo {pages} {240403} (\bibinfo {year}
		{2018})}\BibitemShut {NoStop}%
	\bibitem [{\citenamefont {Gu}\ \emph {et~al.}(2024)\citenamefont {Gu},
		\citenamefont {Liu}, \citenamefont {Ke}, \citenamefont {Li},\ and\
		\citenamefont {Dai}}]{Gu-AP-2024}%
	\BibitemOpen
	\bibfield  {author} {\bibinfo {author} {\bibfnamefont {X.}~\bibnamefont
			{Gu}}, \bibinfo {author} {\bibfnamefont {Y.-Y.}\ \bibnamefont {Liu}},
		\bibinfo {author} {\bibfnamefont {H.-W.}\ \bibnamefont {Ke}}, \bibinfo
		{author} {\bibfnamefont {W.-D.}\ \bibnamefont {Li}},\ and\ \bibinfo {author}
		{\bibfnamefont {W.-S.}\ \bibnamefont {Dai}},\ }\bibfield  {title} {\bibinfo
		{title} {{Exactly solvable time-dependent oscillator family}},\ }\href
	{https://doi.org/10.1016/j.aop.2024.169831} {\bibfield  {journal} {\bibinfo
			{journal} {Ann. Phys. (N. Y).}\ }\textbf {\bibinfo {volume} {470}},\ \bibinfo
		{pages} {169831} (\bibinfo {year} {2024})}\BibitemShut {NoStop}%
	\bibitem [{\citenamefont {Yeon}\ \emph {et~al.}(1993)\citenamefont {Yeon},
		\citenamefont {Lee}, \citenamefont {Um}, \citenamefont {George},\ and\
		\citenamefont {Pandey}}]{Yeon-PRA-1993}%
	\BibitemOpen
	\bibfield  {author} {\bibinfo {author} {\bibfnamefont {K.~H.}\ \bibnamefont
			{Yeon}}, \bibinfo {author} {\bibfnamefont {K.~K.}\ \bibnamefont {Lee}},
		\bibinfo {author} {\bibfnamefont {C.~I.}\ \bibnamefont {Um}}, \bibinfo
		{author} {\bibfnamefont {T.~F.}\ \bibnamefont {George}},\ and\ \bibinfo
		{author} {\bibfnamefont {L.~N.}\ \bibnamefont {Pandey}},\ }\bibfield  {title}
	{\bibinfo {title} {{Exact quantum theory of a time-dependent bound quadratic
				Hamiltonian system}},\ }\href {https://doi.org/10.1103/PhysRevA.48.2716}
	{\bibfield  {journal} {\bibinfo  {journal} {Phys. Rev. A}\ }\textbf {\bibinfo
			{volume} {48}},\ \bibinfo {pages} {2716} (\bibinfo {year}
		{1993})}\BibitemShut {NoStop}%
	\bibitem [{\citenamefont {Cheng}(1985)}]{Cheng-PLA-1985}%
	\BibitemOpen
	\bibfield  {author} {\bibinfo {author} {\bibfnamefont {B.~K.}\ \bibnamefont
			{Cheng}},\ }\bibfield  {title} {\bibinfo {title} {{Exact propagator of the
				harmonic oscillator with a time-dependent mass}},\ }\href
	{https://doi.org/10.1016/0375-9601(85)90166-5} {\bibfield  {journal}
		{\bibinfo  {journal} {Phys. Lett. A}\ }\textbf {\bibinfo {volume} {113}},\
		\bibinfo {pages} {293} (\bibinfo {year} {1985})}\BibitemShut {NoStop}%
	\bibitem [{\citenamefont {Soto-Eguibar}\ \emph {et~al.}(2021)\citenamefont
		{Soto-Eguibar}, \citenamefont {Asenjo}, \citenamefont {Hojman},\ and\
		\citenamefont {Moya-Cessa}}]{Eguibar-JMP-2021}%
	\BibitemOpen
	\bibfield  {author} {\bibinfo {author} {\bibfnamefont {F.}~\bibnamefont
			{Soto-Eguibar}}, \bibinfo {author} {\bibfnamefont {F.~A.}\ \bibnamefont
			{Asenjo}}, \bibinfo {author} {\bibfnamefont {S.~A.}\ \bibnamefont {Hojman}},\
		and\ \bibinfo {author} {\bibfnamefont {H.~M.}\ \bibnamefont {Moya-Cessa}},\
	}\bibfield  {title} {\bibinfo {title} {{Bohm potential for the time dependent
				harmonic oscillator}},\ }\href {https://doi.org/10.1063/5.0044144} {\bibfield
		{journal} {\bibinfo  {journal} {J. Math. Phys.}\ }\textbf {\bibinfo {volume}
			{62}},\ \bibinfo {pages} {122103} (\bibinfo {year} {2021})}\BibitemShut
	{NoStop}%
	\bibitem [{\citenamefont {Tibaduiza}\ \emph
		{et~al.}(2020{\natexlab{a}})\citenamefont {Tibaduiza}, \citenamefont {Pires},
		\citenamefont {Szilard}, \citenamefont {Zarro}, \citenamefont {Farina},\ and\
		\citenamefont {Rego}}]{Tibaduiza-BJP-2020}%
	\BibitemOpen
	\bibfield  {author} {\bibinfo {author} {\bibfnamefont {D.~M.}\ \bibnamefont
			{Tibaduiza}}, \bibinfo {author} {\bibfnamefont {L.}~\bibnamefont {Pires}},
		\bibinfo {author} {\bibfnamefont {D.}~\bibnamefont {Szilard}}, \bibinfo
		{author} {\bibfnamefont {C.~A.~D.}\ \bibnamefont {Zarro}}, \bibinfo {author}
		{\bibfnamefont {C.}~\bibnamefont {Farina}},\ and\ \bibinfo {author}
		{\bibfnamefont {A.~L.~C.}\ \bibnamefont {Rego}},\ }\bibfield  {title}
	{\bibinfo {title} {{A Time-Dependent Harmonic Oscillator with Two Frequency
				Jumps: an Exact Algebraic Solution}},\ }\href
	{https://doi.org/10.1007/s13538-020-00770-x} {\bibfield  {journal} {\bibinfo
			{journal} {Braz. J. Phys.}\ }\textbf {\bibinfo {volume} {50}},\ \bibinfo
		{pages} {634} (\bibinfo {year} {2020}{\natexlab{a}})}\BibitemShut {NoStop}%
	\bibitem [{\citenamefont {Tibaduiza}\ \emph
		{et~al.}(2020{\natexlab{b}})\citenamefont {Tibaduiza}, \citenamefont {Pires},
		\citenamefont {Rego}, \citenamefont {Szilard}, \citenamefont {Zarro},\ and\
		\citenamefont {Farina}}]{Tibaduiza-PS-2020}%
	\BibitemOpen
	\bibfield  {author} {\bibinfo {author} {\bibfnamefont {D.~M.}\ \bibnamefont
			{Tibaduiza}}, \bibinfo {author} {\bibfnamefont {L.}~\bibnamefont {Pires}},
		\bibinfo {author} {\bibfnamefont {A.~L.~C.}\ \bibnamefont {Rego}}, \bibinfo
		{author} {\bibfnamefont {D.}~\bibnamefont {Szilard}}, \bibinfo {author}
		{\bibfnamefont {C.}~\bibnamefont {Zarro}},\ and\ \bibinfo {author}
		{\bibfnamefont {C.}~\bibnamefont {Farina}},\ }\bibfield  {title} {\bibinfo
		{title} {{Efficient algebraic solution for a time-dependent quantum harmonic
				oscillator}},\ }\href {https://doi.org/10.1088/1402-4896/abb254} {\bibfield
		{journal} {\bibinfo  {journal} {Phys. Scr.}\ }\textbf {\bibinfo {volume}
			{95}},\ \bibinfo {pages} {105102} (\bibinfo {year}
		{2020}{\natexlab{b}})}\BibitemShut {NoStop}%
	\bibitem [{\citenamefont {Mart{\'{i}}nez-Tibaduiza}\ \emph
		{et~al.}(2021)\citenamefont {Mart{\'{i}}nez-Tibaduiza}, \citenamefont
		{Pires},\ and\ \citenamefont {Farina}}]{Tibaduiza-JPB-2021}%
	\BibitemOpen
	\bibfield  {author} {\bibinfo {author} {\bibfnamefont {D.}~\bibnamefont
			{Mart{\'{i}}nez-Tibaduiza}}, \bibinfo {author} {\bibfnamefont
			{L.}~\bibnamefont {Pires}},\ and\ \bibinfo {author} {\bibfnamefont
			{C.}~\bibnamefont {Farina}},\ }\bibfield  {title} {\bibinfo {title}
		{{Time-dependent quantum harmonic oscillator: a continuous route from
				adiabatic to sudden changes}},\ }\href
	{https://doi.org/10.1088/1361-6455/ac36ba} {\bibfield  {journal} {\bibinfo
			{journal} {J. Phys. B At. Mol. Opt. Phys.}\ }\textbf {\bibinfo {volume}
			{54}},\ \bibinfo {pages} {205401} (\bibinfo {year} {2021})}\BibitemShut
	{NoStop}%
	\bibitem [{\citenamefont {Winter}\ and\ \citenamefont
		{Ortjohann}(1991)}]{Winter-AJP-1991}%
	\BibitemOpen
	\bibfield  {author} {\bibinfo {author} {\bibfnamefont {H.}~\bibnamefont
			{Winter}}\ and\ \bibinfo {author} {\bibfnamefont {H.~W.}\ \bibnamefont
			{Ortjohann}},\ }\bibfield  {title} {\bibinfo {title} {{Simple demonstration
				of storing macroscopic particles in a ‘‘Paul trap''}},\ }\href
	{https://doi.org/10.1119/1.16830} {\bibfield  {journal} {\bibinfo  {journal}
			{Am. J. Phys.}\ }\textbf {\bibinfo {volume} {59}},\ \bibinfo {pages} {807}
		(\bibinfo {year} {1991})}\BibitemShut {NoStop}%
	\bibitem [{\citenamefont {Rueckner}\ \emph {et~al.}(1995)\citenamefont
		{Rueckner}, \citenamefont {Georgi}, \citenamefont {Goodale}, \citenamefont
		{Rosenberg},\ and\ \citenamefont {Tavilla}}]{Rueckner-AJP-1995}%
	\BibitemOpen
	\bibfield  {author} {\bibinfo {author} {\bibfnamefont {W.}~\bibnamefont
			{Rueckner}}, \bibinfo {author} {\bibfnamefont {J.}~\bibnamefont {Georgi}},
		\bibinfo {author} {\bibfnamefont {D.}~\bibnamefont {Goodale}}, \bibinfo
		{author} {\bibfnamefont {D.}~\bibnamefont {Rosenberg}},\ and\ \bibinfo
		{author} {\bibfnamefont {D.}~\bibnamefont {Tavilla}},\ }\bibfield  {title}
	{\bibinfo {title} {{Rotating saddle Paul trap}},\ }\href
	{https://doi.org/10.1119/1.17983} {\bibfield  {journal} {\bibinfo  {journal}
			{Am. J. Phys.}\ }\textbf {\bibinfo {volume} {63}},\ \bibinfo {pages} {186}
		(\bibinfo {year} {1995})}\BibitemShut {NoStop}%
	\bibitem [{\citenamefont {Ruby}(1996)}]{Ruby-AJP-1996}%
	\BibitemOpen
	\bibfield  {author} {\bibinfo {author} {\bibfnamefont {L.}~\bibnamefont
			{Ruby}},\ }\bibfield  {title} {\bibinfo {title} {{Applications of the Mathieu
				equation}},\ }\href {https://doi.org/10.1119/1.18290} {\bibfield  {journal}
		{\bibinfo  {journal} {Am. J. Phys.}\ }\textbf {\bibinfo {volume} {64}},\
		\bibinfo {pages} {39} (\bibinfo {year} {1996})}\BibitemShut {NoStop}%
	\bibitem [{\citenamefont {Nasse}\ and\ \citenamefont
		{Foot}(2001)}]{Nasse-EJP-2001}%
	\BibitemOpen
	\bibfield  {author} {\bibinfo {author} {\bibfnamefont {M.}~\bibnamefont
			{Nasse}}\ and\ \bibinfo {author} {\bibfnamefont {C.}~\bibnamefont {Foot}},\
	}\bibfield  {title} {\bibinfo {title} {{Influence of background pressure on
				the stability region of a Paul trap}},\ }\href
	{https://doi.org/10.1088/0143-0807/22/6/301} {\bibfield  {journal} {\bibinfo
			{journal} {Eur. J. Phys.}\ }\textbf {\bibinfo {volume} {22}},\ \bibinfo
		{pages} {563} (\bibinfo {year} {2001})}\BibitemShut {NoStop}%
	\bibitem [{\citenamefont {Johnson}\ and\ \citenamefont
		{Rabchuk}(2009)}]{Johnson-AJP-2009}%
	\BibitemOpen
	\bibfield  {author} {\bibinfo {author} {\bibfnamefont {A.~K.}\ \bibnamefont
			{Johnson}}\ and\ \bibinfo {author} {\bibfnamefont {J.~A.}\ \bibnamefont
			{Rabchuk}},\ }\bibfield  {title} {\bibinfo {title} {{A bead on a hoop
				rotating about a horizontal axis: A one-dimensional ponderomotive trap}},\
	}\href {https://doi.org/10.1119/1.3167216} {\bibfield  {journal} {\bibinfo
			{journal} {Am. J. Phys.}\ }\textbf {\bibinfo {volume} {77}},\ \bibinfo
		{pages} {1039} (\bibinfo {year} {2009})}\BibitemShut {NoStop}%
	\bibitem [{\citenamefont {Wang}\ and\ \citenamefont
		{Wang}(2013)}]{Wang-EJP-2013}%
	\BibitemOpen
	\bibfield  {author} {\bibinfo {author} {\bibfnamefont {Y.-R.}\ \bibnamefont
			{Wang}}\ and\ \bibinfo {author} {\bibfnamefont {L.-B.}\ \bibnamefont
			{Wang}},\ }\bibfield  {title} {\bibinfo {title} {{Demonstration experiments
				for the electronic detection of trapped ions in a linear Paul trap}},\ }\href
	{https://doi.org/10.1088/0143-0807/34/3/787} {\bibfield  {journal} {\bibinfo
			{journal} {Eur. J. Phys.}\ }\textbf {\bibinfo {volume} {34}},\ \bibinfo
		{pages} {787} (\bibinfo {year} {2013})}\BibitemShut {NoStop}%
	\bibitem [{\citenamefont {Madsen}\ and\ \citenamefont
		{Skowronski}(2014)}]{Madsen-AJP-2014}%
	\BibitemOpen
	\bibfield  {author} {\bibinfo {author} {\bibfnamefont {M.~J.}\ \bibnamefont
			{Madsen}}\ and\ \bibinfo {author} {\bibfnamefont {A.~D.}\ \bibnamefont
			{Skowronski}},\ }\bibfield  {title} {\bibinfo {title} {{Brownian motion of a
				trapped microsphere ion}},\ }\href {https://doi.org/10.1119/1.4881609}
	{\bibfield  {journal} {\bibinfo  {journal} {Am. J. Phys.}\ }\textbf {\bibinfo
			{volume} {82}},\ \bibinfo {pages} {934} (\bibinfo {year} {2014})}\BibitemShut
	{NoStop}%
	\bibitem [{\citenamefont {Vinitsky}\ \emph {et~al.}(2015)\citenamefont
		{Vinitsky}, \citenamefont {Black},\ and\ \citenamefont
		{Libbrecht}}]{Vinitsky-AJP-2015}%
	\BibitemOpen
	\bibfield  {author} {\bibinfo {author} {\bibfnamefont {E.~A.}\ \bibnamefont
			{Vinitsky}}, \bibinfo {author} {\bibfnamefont {E.~D.}\ \bibnamefont
			{Black}},\ and\ \bibinfo {author} {\bibfnamefont {K.~G.}\ \bibnamefont
			{Libbrecht}},\ }\bibfield  {title} {\bibinfo {title} {{Particle dynamics in
				damped nonlinear quadrupole ion traps}},\ }\href
	{https://doi.org/10.1119/1.4902185} {\bibfield  {journal} {\bibinfo
			{journal} {Am. J. Phys.}\ }\textbf {\bibinfo {volume} {83}},\ \bibinfo
		{pages} {313} (\bibinfo {year} {2015})}\BibitemShut {NoStop}%
	\bibitem [{\citenamefont {Libbrecht}\ and\ \citenamefont
		{Black}(2018)}]{Libbrecht-AJP-2018}%
	\BibitemOpen
	\bibfield  {author} {\bibinfo {author} {\bibfnamefont {K.~G.}\ \bibnamefont
			{Libbrecht}}\ and\ \bibinfo {author} {\bibfnamefont {E.~D.}\ \bibnamefont
			{Black}},\ }\bibfield  {title} {\bibinfo {title} {{Improved microparticle
				electrodynamic ion traps for physics teaching}},\ }\href
	{https://doi.org/10.1119/1.5034344} {\bibfield  {journal} {\bibinfo
			{journal} {Am. J. Phys.}\ }\textbf {\bibinfo {volume} {86}},\ \bibinfo
		{pages} {539} (\bibinfo {year} {2018})}\BibitemShut {NoStop}%
	\bibitem [{\citenamefont {Qian}\ and\ \citenamefont
		{Su}(1994)}]{Qian-PRL-1994}%
	\BibitemOpen
	\bibfield  {author} {\bibinfo {author} {\bibfnamefont {T.-Z.}\ \bibnamefont
			{Qian}}\ and\ \bibinfo {author} {\bibfnamefont {Z.-B.}\ \bibnamefont {Su}},\
	}\bibfield  {title} {\bibinfo {title} {{Spin-orbit interaction and
				Aharonov-Anandan phase in mesoscopic rings}},\ }\href
	{https://doi.org/10.1103/PhysRevLett.72.2311} {\bibfield  {journal} {\bibinfo
			{journal} {Phys. Rev. Lett.}\ }\textbf {\bibinfo {volume} {72}},\ \bibinfo
		{pages} {2311} (\bibinfo {year} {1994})}\BibitemShut {NoStop}%
	\bibitem [{\citenamefont {Shen}\ \emph {et~al.}(2003)\citenamefont {Shen},
		\citenamefont {Zhu},\ and\ \citenamefont {Chen}}]{Shen-EPJD-2003}%
	\BibitemOpen
	\bibfield  {author} {\bibinfo {author} {\bibfnamefont {J.-Q.}\ \bibnamefont
			{Shen}}, \bibinfo {author} {\bibfnamefont {H.-Y.}\ \bibnamefont {Zhu}},\ and\
		\bibinfo {author} {\bibfnamefont {P.}~\bibnamefont {Chen}},\ }\bibfield
	{title} {\bibinfo {title} {{Exact solutions and geometric phase factor of
				time-dependent three-generator quantum systems}},\ }\href
	{https://doi.org/10.1140/epjd/e2003-00043-7} {\bibfield  {journal} {\bibinfo
			{journal} {Eur. Phys. J. D - At. Mol. Opt. Phys.}\ }\textbf {\bibinfo
			{volume} {23}},\ \bibinfo {pages} {305} (\bibinfo {year} {2003})}\BibitemShut
	{NoStop}%
	\bibitem [{\citenamefont {Dutta}\ \emph {et~al.}(2022)\citenamefont {Dutta},
		\citenamefont {Ganguly},\ and\ \citenamefont {Gangopadhyay}}]{Dutta-PS-2022}%
	\BibitemOpen
	\bibfield  {author} {\bibinfo {author} {\bibfnamefont {M.}~\bibnamefont
			{Dutta}}, \bibinfo {author} {\bibfnamefont {S.}~\bibnamefont {Ganguly}},\
		and\ \bibinfo {author} {\bibfnamefont {S.}~\bibnamefont {Gangopadhyay}},\
	}\bibfield  {title} {\bibinfo {title} {{Explicit form of Berry phase for time
				dependent harmonic oscillators in noncommutative space}},\ }\href
	{https://doi.org/10.1088/1402-4896/ac8dca} {\bibfield  {journal} {\bibinfo
			{journal} {Phys. Scr.}\ }\textbf {\bibinfo {volume} {97}},\ \bibinfo {pages}
		{105204} (\bibinfo {year} {2022})}\BibitemShut {NoStop}%
	\bibitem [{\citenamefont {Sen}\ \emph {et~al.}(2024)\citenamefont {Sen},
		\citenamefont {Dutta},\ and\ \citenamefont {Gangopadhyay}}]{Sen-PS-2024}%
	\BibitemOpen
	\bibfield  {author} {\bibinfo {author} {\bibfnamefont {S.}~\bibnamefont
			{Sen}}, \bibinfo {author} {\bibfnamefont {M.}~\bibnamefont {Dutta}},\ and\
		\bibinfo {author} {\bibfnamefont {S.}~\bibnamefont {Gangopadhyay}},\
	}\bibfield  {title} {\bibinfo {title} {{Lewis and berry phases for a
				gravitational wave interacting with a quantum harmonic oscillator}},\ }\href
	{https://doi.org/10.1088/1402-4896/ad1234} {\bibfield  {journal} {\bibinfo
			{journal} {Phys. Scr.}\ }\textbf {\bibinfo {volume} {99}},\ \bibinfo {pages}
		{015007} (\bibinfo {year} {2024})}\BibitemShut {NoStop}%
	\bibitem [{\citenamefont {Gao}\ \emph {et~al.}(1991)\citenamefont {Gao},
		\citenamefont {Xu},\ and\ \citenamefont {Qian}}]{Gao-PRA-1991}%
	\BibitemOpen
	\bibfield  {author} {\bibinfo {author} {\bibfnamefont {X.-C.}\ \bibnamefont
			{Gao}}, \bibinfo {author} {\bibfnamefont {J.-B.}\ \bibnamefont {Xu}},\ and\
		\bibinfo {author} {\bibfnamefont {T.-Z.}\ \bibnamefont {Qian}},\ }\bibfield
	{title} {\bibinfo {title} {{Geometric phase and the generalized invariant
				formulation}},\ }\href {https://doi.org/10.1103/PhysRevA.44.7016} {\bibfield
		{journal} {\bibinfo  {journal} {Phys. Rev. A}\ }\textbf {\bibinfo {volume}
			{44}},\ \bibinfo {pages} {7016} (\bibinfo {year} {1991})}\BibitemShut
	{NoStop}%
	\bibitem [{\citenamefont {Kim}\ and\ \citenamefont
		{Page}(2001)}]{Kim-PRA-2001}%
	\BibitemOpen
	\bibfield  {author} {\bibinfo {author} {\bibfnamefont {S.~P.}\ \bibnamefont
			{Kim}}\ and\ \bibinfo {author} {\bibfnamefont {D.~N.}\ \bibnamefont {Page}},\
	}\bibfield  {title} {\bibinfo {title} {{Classical and quantum action-phase
				variables for time-dependent oscillators}},\ }\href
	{https://doi.org/10.1103/PhysRevA.64.012104} {\bibfield  {journal} {\bibinfo
			{journal} {Phys. Rev. A}\ }\textbf {\bibinfo {volume} {64}},\ \bibinfo
		{pages} {012104} (\bibinfo {year} {2001})}\BibitemShut {NoStop}%
	\bibitem [{\citenamefont {Choi}\ and\ \citenamefont
		{Kim}(2004)}]{Choi-JKPS-2004}%
	\BibitemOpen
	\bibfield  {author} {\bibinfo {author} {\bibfnamefont {J.~R.}\ \bibnamefont
			{Choi}}\ and\ \bibinfo {author} {\bibfnamefont {D.~W.}\ \bibnamefont {Kim}},\
	}\bibfield  {title} {\bibinfo {title} {{Squeezed states for the general
				time-dependent harmonic oscillator with and without a singularity}},\
	}\href@noop {} {\bibfield  {journal} {\bibinfo  {journal} {J. Korean Phys.
				Soc.}\ }\textbf {\bibinfo {volume} {45}},\ \bibinfo {pages} {1426} (\bibinfo
		{year} {2004})}\BibitemShut {NoStop}%
	\bibitem [{\citenamefont {{Ryeol Choi}}\ \emph {et~al.}(2009)\citenamefont
		{{Ryeol Choi}}, \citenamefont {{Hwang Yeon}}, \citenamefont {Maamache},\ and\
		\citenamefont {Menouar}}]{Choi-JPSJ-2009}%
	\BibitemOpen
	\bibfield  {author} {\bibinfo {author} {\bibfnamefont {J.}~\bibnamefont
			{{Ryeol Choi}}}, \bibinfo {author} {\bibfnamefont {K.}~\bibnamefont {{Hwang
					Yeon}}}, \bibinfo {author} {\bibfnamefont {M.}~\bibnamefont {Maamache}},\
		and\ \bibinfo {author} {\bibfnamefont {S.}~\bibnamefont {Menouar}},\
	}\bibfield  {title} {\bibinfo {title} {{Exact Quantum Solutions for
				Time-Dependent Charged Oscillator}},\ }\href
	{https://doi.org/10.1143/JPSJ.78.054001} {\bibfield  {journal} {\bibinfo
			{journal} {J. Phys. Soc. Japan}\ }\textbf {\bibinfo {volume} {78}},\ \bibinfo
		{pages} {054001} (\bibinfo {year} {2009})}\BibitemShut {NoStop}%
	\bibitem [{\citenamefont {Dutta}\ \emph {et~al.}(2020)\citenamefont {Dutta},
		\citenamefont {Ganguly},\ and\ \citenamefont
		{Gangopadhyay}}]{Dutta-IJTP-2020}%
	\BibitemOpen
	\bibfield  {author} {\bibinfo {author} {\bibfnamefont {M.}~\bibnamefont
			{Dutta}}, \bibinfo {author} {\bibfnamefont {S.}~\bibnamefont {Ganguly}},\
		and\ \bibinfo {author} {\bibfnamefont {S.}~\bibnamefont {Gangopadhyay}},\
	}\bibfield  {title} {\bibinfo {title} {{Exact Solutions of a Damped Harmonic
				Oscillator in a Time Dependent Noncommutative Space}},\ }\href
	{https://doi.org/10.1007/s10773-020-04637-4} {\bibfield  {journal} {\bibinfo
			{journal} {Int. J. Theor. Phys.}\ }\textbf {\bibinfo {volume} {59}},\
		\bibinfo {pages} {3852} (\bibinfo {year} {2020})}\BibitemShut {NoStop}%
	\bibitem [{\citenamefont {Dutta}\ \emph {et~al.}(2021)\citenamefont {Dutta},
		\citenamefont {Ganguly},\ and\ \citenamefont {Gangopadhyay}}]{Dutta-PS-2021}%
	\BibitemOpen
	\bibfield  {author} {\bibinfo {author} {\bibfnamefont {M.}~\bibnamefont
			{Dutta}}, \bibinfo {author} {\bibfnamefont {S.}~\bibnamefont {Ganguly}},\
		and\ \bibinfo {author} {\bibfnamefont {S.}~\bibnamefont {Gangopadhyay}},\
	}\bibfield  {title} {\bibinfo {title} {{Investigation of a harmonic
				oscillator in a magnetic field with damping and time dependent
				noncommutativity}},\ }\href {https://doi.org/10.1088/1402-4896/ac2b4c}
	{\bibfield  {journal} {\bibinfo  {journal} {Phys. Scr.}\ }\textbf {\bibinfo
			{volume} {96}},\ \bibinfo {pages} {125224} (\bibinfo {year}
		{2021})}\BibitemShut {NoStop}%
	\bibitem [{\citenamefont {Dutta}\ \emph {et~al.}(2024)\citenamefont {Dutta},
		\citenamefont {Ganguly},\ and\ \citenamefont
		{Gangopadhyay}}]{Dutta-IJTP-2024}%
	\BibitemOpen
	\bibfield  {author} {\bibinfo {author} {\bibfnamefont {M.}~\bibnamefont
			{Dutta}}, \bibinfo {author} {\bibfnamefont {S.}~\bibnamefont {Ganguly}},\
		and\ \bibinfo {author} {\bibfnamefont {S.}~\bibnamefont {Gangopadhyay}},\
	}\bibfield  {title} {\bibinfo {title} {{Quantum Harmonic Oscillator in a Time
				Dependent Noncommutative Background}},\ }\href
	{https://doi.org/10.1007/s10773-024-05707-7} {\bibfield  {journal} {\bibinfo
			{journal} {Int. J. Theor. Phys.}\ }\textbf {\bibinfo {volume} {63}},\
		\bibinfo {pages} {169} (\bibinfo {year} {2024})}\BibitemShut {NoStop}%
	\bibitem [{\citenamefont {Sobhani}\ and\ \citenamefont
		{Hassanabadi}(2016)}]{Sobhani-JKPS-2016}%
	\BibitemOpen
	\bibfield  {author} {\bibinfo {author} {\bibfnamefont {H.}~\bibnamefont
			{Sobhani}}\ and\ \bibinfo {author} {\bibfnamefont {H.}~\bibnamefont
			{Hassanabadi}},\ }\bibfield  {title} {\bibinfo {title} {{Investigation of a
				time-dependent two-body system via the Lewis-Riesenfeld dynamical invariant
				method}},\ }\href {https://doi.org/10.3938/jkps.69.1509} {\bibfield
		{journal} {\bibinfo  {journal} {J. Korean Phys. Soc.}\ }\textbf {\bibinfo
			{volume} {69}},\ \bibinfo {pages} {1509} (\bibinfo {year}
		{2016})}\BibitemShut {NoStop}%
	\bibitem [{\citenamefont {Prykarpatskyy}(2018)}]{Prykarpatskyy-JMS-2018}%
	\BibitemOpen
	\bibfield  {author} {\bibinfo {author} {\bibfnamefont {Y.}~\bibnamefont
			{Prykarpatskyy}},\ }\bibfield  {title} {\bibinfo {title}
		{{Steen–Ermakov–Pinney Equation and Integrable Nonlinear Deformation of
				the One-Dimensional Dirac Equation}},\ }\href
	{https://doi.org/10.1007/s10958-018-3851-8} {\bibfield  {journal} {\bibinfo
			{journal} {J. Math. Sci.}\ }\textbf {\bibinfo {volume} {231}},\ \bibinfo
		{pages} {820} (\bibinfo {year} {2018})}\BibitemShut {NoStop}%
	\bibitem [{\citenamefont {Pinney}(1950)}]{Pinney-PAMS-1950}%
	\BibitemOpen
	\bibfield  {author} {\bibinfo {author} {\bibfnamefont {E.}~\bibnamefont
			{Pinney}},\ }\bibfield  {title} {\bibinfo {title} {{The nonlinear
				differential equation $y^{\prime\prime}+p(x)y+cy^{-3}=0$}},\ }\href
	{https://doi.org/10.1090/S0002-9939-1950-0037979-4} {\bibfield  {journal}
		{\bibinfo  {journal} {Proc. Am. Math. Soc.}\ }\textbf {\bibinfo {volume}
			{1}},\ \bibinfo {pages} {681} (\bibinfo {year} {1950})}\BibitemShut {NoStop}%
	\bibitem [{\citenamefont {de~Lima}\ \emph {et~al.}(2009)\citenamefont
		{de~Lima}, \citenamefont {Rosas},\ and\ \citenamefont
		{Pedrosa}}]{Lima-JMO-2009}%
	\BibitemOpen
	\bibfield  {author} {\bibinfo {author} {\bibfnamefont {A.~L.}\ \bibnamefont
			{de~Lima}}, \bibinfo {author} {\bibfnamefont {A.}~\bibnamefont {Rosas}},\
		and\ \bibinfo {author} {\bibfnamefont {I.}~\bibnamefont {Pedrosa}},\
	}\bibfield  {title} {\bibinfo {title} {{Quantum dynamics of a particle
				trapped by oscillating fields}},\ }\href
	{https://doi.org/10.1080/09500340802495834} {\bibfield  {journal} {\bibinfo
			{journal} {J. Mod. Opt.}\ }\textbf {\bibinfo {volume} {56}},\ \bibinfo
		{pages} {75} (\bibinfo {year} {2009})}\BibitemShut {NoStop}%
	\bibitem [{\citenamefont {Cari{\~{n}}ena}\ and\ \citenamefont
		{de~Lucas}(2009)}]{Carinena-IJGMMP-2009}%
	\BibitemOpen
	\bibfield  {author} {\bibinfo {author} {\bibfnamefont {J.~F.}\ \bibnamefont
			{Cari{\~{n}}ena}}\ and\ \bibinfo {author} {\bibfnamefont {J.}~\bibnamefont
			{de~Lucas}},\ }\bibfield  {title} {\bibinfo {title} {{Applications of Lie
				Systems in Dissipative Milne-Pinney Equations}},\ }\href
	{https://doi.org/10.1142/S0219887809003758} {\bibfield  {journal} {\bibinfo
			{journal} {Int. J. Geom. Methods Mod. Phys.}\ }\textbf {\bibinfo {volume}
			{06}},\ \bibinfo {pages} {683} (\bibinfo {year} {2009})}\BibitemShut
	{NoStop}%
	\bibitem [{\citenamefont {Weber}\ and\ \citenamefont
		{Arfken}(2003)}]{Arfken-Mathematical-Physics-2003}%
	\BibitemOpen
	\bibfield  {author} {\bibinfo {author} {\bibfnamefont {H.~J.}\ \bibnamefont
			{Weber}}\ and\ \bibinfo {author} {\bibfnamefont {G.~B.}\ \bibnamefont
			{Arfken}},\ }\href@noop {} {\emph {\bibinfo {title} {{Essential mathematical
					methods for physicists}}}},\ \bibinfo {edition} {6th}\ ed.\ (\bibinfo
	{publisher} {Academic Press},\ \bibinfo {address} {San Diego, USA},\ \bibinfo
	{year} {2003})\BibitemShut {NoStop}%
	\bibitem [{\citenamefont {Pepore}\ \emph {et~al.}(2006)\citenamefont {Pepore},
		\citenamefont {Winotai}, \citenamefont {Osotchan},\ and\ \citenamefont
		{Robkob}}]{Pepore-ScienceAsia-2006}%
	\BibitemOpen
	\bibfield  {author} {\bibinfo {author} {\bibfnamefont {S.}~\bibnamefont
			{Pepore}}, \bibinfo {author} {\bibfnamefont {P.}~\bibnamefont {Winotai}},
		\bibinfo {author} {\bibfnamefont {T.}~\bibnamefont {Osotchan}},\ and\
		\bibinfo {author} {\bibfnamefont {U.}~\bibnamefont {Robkob}},\ }\bibfield
	{title} {\bibinfo {title} {{Path integral for a harmonic oscillator with
				time-dependent mass and frequency}},\ }\href
	{https://doi.org/10.2306/scienceasia1513-1874.2006.32.173} {\bibfield
		{journal} {\bibinfo  {journal} {ScienceAsia}\ }\textbf {\bibinfo {volume}
			{32}},\ \bibinfo {pages} {173} (\bibinfo {year} {2006})}\BibitemShut
	{NoStop}%
	\bibitem [{\citenamefont {Xin}\ \emph {et~al.}(2021)\citenamefont {Xin},
		\citenamefont {Leong}, \citenamefont {Chen}, \citenamefont {Wang},\ and\
		\citenamefont {Lan}}]{Xin-PRL-2021}%
	\BibitemOpen
	\bibfield  {author} {\bibinfo {author} {\bibfnamefont {M.}~\bibnamefont
			{Xin}}, \bibinfo {author} {\bibfnamefont {W.~S.}\ \bibnamefont {Leong}},
		\bibinfo {author} {\bibfnamefont {Z.}~\bibnamefont {Chen}}, \bibinfo {author}
		{\bibfnamefont {Y.}~\bibnamefont {Wang}},\ and\ \bibinfo {author}
		{\bibfnamefont {S.-Y.}\ \bibnamefont {Lan}},\ }\bibfield  {title} {\bibinfo
		{title} {{Rapid Quantum Squeezing by Jumping the Harmonic Oscillator
				Frequency}},\ }\href {https://doi.org/10.1103/PhysRevLett.127.183602}
	{\bibfield  {journal} {\bibinfo  {journal} {Phys. Rev. Lett.}\ }\textbf
		{\bibinfo {volume} {127}},\ \bibinfo {pages} {183602} (\bibinfo {year}
		{2021})}\BibitemShut {NoStop}%
	\bibitem [{\citenamefont {Guerry}\ and\ \citenamefont
		{Knight}(2005)}]{Guerry-Quantum-Optics-2005}%
	\BibitemOpen
	\bibfield  {author} {\bibinfo {author} {\bibfnamefont {C.~C.}\ \bibnamefont
			{Guerry}}\ and\ \bibinfo {author} {\bibfnamefont {P.~L.}\ \bibnamefont
			{Knight}},\ }\href@noop {} {\emph {\bibinfo {title} {{Introductory Quantum
					Optics}}}},\ \bibinfo {edition} {1st}\ ed.\ (\bibinfo  {publisher} {Cambridge
		University Press},\ \bibinfo {address} {Cambridge, UK},\ \bibinfo {year}
	{2005})\ pp.\ \bibinfo {pages} {150--165}\BibitemShut {NoStop}%
	\bibitem [{\citenamefont {Daneshmand}\ and\ \citenamefont
		{Tavassoly}(2017)}]{Daneshmand-CTP-2017}%
	\BibitemOpen
	\bibfield  {author} {\bibinfo {author} {\bibfnamefont {R.}~\bibnamefont
			{Daneshmand}}\ and\ \bibinfo {author} {\bibfnamefont {M.~K.}\ \bibnamefont
			{Tavassoly}},\ }\bibfield  {title} {\bibinfo {title} {{Dynamics of
				Nonclassicality of Time- and Conductivity-Dependent Squeezed States and
				Excited Even/Odd Coherent States}},\ }\href
	{https://doi.org/10.1088/0253-6102/67/4/365} {\bibfield  {journal} {\bibinfo
			{journal} {Commun. Theor. Phys.}\ }\textbf {\bibinfo {volume} {67}},\
		\bibinfo {pages} {365} (\bibinfo {year} {2017})}\BibitemShut {NoStop}%
	\bibitem [{\citenamefont {Kim}\ \emph {et~al.}(1989)\citenamefont {Kim},
		\citenamefont {de~Oliveira},\ and\ \citenamefont {Knight}}]{Kim-OC-1989}%
	\BibitemOpen
	\bibfield  {author} {\bibinfo {author} {\bibfnamefont {M.}~\bibnamefont
			{Kim}}, \bibinfo {author} {\bibfnamefont {F.}~\bibnamefont {de~Oliveira}},\
		and\ \bibinfo {author} {\bibfnamefont {P.}~\bibnamefont {Knight}},\
	}\bibfield  {title} {\bibinfo {title} {{Photon number distributions for
				squeezed number states and squeezed thermal states}},\ }\href
	{https://doi.org/10.1016/0030-4018(89)90263-0} {\bibfield  {journal}
		{\bibinfo  {journal} {Opt. Commun.}\ }\textbf {\bibinfo {volume} {72}},\
		\bibinfo {pages} {99} (\bibinfo {year} {1989})}\BibitemShut {NoStop}%
	\bibitem [{\citenamefont {Janszky}\ and\ \citenamefont
		{Yushin}(1989)}]{Janszky-PRA-1989}%
	\BibitemOpen
	\bibfield  {author} {\bibinfo {author} {\bibfnamefont {J.}~\bibnamefont
			{Janszky}}\ and\ \bibinfo {author} {\bibfnamefont {Y.}~\bibnamefont
			{Yushin}},\ }\bibfield  {title} {\bibinfo {title} {{Comment on
				‘‘Squeezing and frequency jump of a harmonic oscillator''}},\ }\href
	{https://doi.org/10.1103/PhysRevA.39.5445} {\bibfield  {journal} {\bibinfo
			{journal} {Phys. Rev. A}\ }\textbf {\bibinfo {volume} {39}},\ \bibinfo
		{pages} {5445} (\bibinfo {year} {1989})}\BibitemShut {NoStop}%
	\bibitem [{\citenamefont {Leibfried}\ \emph {et~al.}(2003)\citenamefont
		{Leibfried}, \citenamefont {Blatt}, \citenamefont {Monroe},\ and\
		\citenamefont {Wineland}}]{Leibfried-RMP-2003}%
	\BibitemOpen
	\bibfield  {author} {\bibinfo {author} {\bibfnamefont {D.}~\bibnamefont
			{Leibfried}}, \bibinfo {author} {\bibfnamefont {R.}~\bibnamefont {Blatt}},
		\bibinfo {author} {\bibfnamefont {C.}~\bibnamefont {Monroe}},\ and\ \bibinfo
		{author} {\bibfnamefont {D.}~\bibnamefont {Wineland}},\ }\bibfield  {title}
	{\bibinfo {title} {{Quantum dynamics of single trapped ions}},\ }\href
	{https://doi.org/10.1103/RevModPhys.75.281} {\bibfield  {journal} {\bibinfo
			{journal} {Rev. Mod. Phys.}\ }\textbf {\bibinfo {volume} {75}},\ \bibinfo
		{pages} {281} (\bibinfo {year} {2003})}\BibitemShut {NoStop}%
	\bibitem [{\citenamefont {L{\"{o}}fgren}\ \emph {et~al.}(2023)\citenamefont
		{L{\"{o}}fgren}, \citenamefont {Fragoso}, \citenamefont {Weidow},\ and\
		\citenamefont {Enger}}]{Lofgren-PT-2023}%
	\BibitemOpen
	\bibfield  {author} {\bibinfo {author} {\bibfnamefont {S.~K.}\ \bibnamefont
			{L{\"{o}}fgren}}, \bibinfo {author} {\bibfnamefont {R.~M.}\ \bibnamefont
			{Fragoso}}, \bibinfo {author} {\bibfnamefont {J.}~\bibnamefont {Weidow}},\
		and\ \bibinfo {author} {\bibfnamefont {J.}~\bibnamefont {Enger}},\ }\bibfield
	{title} {\bibinfo {title} {{The Mechanical Paul Trap: Introducing the
				Concept of Ion Trapping}},\ }\href {https://doi.org/10.1119/5.0106359}
	{\bibfield  {journal} {\bibinfo  {journal} {Phys. Teach.}\ }\textbf {\bibinfo
			{volume} {61}},\ \bibinfo {pages} {762} (\bibinfo {year} {2023})}\BibitemShut
	{NoStop}%
	\bibitem [{\citenamefont {Leach}\ and\ \citenamefont
		{Andriopoulos}(2008)}]{Leach-AADM-2008}%
	\BibitemOpen
	\bibfield  {author} {\bibinfo {author} {\bibfnamefont {P.}~\bibnamefont
			{Leach}}\ and\ \bibinfo {author} {\bibfnamefont {S.}~\bibnamefont
			{Andriopoulos}},\ }\bibfield  {title} {\bibinfo {title} {{The Ermakov
				equation: A commentary}},\ }\href {https://doi.org/10.2298/AADM0802146L}
	{\bibfield  {journal} {\bibinfo  {journal} {Appl. Anal. Discret. Math.}\
		}\textbf {\bibinfo {volume} {2}},\ \bibinfo {pages} {146} (\bibinfo {year}
		{2008})}\BibitemShut {NoStop}%
	\bibitem [{\citenamefont {Fiore}(2025)}]{Fiore-JPA-2025}%
	\BibitemOpen
	\bibfield  {author} {\bibinfo {author} {\bibfnamefont {G.}~\bibnamefont
			{Fiore}},\ }\bibfield  {title} {\bibinfo {title} {{The time-dependent
				harmonic oscillator revisited}},\ }\href
	{https://doi.org/10.1088/1751-8121/adaab5} {\bibfield  {journal} {\bibinfo
			{journal} {J. Phys. A Math. Theor.}\ }\textbf {\bibinfo {volume} {58}},\
		\bibinfo {pages} {055202} (\bibinfo {year} {2025})}\BibitemShut {NoStop}%
	\bibitem [{\citenamefont {Dabrowski}\ and\ \citenamefont
		{Dunne}(2016)}]{Dabrowski-PRD-2016}%
	\BibitemOpen
	\bibfield  {author} {\bibinfo {author} {\bibfnamefont {R.}~\bibnamefont
			{Dabrowski}}\ and\ \bibinfo {author} {\bibfnamefont {G.~V.}\ \bibnamefont
			{Dunne}},\ }\bibfield  {title} {\bibinfo {title} {{Time dependence of
				adiabatic particle number}},\ }\href
	{https://doi.org/10.1103/PhysRevD.94.065005} {\bibfield  {journal} {\bibinfo
			{journal} {Phys. Rev. D}\ }\textbf {\bibinfo {volume} {94}},\ \bibinfo
		{pages} {065005} (\bibinfo {year} {2016})}\BibitemShut {NoStop}%
\end{thebibliography}

%

\end{document}